\documentclass[12pt]{article}
\pdfoutput=1
\usepackage{graphicx}
\usepackage{epstopdf}
\usepackage{amsmath}
\usepackage{amsfonts}
\usepackage{amssymb}
\usepackage{color}
\usepackage{mathrsfs}   

\usepackage[font={footnotesize}]{caption}

\pdfoutput=1

\usepackage[english]{babel}

\usepackage[colorlinks,bookmarks]{hyperref}
\definecolor{linkblue}{rgb}{0,0,0.8}
\definecolor{linkgreen}{rgb}{0,0.5,0}

\hypersetup{pdfpagemode=UseNone, pdfstartview=FitH, linkcolor=linkblue, %
            citecolor=linkgreen, urlcolor=linkblue}

\numberwithin{equation}{section}

\definecolor{darkgreen}{rgb}{0,0.3,0}
\definecolor{darkblue}{rgb}{0,0,0.3}
\definecolor{darkred}{rgb}{0.7,0,0}

\newcommand{\be}{\begin{equation}}
\newcommand{\bse}{\begin{subequations}}
\newcommand{\ese}{\end{subequations}}
\newcommand{\bea}{\begin{eqnarray}}
\newcommand{\eea}{\end{eqnarray}}
\newcommand{\ba}{\begin{array}}
\newcommand{\ea}{\end{array}}
\newcommand{\ee}{\end{equation}}
\newcommand{\nn}{\nonumber}

\newcommand{\mpl}{M_{\rm Pl}}

\newcommand{\gmn}{g_{\mu \nu}}

\newcommand{\css}{c_s^2}

\newcommand{\pd}{\partial}

\newcommand{\half}{\frac{1}{2}}
\newcommand{\eq}[1]{Eq. (\ref{#1})}

\newcommand{\motb}{\bar{m}_1^3}
\newcommand{\eqn}[1]{Eq.~(\ref{#1})}
\newcommand{\xvec}{\vec{x}}
\newcommand{\kvec}{\vec{k}}
\newcommand{\cH}{\mathcal{H}}
\newcommand{\fr}{\frac}
\newcommand{\cL}{\mathcal{L}}
\newcommand{\qvec}{\vec{q}}

\newcommand{\mm}{\bar m_1^3}
\newcommand{\M}{M_2^4}
\newcommand{\rhod}{\bar \rho_D}
\newcommand{\rhom}{\bar \rho_m}  
\newcommand{\rhomi}{ \rho_{m,0}}   
\newcommand{\rhodi}{ \rho_{D,0}}
\newcommand{\omegam}{\Omega_{m,0}}  
\newcommand{\omegad}{\Omega_{D,0}}
\newcommand{\deltaku}{\delta K_{\rm u}}
\newcommand{\knl}{k_{\rm NL}}
\newcommand{\la}{\langle} 
\newcommand{\ra}{\rangle} 
\newcommand{\n}{\nonumber \\}
\newcommand{\unitsk}{\, h { \rm Mpc^{-1}}}  % NEW
\newcommand{\deltagu}{\delta g^{00}_{\rm u}}

\newcommand{\tT}{\Theta}
\newcommand{\td}{\delta}

\newcommand{\ta}{\tilde{a}}
\newcommand{\tp}{\tilde{\pi}}

\newcommand{\mG}{\mathcal{G}}

\newcommand{\mU}{\mathcal{U}}
\newcommand{\mV}{\mathcal{V}}

\newcommand{\A}{\mathcal{A}}
\newcommand{\B}{\mathcal{B}}
\newcommand{\C}{\mathcal{C}}
\newcommand{\vk}{\vec{k}}
\newcommand{\vq}{\vec{q}}
\newcommand{\vp}{\vec{p}}
\newcommand{\vx}{\vec{x}}

\newcommand{\website}{\url{http://web.stanford.edu/~senatore/}}

\makeatletter \@addtoreset{equation}{section}

\makeatletter\renewcommand\section{\@startsection {section}{1}{\z@}%
                                   {-3.5ex \@plus -1ex \@minus -.2ex}%nn
                                   {2.3ex \@plus.2ex}%
                                   {\normalfont\large\bfseries}}
\renewcommand\subsection{\@startsection{subsection}{2}{\z@}%
                                     {-3.25ex\@plus -1ex \@minus -.2ex}%
                                     {1.5ex \@plus .2ex}%
                                     {\normalfont\bfseries}}

\begin{document}
%\begin{titlepage}
%\begin{flushright}\vspace{-3cm}
%{\small
%\today }\end{flushright}
%\vspace{-.5cm}

\begin{flushright}
%hep-th/yymmnnn\\
\end{flushright}

\begin{center}

{\LARGE{\bf An effective description of dark matter and dark energy in the mildly non-linear regime}}
\\[0.7cm]
{\large Matthew Lewandowski${}^{1,2,3}$, Azadeh Maleknejad${}^{4}$,  Leonardo Senatore${}^{1,2}$}
\\[0.7cm]
\vspace{.3cm}
{\normalsize { \sl $^{1}$ Stanford Institute for Theoretical Physics,\\ Stanford University, Stanford, CA 94306}}\\
\vspace{.3cm}

{\normalsize { \sl $^{2}$ Kavli Institute for Particle Astrophysics and Cosmology, \\
Physics Department and SLAC, Menlo Park, CA 94025}}\\
\vspace{.3cm}

{\normalsize { \sl $^{3}$  Institut de Physique Th\'eorique, Universit\'e Paris Saclay,\\ CEA, CNRS, 91191 Gif-sur-Yvette, France}}\\

\vspace{.3cm}
{\normalsize { \sl $^{4}$ School of Physics, Institute for Research in Fundamental Sciences (IPM), \\
P. Code. 19538-33511, Tehran, Iran}}\\
\vspace{.3cm}

%\date{\today}

\end{center}

\vspace{8mm}
\setcounter{page}{1} \baselineskip=15.5pt \thispagestyle{empty}

\hrule \vspace{0.3cm}
{\small  \noindent \textbf{Abstract} \\[0.3cm]
\noindent  
In the next few years, we are going to probe the low-redshift universe with unprecedented accuracy. Among the various fruits that this will bear, it will greatly improve our knowledge of the dynamics of dark energy, though for this there is a strong theoretical preference for a cosmological constant.  We assume that dark energy is described by the so-called Effective Field Theory of Dark Energy, which assumes that dark energy is the Goldstone boson of time translations. Such a formalism makes it easy to ensure that our signatures are consistent with well-established principles of physics. Since most of the information resides at high wavenumbers, it is important to be able to make predictions at the highest wavenumber that is possible. The Effective Field Theory of Large-Scale Structure (EFTofLSS) is a theoretical framework that has allowed us to make accurate predictions in the mildly non-linear regime. In this paper, we derive the non-linear equations that extend the EFTofLSS to include the effect of dark energy both on the matter fields and on the biased tracers. For the specific case of clustering quintessence, we then perturbatively solve to cubic order the resulting non-linear equations and construct the one-loop power spectrum of the total density contrast.

 \vspace{0.3cm}
\hrule

 \vspace{0.3cm}

%\end{titlepage}
\newpage

\tableofcontents

%%%%%%%%%%%%%%%%%%%%%%%%%%%%%%%
%
%         introduction
%
%%%%%%%%%%%%%%%%%%%%%%%%%%%%%
\section{Introduction}\label{Inflation-section}
One of the most unexpected discoveries of modern cosmology is the observation of the accelerated expansion of the Universe in 1998. It had been first observed by supernovae Ia (SnIa) surveys \cite{Riess:1998cb, Perlmutter:1998np, Riess:1998dv} and then it was confirmed by other observations including large-scale structure (LSS) \cite{Tegmark:2003ud, Tegmark:2006az}, cosmic microwave background (CMB) \cite{Spergel:2003cb, Ade:2015rim} and baryon acoustic oscillations (BAO) \cite{Percival:2007yw, Aubourg:2014yra} that about 70$\%$ of the Universe today is made of an unknown component called dark energy (DE). Concerning the background evolution, current observations restrict the value of the equation-of-state of DE to be very close to $-1$ at low redshifts, while present constraints on the time evolution of $w$ and DE energy density at higher redshifts are still very weak \cite{Ade:2015rim}. Even less constrained is the behavior of the fluctuations in DE. 

Contrary to the case of inflation, it is {\it relatively} easy to make progress in our observational knowledge of dark energy with respect to the one of inflation. In fact, phenomena that left significant signatures in the early universe have already been exposed to being probed by the CMB. This has provided very accurate measurements in the last three decades of the universe at the recombination epoch, significantly constraining all processes that affected that epoch (including the initial conditions for the fluctuations). It is expected that the CMB will make further progress in the measurement of the polarization, but most luckily the largest gain in information will be associated to measurements of large-scale structure {\it through} the CMB. Since dark energy is mainly important at low redshifts, where our knowledge is much less accurate than at higher redshifts,\footnote{Though there can be constraints originating from the behavior of dark energy at high redshift, which is, however, model dependent~\cite{Ade:2015rim}. Of course, when this is the case, the high redshift measurements provide a very strong constraint.} in the next few years our improvement has the chance to be quite spectacular. In fact, a number of upcoming probes, both through CMB and Large-Scale Structure surveys, will improve our knowledge of the low-redshift universe.  Among them are the space missions Euclid \cite{Amendola:2016saw} and Wide Field Infrared Survey Telescope (WFIRST) \cite{Spergel:2015sza} as well as ground-based experiments such as the Dark Energy Spectroscopic Instrument (DESI), the Large Synoptic Survey Telescope (LSST), the Ground-Based Stage IV BAO Experiment (BigBOSS) and the Hobby-Eberly Telescope Dark Energy Experiment (HETDEX). Also, CMB probes will keep measuring with greater and greater accuracy the LSS through the induced lensing on the CMB (see for example~\cite{Das:2013zf, vanEngelen:2012va}). For some exhaustive reviews on the subject see \cite{Weinberg:2012es}, \cite{Joyce:2016vqv} and the references within.

Let us now pass to the theoretical side. By a very very large amount, an amount that it is difficult to overstate, the current preferred model for dark energy is a cosmological constant. In fact, the cosmological constant is already part of our laws of physics. We just do not know its size. Historically, it has been extremely difficult to tame the large quantum corrections that affect its size, and that are expected to make it huge. However, a beautiful possible explanation of its smallness was provided by Weinberg in 1987~\cite{Weinberg:1987dv}, based on anthropic reasoning. Weinberg reflected that if the cosmological constant were to be very large, then it would dominate the energy density of the universe before any structure such as planets could have formed. In such a universe, there would be no observers to measure a large value of a cosmological constant. Therefore, Weinberg inferred that if we live in a multiverse where the cosmological constant can take several different values, our observed value of the cosmological constant must be below an upper bound so that structures could have formed. Furthermore, he argued that, given that it is famously hard to make the cosmological constant small, the observed value will most likely be close to the upper bound given by the requirement of structure formation. Subsequently, the landscape of string theory and the inflationary paradigm have provided a natural theoretical framework of this anthropic explanation.

Weinberg's line of reasoning predicted that our universe should be currently accelerating with $w$ close to $-1$, driven by a non-fluctuating component which is a cosmological constant with a certain value. Within the uncertainties of this theoretical argument, these predictions were matched by the observations made one decade later and the subsequent ones.

So, if Weinberg's explanation is so compelling, why do we not declare dark energy to be a cosmological constant? While many authors would agree in doing this, including at least one of the authors of this paper, it is true that in the next few years we are going to make such a tremendous observational progress that it is worth giving a further look at the problem, both observationally and theoretically. Of course, while we do this, we have the chance of making discoveries even greater than the one of the cosmological constant.

A quite general approach is to assume that the current acceleration of the universe is associated to the presence of a new light degree of freedom, called DE. It should be stressed that this hypothesis does not necessarily imply that the universe is accelerating, nor offers automatically an explanation of the smallness of the cosmological constant. However, it is conceivable, at least as a matter of principle, that the presence of this degree of freedom is associated to the acceleration of the universe, and since we are going to test this hypothesis with unprecedented precision, then it is worth studying this hypothesis.

We will parametrize the generic signatures of DE by assuming that this light degree of freedom is associated to the breaking of time translations, which is quite a general phenomenon in an FRW universe. In this case, the new degree of freedom is the Goldstone boson of time translations, whose action can be constructed without specific knowledge of the dynamics that leads to the onset of the background cosmology. This approach to describe dark energy is called `The Effective Field Theory of Dark Energy' (EFTofDE) and was originally developed in~\cite{Creminelli:2006xe}. Then, it was further developed and applied to describe Inflation in~\cite{Cheung:2007st} and then further developed in the context of dark energy (where the name effective field theory of dark energy was actually introduced, and the research program on the phenomenology of dark energy more systematically initiated) in~\cite{Creminelli:2008wc,Gubitosi:2012hu,Gleyzes:2013ooa} and a large subsequent literature (see for example,~\cite{Bloomfield:2012ff, Gleyzes:2014rba, Gleyzes:2014qga, Hu:2013twa, Silvestri:2013ne, Piazza:2013pua,Perenon:2015sla,Frusciante:2016xoj}). 

This approach to describe dark energy has the advantage of being very general. Maybe even more important is the fact that the signatures derive from a Lagrangian. This simple fact guarantees us that the system respects our generally accepted principle of physics, such as locality, causality, unitarity, etc.\footnote{Not all values of the parameters of the EFTofDE are allowed by these same principles, as for example, some values can lead to non-analyticity of the $S$-matrix or to superluminal propagation~\cite{Adams:2006sv}. For a discussion about some of the constraints on the parameters of the EFTofDE imposed by these issues, see \cite{Perenon:2015sla} and \cite{Frusciante:2016xoj}.} This is the main difference between using a formalism such as the EFTofDE, versus some more phenomenological approaches, that parametrize the equation of state, $\delta P/\delta\rho$, the difference between the gravitational potentials, $\Phi-\Psi$, and the modifications of the Poisson equation, in some general form. The latter approach runs the uncontrollable risk of including regimes that are incompatible with the currently accepted principles of physics.

Observationally, since the number of modes is dominated by the shortest wavenumbers, most of the information about dark energy (and pretty much everything else), will be stored at those wavenumbers where the non-linearities of the LSS will be sizable. This makes it important to be able to describe the mildly non-linear regime both for dark matter and dark energy. 

In the last few years, remarkable progress has occurred in our capability to describe the quasi-linear clustering of large-scale structures in the absence of dark energy, through the introduction of the so-called Effective Field Theory of Large-Scale Structure (EFTofLSS)~\cite{Baumann:2010tm,Carrasco:2012cv,Porto:2013qua,Senatore:2014via}. The availability of a satisfactory analytical treatment for large-scale structure has been delayed for about three decades because of the difficulty in dealing with the strong non-linearities at short distances that affect long wavelength perturbations. Since short distance fluctuations are not under perturbative control, it appeared that it was naively impossible to parametrize their effect at long distances. Instead, in the EFTofLSS, such an effect is accurately accounted for by the inclusion of suitable counterterms, that, after the related coupling constants are fitted to observations, can correctly include the effect of short distance fluctuations at long distances. In recent years, a large activity has occurred in this small field, see for example~\cite{Baumann:2010tm,Carrasco:2012cv,Porto:2013qua,Senatore:2014via,Carrasco:2013sva,Carrasco:2013mua,Pajer:2013jj,Carroll:2013oxa,Mercolli:2013bsa,Angulo:2014tfa,Baldauf:2014qfa,Senatore:2014eva,Senatore:2014vja,Lewandowski:2014rca,Mirbabayi:2014zca,Foreman:2015uva,Angulo:2015eqa,McQuinn:2015tva,Assassi:2015jqa,Baldauf:2015tla,Baldauf:2015xfa,Foreman:2015lca,Baldauf:2015aha,Baldauf:2015zga,Bertolini:2015fya,Bertolini:2016bmt,Assassi:2015fma,Lewandowski:2015ziq,Cataneo:2016suz,Bertolini:2016hxg,Fujita:2016dne,Perko:2016puo}. This community, as we review later, has been able to satisfactory show that the clustering of large-scale structures can be reproduced  with great accuracy both for dark matter, galaxies, and in redshift space up to relatively high wavenumbers.

The purpose of this paper will be to develop a formalism that allows us to treat the mildly non-linear dynamics of large-scale structure in the presence of dark energy. We will achieve this by extending the EFTofLSS to include the presence of an additional species, dark energy, whose dynamics is described by the EFTofDE. For simplicity, we will focus on some specific choices of parameters of the EFTofDE, which amounts to studying the so-called clustering quintessence, though our methods are straightforwardly extendable to other choices of parameters, that allow one, for example, to describe the so-called Horndeski models\footnote{Horndeski models are the most generic scalar-tensor theories, universally coupled to gravity, with second-order equations of motion. } and other models of modified gravity. After formulating the set of coupled non-linear equations, including the relevant counterterms, we will compute the power spectrum of the total density at one-loop order.  Throughout the paper, we will use the notation $\partial^2 = \sum_{i =1}^3 \partial_i \partial_i$, $\dot F = d F / d t $ and $F' = d F / d a$.

% \input{notation_v21}

%%%%%%%%%%%%%%%%%%
%
%       Section 2
%
%%%%%%%%%%%%%%%%%%%%%

\section{Review of EFTs of Dark Energy and Large-Scale Structure}\label{EFT-DE-section}

\subsection{Effective Field Theory of Dark Energy}

In this subsection, we review the effective field theory of dark energy developed in \cite{Creminelli:2006xe}, which was applied to inflation in \cite{Cheung:2007st} and further developed for dark energy in \cite{Gubitosi:2012hu}.  The basic idea is that we would like to describe the most general low-energy theory of fluctuations around a time-dependent background solution, which necessarily spontaneously breaks time diffeomorphisms by providing a preferred time slicing of space-time.  In the context of inflation, such a scenario is highly motivated because inflation must end and be smoothly connected to a hot big bang phase. The time slicing in this case is usually, but not necessarily, achieved by the evolution of a scalar field $\bar \phi ( t)$ which acts as the clock for the system.  Because of the new field $\phi ( \xvec , t )$, the system now has, in addition to gravitational degrees of freedom, an additional scalar degree of freedom $\delta \phi ( \xvec , t ) = \phi ( \xvec , t ) - \bar \phi ( t)$ which describes the fluctuations around the background solution.  Although time diffeomorphisms  $t \rightarrow t + \xi^0 ( \xvec, t)$ are not realized linearly on $\delta \phi$, they are realized non-linearly, through $\delta \phi~\rightarrow~\delta \phi - \dot{ \bar \phi} \xi^0$, because the original theory was invariant.  

Unitary gauge is the one in which we choose the time coordinate such that $\delta \phi ( \xvec , t ) =0$ on the constant time surfaces, and the scalar degree of freedom appears in the metric.  One then has a theory of three degrees of freedom, the two standard ones from the metric and the new scalar which in unitary gauge appears in the metric as well.  To build the most general theory in this gauge, we write all of the operators in terms of the metric that are invariant under the remaining time-dependent spatial diffeomorphisms $x^i \rightarrow x^i + \xi^i ( \xvec , t)$, but that do not have to be invariant under time diffeomorphisms.  The new scalar degree of freedom can then be introduced by performing a broken time diffeomorphism on this action via the St\"{u}ckelberg trick: $t \rightarrow \tilde t = t + \xi^0 ( \xvec, t)$.  Then we make the replacement $\xi^0 (x ( \tilde x)) \rightarrow - \tilde \pi ( \tilde x)$, where $\pi$ is the Goldstone boson that non-linearly realizes the time diffeomorphism symmetry, which is restored if $\pi$ transforms like $\pi ( \xvec , t ) \rightarrow \pi( \xvec, t) - \xi^0( \xvec , t)$.

The situation is similar for dark energy where we consider a general Friedmann-Robertson-Walker (FRW) background close to de Sitter.  We know that the universe is close to $\Lambda$CDM, which has a constant cosmological constant $\Lambda$ and is fully diffeomorphism invariant, so it makes sense to describe deviations from $\Lambda$CDM by assuming that time diffeomorphisms are spontaneously broken.  As in the inflationary case, in this case there will be a Goldstone mode related to this symmetry breaking.  The main new ingredient with respect to inflation is that this theory is coupled to matter.  To get the most general theory, we write the actions for the metric and matter in unitary gauge, and we allow the inclusion of operators that break time diffeomorphism invariance, but are invariant under time-dependent spatial diffeomorphisms.  This allows the inclusion in the action of $n_\mu$, the unit normal to equal time hypersurfaces, and covariant derivatives of $n_\mu$.  This implies that the gravitational action, $S_G$, can depend on gauge invariant operators like the cosmological constant and contractions of the Reimann tensor, and can also depend on operators which break time diffeomorphisms, like a time-dependent cosmological constant (and other time-dependent couplings), $g^{00}$ (or any other $4$-dimensional tensor with upper $0$ indices), and $K_{ij}$ (the extrinsic curvature of equal-time slices).  For a more complete discussion of the fields which break time diffeomorphisms, see Section \ref{pert-DE} and Appendix \ref{Gold-appendix}.  Thus, the gravitational action has the following form \cite{Creminelli:2006xe, Cheung:2007st, Gubitosi:2012hu}
\be \label{sgexamp}
S_G = \int d^4 x \sqrt{-g} \,  F_G \left( R_{\mu \nu \rho \sigma} , g^{00} , K_{ij} , \nabla_\mu ; t \right) \ . 
\ee
The matter action, $S_M$, can also in principle depend on all of the aforementioned fields and the matter fields, $\chi_a$, coupled in such a way that allows operators which break time diffeomorphisms.  Thus, the generic form is  (see \cite{Senatore:2010wk} for example)
\be
S_M = \int d^4 x \sqrt{-g} \, F_M \left(R_{\mu \nu \rho \sigma} , g^{00} , K_{ij} , \nabla_\mu , \chi_a ; t \right), \ 
\ee
with the same rule that for any covariant object, it is allowed to appear with an upper $0$ index.  For example, one can generically have couplings like $( g^{00})^2 \chi_a^2$ and $\partial^0 \chi_a \, \partial^0 \chi_a$ in $F_M$.  From the unitary gauge action, one can introduce $\pi$, the Goldstone mode related to the broken time diffeomorphisms, in the standard way using the St\"{u}ckelberg trick.

In this work, where we concentrate on correctly joining the EFT of dark matter and the EFT of dark energy, we choose a simplified setup for illustration purposes.  We assume the existence of a frame, called the Jordan frame, where each matter species is minimally coupled to the same metric.  In addition to simplifying our computations below, this assumption also ensures that the weak equivalence principle (WEP) holds (since all matter follows geodesics of the same metric).  This is not a necessary assumption, and our results can be extended to WEP violating theories, but experiments strongly constrain the amount of WEP violation (see for example \cite{Schlamminger:2007ht}).  Then, the action in the Jordan frame in unitary gauge reads 
\be \label{generalaction}
S = S_{G}[g_{\mu\nu}]+ S_M [g_{\mu \nu} , \chi_a],
\ee
where $S_G$ is as in \eq{sgexamp}, but $S_M$ is fully diffeomorphism invariant.  Thus, when the Goldstone mode $\pi$ is introduced, there is no direct coupling between $\pi$ and the matter sector.
More specifically, we will consider the example 
\be\label{action}
S_G = \int d^4 x \sqrt{-g}\bigg[ \frac{\mpl^2}{2} f(t) R - \Lambda(t) - c(t) g_{\rm u}^{00} \bigg] + S_{DE}^{(2)}  \,,
\ee
where $S_{DE}^{(2)}$ only contains terms quadratic and higher in the perturbations, so that the other operators shown are the only ones containing linear perturbations. In this paper, for convenience, we choose the following form for $S_{DE}^{(2)}$
\be
S_{DE}^{(2)} = \int d^4 x \sqrt{-g} \bigg[ \frac{M_2^4(t)}{2} ( \delta g_{\rm u}^{00} )^2 - \frac{\bar{m}_1^3(t) }{2 } \delta g_{\rm u}^{00} \delta K_{\rm u} \bigg] \ , 
\ee
and in fact we will take $f(t) = 1$ for simplicity.  In the above equation, $\delta g^{00}_{\rm u} = 1 + g^{00}_{\rm u}$ and $\delta K_{\rm u}$ is the variation of the trace of the extrinsic curvature \cite{Creminelli:2006xe, Cheung:2007sv, Gubitosi:2012hu} (for which we present a detailed computation to second order in Appendix \ref{Gold-appendix}).  The two functions $c(t)$ and $\Lambda(t)$ are chosen to fix the background equations, i.e. to eliminate the tadpole terms, and $f(t)$, $M_2^4(t)$ and $\mm(t)$ encode the different theories of fluctuations in this particular setup.  The $``\rm{u}$'' subscript on the operators that break time diffeomorphisms emphasize the fact that they are presented in unitary gauge, where the new scalar degree of freedom is contained in the metric.  Later in \ref{pert-DE}, by means of the St\"{u}ckelberg trick, we restore the diffeomorphism and re-introduce the scalar field fluctuations.  In this paper, we will study the linear equations for both of the operators $( \deltagu)^2$ and $ \deltagu \, \delta K_{\rm u} $, which we present in Appendix \ref{lineareqs}, and we will study the non-linear system with $\bar m_1^3 = 0$ in the rest of the text.

%%%%%%%%%%%%%%%%%%%%%%%%%%%%%%%%%%%%%%

\subsubsection{Background equations}
The matter in our theory is cold dark matter (CDM), and so its background equations are described by a time dependent energy density $\bar \rho_m(t)$ and a pressure $\bar p_m(t)$.  Then, the zeroth order Einstein equations (Friedman equations) for the background FRW metric are 
\begin{align}
c(t) &  = - \dot H \mpl^2 - \half \left( \rhom + \bar p_m \right) \label{cequation}\\
\Lambda(t) & = ( \dot H + 3 H^2 ) \mpl^2  - \half \left( \rhom - \bar p_m \right) \label{lambdaequation}\ .
\end{align}
Instead of using $c(t)$ and $\Lambda(t)$ to describe the background, it is useful to change to two new functions $\bar \rho_D(t)$ and $\bar p_D (t)$ such that 
\begin{align} \label{cequation1}
c(t) & = \half \left( \bar{\rho}_D + \bar{p}_D \right), \\
\Lambda(t)& = \half \left( \bar{\rho}_D - \bar{p}_D \right) \label{lambdaequation1}\ , 
\end{align}
after which \eqn{cequation} and \eqn{lambdaequation} become 
\begin{align}
- 2 \dot H \mpl^2  & = \rhom + \bar p_m + \bar \rho_D + \bar p_D \\
3 H^2 \mpl^2 & = \bar \rho_m + \bar \rho_D  \label{anotherfriedman}\ .
\end{align}
These are the Friedman equations in a much more recognizable form, written in terms of the background dark-energy energy density $\bar \rho_D (t)$ and pressure $\bar p_D ( t)$.  In order to describe normal cold dark matter, we assume the background continuity equation $\dot{\bar \rho}_m + 3 H ( \bar \rho_m + \bar p_m ) = 0$.  For the rest of this paper, we will take $\bar p_m = 0$ because we are describing CDM.   Thus we have 
\be
\rhom = \rhomi \left( \frac{a}{a_0} \right)^{-3} \ . 
\ee
Hereafter, the subscript $0$ denotes the present time value.  Then, taking the time derivative of \eqn{anotherfriedman}, we find that $\dot{ \bar \rho}_D + 3 H ( \rhod + \bar p_D) = 0$.  In this work, for simplicity, we consider a dark-energy component whose background is described by a constant equation of state $\bar p_D = w \rhod$.  This gives a background solution 
\be
\rhod = \rhodi \left( \frac{a}{a_0} \right)^{-3  (1+w)} \ . 
\ee
It is also useful to write the Friedman equation as
\begin{align}
\frac{ H^2}{ H_0^2}&= \Omega_{m,0} \left( \frac{a}{a_0} \right)^{-3} + \Omega_{D,0} \left( \frac{a}{a_0} \right)^{-3(1+w)} \ ,   \label{mfried} 
\end{align}
where $\omegam \equiv \frac{ \rhomi}{\rhomi + \rhodi}$ and $\omegad = \frac{ \rhodi }{ \rhomi + \rhodi}$ are the current day energy density fractions of CDM and dark energy, respectively.

%%%%%%%%%%%%%%%%%%%%%%%%%%%%%%%%%%%%%%%%%%%

\subsubsection{Perturbations in the dark-energy sector}\label{pert-DE}

With the action in unitary gauge, it is useful to introduce the Goldstone mode $\pi$ using the St\"{u}ckelberg trick.  In order to do that, we perform a time diffeomorphism $ x^0 \rightarrow x^0 + \xi^0(\xvec , t)$ and $x^i \rightarrow x^i$ on the action \eqn{generalaction}.  Under a general diffeomorphism $x^\mu \rightarrow \tilde x^\mu ( x )$, the metric changes in the standard way 
\be
\tilde g^{\mu \nu} ( \tilde x ( x) ) =  \frac{ \partial \tilde x^\mu}{\partial x^\sigma} \frac{ \partial \tilde x^\nu }{\partial x^\rho} g^{\sigma \rho} ( x ) \ , 
\ee
which means that, following \cite{Cheung:2007st}, after changing variables of integration in the action, replacing $\xi^0 ( x ( \tilde x) ) \rightarrow - \tilde \pi ( \tilde x)$, and then dropping all of the tildes, we should make the replacements 
\begin{align}
g^{\mu \nu}(x)  \rightarrow P^{\mu}{}_\rho P^\nu{}_\sigma \, g^{\rho \sigma} (x)  \hspace{ .5in}  g_{\mu \nu} (x) \rightarrow P^{-1 }{}^\rho{}_\mu  P^{-1}{}^\sigma{}_\nu\, g_{\rho \sigma} (x) \ ,
\end{align}
in the action, where the transformation matrices are given by  
\be
P^\mu{}_\rho = \begin{pmatrix} 1 + \dot \pi & \partial_i \pi \\ 0 & 1 \end{pmatrix}_{\mu \rho} \hspace{.5in} P^{-1}{}^\rho{}_\mu = \begin{pmatrix} \frac{1}{1 + \dot \pi} & - \frac{\partial_i \pi}{1 + \dot \pi} \\ 0 & 1 \end{pmatrix}_{ \rho \mu} \ ,
\ee
and all of the $\pi$ fields are evaluated at the point $x$.  The arguments of the time dependent coefficients in the action, like $f(t)$, $c(t)$, $\Lambda(t)$, $\M(t)$, and $\mm(t)$, shift like
\be\label{c-expand}
c(t) \rightarrow c(t + \pi) = c(t) + \dot c(t) \pi + \half \ddot c(t) \pi^2 + \dots \ .
\ee
 Finally, the replacement rule for derivatives is 
\be
\partial_\mu \rightarrow P^{-1}{}^\rho{}_\mu \,  \partial_\rho \ . 
\ee
Some specific examples that we will need are
\begin{align}\label{metric-expand}
 g^{00}_{\rm u}&  \rightarrow P^0{}_\mu P^0{}_\nu  g^{\mu \nu}  = g^{00} + 2 g^{0 \mu } \partial_\mu \pi + g^{\mu \nu} \partial_\mu \pi \partial_\nu \pi \\
 g^{0i}_{\rm u} & \rightarrow P^0{}_\mu P^i{}_\nu  g^{\mu \nu} = g^{0i} + g^{i \mu } \partial_\mu \pi \\
 g^{ij}_{\rm u} & \rightarrow g^{ij} \\
\delta K_{\rm u} & \rightarrow  \delta K_g - a^{-2} \partial^2 \pi - 3 \dot H \pi  \  ,
\end{align}
where $\delta K_g$ depends only on the metric.  The first three expressions above are fully expanded in terms of $\pi$, but the metrics appearing can still be expanded in perturbations.  In the last line, we have only presented our expression for $\delta K_{\rm u}$ to linear order because these are the most important terms.  In Appendix \ref{Gold-appendix}, we present a detailed computation of $\delta K_{\rm u}$, including a discussion of higher order terms, and thus extend the computations done in \cite{Creminelli:2006xe, Cheung:2007sv, Gubitosi:2012hu}.

\subsection{Effective Field Theory of Large-Scale Structure} \label{eftreviewsection}

We now review the other main ingredient of our study, The Effective Field Theory of Large-Scale Structure,\footnote{Formerly known as the Effective Field Theory of Large Scale Structures.} which describes the dynamics of collisionless dark matter on large scales in the $\Lambda$CDM universe.  The EFTofLSS community has studied the dark matter density two-point function~\cite{Carrasco:2012cv,Senatore:2014via,Carrasco:2013mua,Foreman:2015lca,Baldauf:2015aha}, three-point function~\cite{Angulo:2014tfa,Baldauf:2014qfa}, four-point function~\cite{Bertolini:2015fya,Bertolini:2016bmt}, the dark matter momentum power spectrum~\cite{Senatore:2014via,Baldauf:2015aha},  the displacement field~\cite{Baldauf:2014qfa}, and the vorticity slope~\cite{Carrasco:2013mua,Hahn:2014lca}.  Additionally, baryonic effects on the matter correlation functions have been described within the EFTofLSS in~\cite{Lewandowski:2014rca}.  The extension of the EFTofLSS to describe biased tracers has been performed in~\cite{Senatore:2014vja}, and predictions have been compared to data for the power spectrum and bispectrum (including all mixed correlation functions between matter and halos) in~\cite{Angulo:2015eqa,Fujita:2016dne}.  The EFTofLSS was used to describe redshift-space distortions in~\cite{Senatore:2014vja}, and predictions have been compared to numerical data for matter power spectra in~\cite{Lewandowski:2015ziq}.  Methods to measure the parameters of the EFTofLSS in small numerical simulations have been developed in~\cite{Carrasco:2012cv,Baldauf:2011bh,Baldauf:2015vio,Lazeyras:2015lgp,Bertolini:2016hxg}.  The IR-resummation was implemented and compared to numerical data of dark matter clustering in~\cite{Senatore:2014via}, extended to halos in~\cite{Senatore:2014vja} and compared to halo datasets in~\cite{Angulo:2015eqa}, recently extended to dark matter in redshift space and compared to simulated datasets in~\cite{Senatore:2014vja,Lewandowski:2015ziq}, and finally extended to halos in redshift space in~\cite{Perko:2016puo}.  The signature of primordial non-Gaussianity on large-scale structure observables~\cite{Angulo:2015eqa,Assassi:2015jqa,Assassi:2015fma,Lewandowski:2015ziq} has also been recently included.  Recently, fast implementations of the predictions of the EFTofLSS, which allows us to efficiently explore their dependence on various cosmological parameters, have been developed in~\cite{Cataneo:2016suz}, with public codes available at the following website\footnote{\label{website}\website} (including the Mathematica notebook used in this paper).

 In the rest of this section, we briefly review some of the results and findings of the EFTofLSS in $\Lambda$CDM.  The relevant long wavelength degrees of freedom are the overdensity $\delta_m ( \xvec , t) \equiv ( \rho_m ( \xvec , t) - \rhom(t) ) / \rhom(t)$ and the velocity divergence $\theta_m (\xvec , t ) \equiv \partial_i v^i_m ( \xvec , t)$.\footnote{Vorticity is generated in the EFTofLSS at a high order, but it can be ignored for the one-loop discussion that we present here \cite{Carrasco:2013mua}.}  After integrating out the effects of short scale (UV) physics below some non-linear wavenumber scale $\knl$, the equations for the long wavelength fields take the form
\begin{align} \label{fluid}
&\dot{\delta}_m + \frac{1}{a} \partial_i ( (1 + \delta_m)v^i_m)  =0 \\
\label{fluid2} &\partial_i \dot{v}^i_m +  H \partial_i v^i_m + \frac{1}{a}  \partial_i ( v^j_m \partial_j v^i_m ) + \frac{1}{a}   \partial^2 \Phi = - \frac{1}{a} \partial_i \left( \frac{1}{\rho_m} \partial_j \tau^{ij} \right)_s  \\
& a^{-2} \partial^2 \Phi  = \frac{3}{2 } \frac{ \Omega_{m,0} \cH_0^2 a_0}{a^3} \delta_m \label{poissonfish}\ . 
\end{align}
The effects of UV physics on long distances are encoded in the effective stress tensor $\left( \frac{1}{\rho_m} \partial_j \tau^{ij} \right)_s$, which depends on short modes.  Since we cannot describe the short modes exactly, we expand the stress tensor in powers and derivatives of the long wavelength fields, and we include all operators, called counterterms, that are consistent with the equivalence principle.  As has been discussed \cite{Senatore:2014eva, Carrasco:2013mua, Carroll:2013oxa}, the EFTofLSS is non-local in time.  This means that, after taking the expectation value over the short modes in the background of the long modes, the effective stress tensor can be written as an integral over some unknown kernel of time of an expansion in powers and derivatives of $\partial_i \partial_j \Phi$ and $\partial_i v^j_m$, evaluated along the fluid line element.  The lowest order terms in this expansion are
\begin{align}
-  \left( \frac{1}{\rho_m} \partial_j \tau^{ij} \right)_s ( a , \xvec)  & =   
\int d a' \Bigg[\kappa^{(1)}( a , a' )\,    \partial^i \partial^2 \Phi ( a' , \xvec_{\rm fl}( \xvec ; a,a') )   \nonumber \\
&  \hspace{1in} +\kappa^{(2)}( a , a' )\, \fr{1}{H}\partial^i \partial_j v^j_m ( a' , \xvec_{\rm fl}( \xvec ; a,a') )   \nonumber \\
& \hspace{1in} + \kappa^{(\rm stoch.)} ( a , a' ) \partial^i \bar \Delta_{\rm stoch.} ( a ' , \xvec_{\rm fl}( \xvec ; a,a') )
+ \dots \Bigg] \ ,
\label{str}
\end{align}
where the various $\kappa( a , a') $ are the kernels encoding UV physics, $\bar \Delta_{\rm stoch.}$ is the stochastic counterterm not proportional to the long wavelength fields (which we will ignore in this paper, but is explained in more detail below), the fluid line element $\xvec_{\rm fl}$ is defined implicitly as~\cite{Carrasco:2013mua} 
\be
\xvec_{\rm fl} ( \xvec ; a , a' ) = \xvec - \int_{a'}^a d a''  \;\frac{d \tau }{d a} ( a'' ) \; \vec{v}_m ( a'' , \xvec_{\rm fl} ( \xvec ; a , a'') )\ ,
\ee 
and $\tau$ is conformal time.  The terms associated with the past trajectory, i.e. expanding in $\xvec_{\rm fl}$,  appear at higher orders in the expansion, and we ignore them in this study.  As discussed in~\cite{Carrasco:2013mua}, the non-locality in time can be written such that the counterterms appear as local-in-time.  For example, using the Poisson equation \eqn{poissonfish} and the fact that the linear solution is  $ \delta_m^{(1)} ( a , \xvec ) = D(a) \delta_m^{(1)} ( a_i , \xvec ) / D(a_i)$, we can write 
\begin{align}
\int d a' \kappa^{(1)}( a , a' )\,    \partial^i \partial^2 \Phi^{(1)} ( a' , \xvec ) = \left(  \int d a' \tilde \kappa^{(1)} ( a , a')  \frac{D(a')}{D(a)} \right) \partial^i \delta^{(1)}_m ( a , \xvec)  \label{rewrite1}\ , 
\end{align}
where $\kappa^{(1)}$ and $\tilde \kappa^{(1)}$ are related by the factors in the Poisson equation.  We can then define the local-in-time speed-of-sound parameters by symbolically performing the $a'$ integral, thus leaving us with an unknown function of one variable $a$.  In fact, as a function of the fields, \eqn{rewrite1} is the generic form for a counterterm at one loop, since the term proportional to $\kappa^{(2)}$ in \eqn{str} can be written as proportional to $\partial^i \delta_m^{(1)}$ by using \eqn{fluid}, where the integrand involves some rescaled $\tilde \kappa^{(2)}$ as in \eqn{rewrite1}.  Putting this all together gives us the final expression for the stress tensor at this order
\be
-   \left( \frac{1}{\rho_m} \partial_j \tau^{ij} \right)_s ( a , \xvec) \sim  \left( \int d a' K (a , a' ) \frac{D(a')}{D(a)}  \right)  \partial^i \delta_m^{(1)} ( a , \xvec)  \ ,
 \ee
where we neglect a factor of $e^{i \kvec \cdot \left( \xvec_{\rm fl} - \xvec \right)}\simeq 1$ at the order that we work, and $K = \tilde \kappa^{(1)} + \tilde \kappa^{(2)}$.  In Fourier space, using the conventions $F(\xvec) = \int \frac{d^3 k}{(2 \pi)^3}  e^{-i \kvec \cdot \xvec} F ( \kvec  )$, and switching to the scale factor $a$ as the time variable, we finally have 
\begin{align} \label{nlcontinuity}
& a \cH  \delta_m ( a, \kvec  )'  +  \theta_m ( a ,  \kvec )  = -  \int \frac{d^3 q}{(2 \pi)^3}  \, \alpha ( \vq,\vk-\vq)  \theta_m( a ,  \vq  ) \delta_m(a , \vk-\vq) \\
& a \cH  \theta_m ( a , \kvec  ) '  + \cH \theta_m ( a , \kvec  ) + \frac{3}{2} \frac{ \Omega_{m,0} \cH_0^2 a_0 }{a}  \delta_m ( a ,  \kvec  )  =   9 \, ( 2 \pi) \, c_{s,m}^2 (a) H(a)^2 \frac{k^2}{\knl^2} \delta_m ( a , \kvec  )   \nn \\
& \hspace{2in} -  \int \frac{d^3 q}{(2 \pi)^3} \beta(\vq ,  \vk-\vq ) \theta_m ( a , \vk-\vq  ) \theta_m ( a , \vq )   \label{nleuler}
\end{align}    
where
\begin{align} 
\alpha ( \qvec_1 , \qvec_2 ) & = 1 + \frac{\qvec_1 \cdot \qvec_2}{q_1^2} \label{alphadef} \\
\beta( \qvec_1 , \qvec_2 ) & = \frac{ | \qvec_1 + \qvec_2 |^2 \qvec_1 \cdot \qvec_2}{2 q_1^2 q_2^2}    \label{betadef} \ ,
 \end{align}
$\cH = a H$, and we have included the one-loop counterterm, proportional to $\left( k / \knl \right)^2$.  The effective field theory is a controlled expansion in $k / \knl$, and is valid for $ k / \knl \ll 1$.  For $k / \knl \ll 1$, observables can be computed to arbitrary precision, apart from non-perturbative effects, by including more and more loops and counterterms.  On the right hand side of \eqn{nleuler}, we should also include a stochastic counterterm $\frac{k^2}{\knl^2} \Delta_{\rm stoch.} ( \kvec)$.  This field does not correlate with the matter fields, but it does correlate with itself like 
\be
\langle \Delta_{\rm stoch.} ( \kvec) \Delta_{\rm stoch.} ( \kvec ' ) \rangle = \frac{(2 \pi)^3}{\knl^3} \delta( \kvec + \kvec'),
\ee
and so contributes a term like $k^4 / \knl^4$ to the power spectrum.  This term is negligible in a one-loop computation, so we ignore it for now.  

We then seek a perturbative solution to \eqn{nlcontinuity} and \eqn{nleuler} in the form $\delta_m = \delta_m^{(1)} + \delta_m^{(2)} + \delta_m^{(3)} + \delta_m^{( ct )} + \cdots$, where $\delta_m^{(n)}$ is sourced by $n$ powers of the linear solution $\delta_m^{(1)}$, i.e. $\delta^{(n)}_m \sim \left[ \delta_m^{(1)} \right]^{n}$, and $\delta^{(ct)}_m$ is the same order as $\delta_m^{(3)}$.  The linear solution that grows fastest with time is called the growth factor $D(a)$, so that $ \delta_m^{(1)} ( a , \xvec ) = D(a) \delta_m^{(1)} ( a_i , \xvec ) / D(a_i)$, and is given by 
\be \label{growthd1}
D( a ) = \frac{5}{2} \cH_0^2 \,  \Omega_{m,0} \frac{\cH(a) a_0 }{a} \int_0^a  \, \frac{ d \tilde a }{\cH(\tilde a)^3} \ . 
\ee
The linear power spectrum is defined by 
\be
\langle \delta_m^{(1)} ( a , \kvec )  \delta_m^{(1)} ( a , \kvec'  ) \rangle = ( 2 \pi )^3 \delta ( \kvec + \kvec' )  \left(\frac{D(a)}{D(a_i)} \right)^2 P_{11}( a_i, k ) \ ,
\ee
and the initial power spectrum defined at some initial time $a_i$ is taken from CAMB \cite{Lewis:1999bs}, for example.  Then, to solve for the higher order fields, we can use the Green's function for the system \eqn{nlcontinuity} and \eqn{nleuler}.  We will use this method later in the paper, but for now we present an approximate solution called the EdS approximation, which is exact in the matter era, but in general relies on $ ( \Omega_{m,0} \cH_0^2  a_0 / ( a \cH^2) ) / (a D' / D)^2 $ being close to unity.  This ratio is one at early times and is $1.15$ at $a=1$ \cite{Carrasco:2012cv}, but is close to one for most of the time evolution.  

The one-loop power spectra are defined by
\bea
&& 2  \la \delta^{(1)} ( a , \kvec ) \delta^{(3)}(  a , \kvec ' ) \ra = (2 \pi)^3 \delta( \kvec + \kvec' ) \left( \frac{D(a)}{D(a_i)} \right)^4 P_{13} (a_i ,   k)  \\
&& \la \delta^{(2)}  ( a , \kvec )  \delta^{(2)} ( a , \kvec ' )  \ra = (2 \pi)^3 \delta( \kvec + \kvec' )  \left( \frac{D(a)}{D(a_i)} \right)^4  P_{22} (a_i ,  k ) \\
 && 2  \la \delta^{(1)} ( a , \kvec ) \delta^{(ct)}(  a , \kvec ' ) \ra = (2 \pi)^3 \delta( \kvec + \kvec' ) \, P_{13}^{ct} ( a , k ),
 \eea
and $P_{13}$ and $P_{22}$ are the standard one-loop expressions for dark matter.\footnote{The standard expressions for the loop integrals are
 \begin{align}
P_{22}( a_i , k)& =\fr{k^3}{392 \pi^2}\int_0^{\Lambda/k}  dr \int_{-1}^1dx \fr{(-10 r x^2+3 r+7 x)^2}{(r^2-2 r x+1)^2}P_{11}(a_i , k r)P_{11}(a_i , k \sqrt{r^2-2 r x +1}) \n
P_{13}(a_i , k) & =\fr{k^3}{1008 \pi^2}P_{11}( a_i, k ) \n
& \hspace{.2in} \int_0^{\Lambda/k} dr \left( \fr{3}{r^3}(r^2 - 1)^3(7 r^2+2)  {\rm log } \Bigl  | \fr{1+r}{1-r} \Bigr  | -42 r^4+100r^2+\fr{12}{r^2}-158 \right) P_{11}(a_i , kr) \ .
\label{loops}
\end{align}
The above loop integrals are cut off (or smoothed over) at a scale $\Lambda > \knl$ because the theory is not under perturbative control at such high momenta.  As is thoroughly discussed in previous work \cite{Carrasco:2012cv, Carrasco:2013mua}, the speed of sound parameters, like $\bar c_m^2$, depend on $\Lambda$ in such a way as to cancel the final dependence of any physical observable on $\Lambda$.}  The counterterm power spectrum is given by 
\be
P_{13}^{ct} ( a , k ) = - 2 \,  ( 2 \pi ) \,  \bar c_m^2 ( a ) \frac{k^2}{\knl^2} \left( \frac{D(a)}{D(a_i) }\right)^2 P_{11} ( a_i , k ) \ , 
\ee
where we have redefined the speed of sound parameter for convenience.\footnote{In the power spectrum, the relevant parameter is the following integral:
 \bea\label{eq:green}
&& \bar{c}_m^2 (a ) = \int^a d a'  \, G ( a , a' ) \frac{D(a' ) }{D(a) }\,9\,   H(a')^2 c_{s,m}^2 ( a' ) \ ,
\label{csai}
\eea 
where $ G $ is the retarded Green's function for the linear equation 
\bea
&& - a^2 \cH^2 G''  - \left( 2 a \cH^2 + a^2 \cH \cH' \right)G' + \fr{3 \cH_0^2 a_0 \Omega_m }{2a}G = \delta^{(1)}_D(a-\tilde a) \ ,\n
&& G(a,a)=0\ , \qquad \left.\pd_aG(a,\tilde a)\right|_{ a=\tilde a}=\frac{1}{\tilde a^2\cH(\tilde a)^2}\ .
\eea
In order to estimate the numerical size of the integration over the Green's function, we approximate the integral in (\ref{eq:green}) with the corresponding EdS form and choose $c_{s,m}^2 \propto a^4$ as an example.  This gives
\bea\label{eq:simple_matching}
&&\bar{c}_m^2 (a _0) \simeq c_{s,m}^2 ( a_0 )   \ .
\eea
and explains the factor of 9 that is present in the definition of $c_{s,m}^2$ in \eq{nleuler}.
}

%%%%%%%%%%%%%%%%%%%%%%%%%%%%
%
%            Section 3
%    
%
%%%%%%%%%%%%%%%%%%%%%%%%%%%

\section{EFTofLSS with DE: clustering quintessence example}\label{cosmic-pert-section}

In this section, in order to provide an explicit computation, we consider the following action for the dark-energy degree of freedom
\be \label{fullaction}
 \int d^4 x \, \sqrt{-g}\bigg[ \frac{\mpl^2}{2}  R - \Lambda(t) - c(t) g_{\rm u}^{00}  +  \frac{M_2^4(t)}{2} ( \delta g_{\rm u}^{00} )^2  \bigg]   \ .
\ee
This is coupled to the dark-matter field through gravity in the following way
\begin{align} \label{fluid1}
&\dot{\delta}_m + \frac{1}{a} \partial_i ( (1 + \delta_m)v^i_m)  =0 \\
\label{fluid12} &\partial_i \dot{v}^i_m +  H \partial_i v^i_m + \frac{1}{a}  \partial_i ( v^j_m \partial_j v^i_m ) + \frac{1}{a}   \partial^2 \Phi = - \frac{1}{a} \partial_i \left( \frac{1}{\rho_m} \partial_j \tau^{ij} \right)_s  + \frac{1}{a} \partial_i \gamma^i_s \ ,
\end{align}
where $\gamma^i_s$ is the effective force which accounts for the fact that the two species can exchange momentum \cite{Lewandowski:2014rca}.  Furthermore, we will consider this system in the non-relativistic limit (i.e. on sub-horizon scales where gravitational non-linearities are most important), and when the dark-energy has a small speed of sound (which we discuss in more detail below).  This scenario is equivalent to the clustering quintessence model studied in \cite{Creminelli:2008wc, Creminelli:2009mu, Sefusatti:2011cm}.  We will also extend (and make a slight correction to) the computation in \cite{Sefusatti:2011cm} to include the third-order density fluctuation and counterterm operators, thus treating the dark matter sector as provided by the EFTofLSS.  One can see also~\cite{D'Amico:2011pf, Anselmi:2011ef, Anselmi:2014nya} for other approaches to non-linear clustering quintessence, which differ from our approach in the treatment of UV modes.\footnote{{ In particular, references~\cite{D'Amico:2011pf, Anselmi:2011ef, Anselmi:2014nya} claim to apply some resummation schemes for the infrared modes below the non-linear scale, but ignore contaminations from non-linear modes. This is radically different from our approach: while references~\cite{D'Amico:2011pf, Anselmi:2011ef, Anselmi:2014nya}  attempt to provide a more accurate solution to the equations of a perfect pressureless fluid coupled to dark energy, here we are rather changing the equations of motion for the fluid. Therefore, there is a major difference already at the level of the equations to be solved. We change the equations because we need to  include counterterms to consistently describe the effect of non-linear modes at long distances. The requirement of these terms has by now been quite well established by the literature on the EFTofLSS, and therefore we consider it inconsistent to not include these terms. Therefore, we do not find the need to perform an explicit comparison with the results of~\cite{D'Amico:2011pf, Anselmi:2011ef, Anselmi:2014nya}. In fact,  in this paper we are treating the modes below the non-linear scale perturbatively (i.e. no resummation).  Readers interested in these resummation schemes~\cite{D'Amico:2011pf, Anselmi:2011ef, Anselmi:2014nya} may find it interesting to apply them within the EFTofLSS with DE, and see if they can improve the results. However, this goes beyond the scope of the present paper. }}  In this paper, we work in a spatially flat FRW background in the Newtonian gauge and ignore tensor fluctuations of the metric.  This means that we can write the metric as\footnote{From the constraint equation \eqn{eq_traceless}, we see that $\Phi = \Psi$ for our study.  We will usually keep track of the fields separately, but in the end we will always set $\Phi = \Psi$.} 
\be\label{metric-pert}%
ds^2=-(1+2\Phi)dt^2+a(t)^2(1-2\Psi)\delta_{ij}dx^idx^j\,.
\ee

\subsection{Linear equations}

Using \eqn{metric-expand}, we see that the kinetic part of the action for $\pi$ is 
\be
S_{\rm kin.} = \int d^4 x \sqrt{-g} \left(  \left( c(t) + 2 M_2^4(t) \right) \dot \pi^2  - c ( t ) a^{-2} \partial^2 \pi \right) \ ,
\ee
from which we can read off the speed of sound of $\pi$ fluctuations to be 
\be \label{speedofsounddeffull}
\css = \frac{c(t)}{c(t) + 2 M_2^4(t)} \ . 
\ee
As discussed in \cite{Creminelli:2006xe}, there is a range of parameters for which the effective field theory is unstable.  In order to prevent the presence of ghosts, we should assume $c(t) + 2 M_2^4 ( t ) > 0$.  The presence of ghost fields is dangerous because the vacuum is unstable against the spontaneous production of positive energy matter particles and negative energy $\pi$ particles.  Then, notice that it is possible to have $c(t)  \equiv \bar \rho_D ( t ) ( 1 + w) / 2 <  0$ and still satisfy the no-ghost condition.  From \eqn{speedofsounddeffull} we see that this makes $c_s^2 < 0$, which seems to signal that the system has a gradient instability.  However, as shown in \cite{Creminelli:2006xe}, higher derivative terms in the action like $( \partial^2 \pi )^2$ stabilize the system at short scales, where the dispersion relation becomes $\omega^2 \approx k^4 / M_2^2$ and the system behaves like the Ghost Condensate \cite{ArkaniHamed:2003uy}.  The main point, though, is that on cosmological scales these higher order terms are highly suppressed unless $| 1 + w | \Omega_D \lesssim 10^{-34}$ \cite{Creminelli:2006xe, Creminelli:2008wc}, which means that for any values of $w$ distinguishable from the cosmological constant, the system only behaves like the Ghost Condensate on very short scales which are irrelevant for cosmology.  Taking this short scale stabilization into account, there is no problem with $w < -1$, but in that case one needs $ - c_s^2 \lesssim 10^{-30}$ in order to make the remaining gradient instability timescale longer than $H^{-1}$.  The bottom line is that for $w >-1$, any value of $c_s^2 \leq 1$ is allowed, but $c_s^2 \rightarrow 0 $ as $w \rightarrow -1$, and for $w<-1$ we must have $-c_s^2 \lesssim 10^{-30}$.  As a final point also noted in \cite{Creminelli:2006xe, Cheung:2007st}, $|c_s^2| \ll 1$ is technically natural, i.e. is not significantly renormalized by higher order operators, because $c_s^2 = 0$ is protected by the shift symmetry in the Ghost Condensate theory.  We will explicitly verify this in Section~\ref{nonlinearequationssubsec} when we consider non-linear terms.

For clustering quintessence, we are interested in the limit $\css \rightarrow 0$.  Then, assuming that $\css$ is constant for simplicity, and keeping in mind that $2 \, c(t)  = \rhod (t) ( 1 + w ) $, we have 
\be
M_2^4 (t)\approx \frac{ \rhod (t) ( 1 + w) }{4  \, \css} \label{speedofsounddef}\ . 
\ee 
Thus, in the $\css \rightarrow 0$ limit, the full linear equation for $\pi$ \eqn{linearpieom} becomes (see for example \cite{Creminelli:2008wc, Creminelli:2009mu, Gleyzes:2013ooa} 
\begin{align} \label{eom11}
 \ddot \pi - \dot \Phi + \frac{\partial_t M_2^4}{M_2^4} \left( \dot \pi - \Phi \right) + 3 H ( \dot \pi - \Phi) - \css a^{-2} \partial^2 \pi & = 0 \ ,
\end{align}
or
\be \label{eom12}
 \frac{1}{a^3 M_2^4} \frac{d}{dt } \left\{ a^3 M_2^4 ( \dot \pi - \Phi) \right\}   =  \css a^{-2} \partial^2 \pi   \  .
\ee
Without solving this equation, we can immediately find an important property of the solution, namely that $\dot \pi - \Phi \sim c_s^2 \,  \partial^2 \Phi / H^2$.  To see this, write $\pi = \pi_0 + c_s^2  \pi_{c_s}$, plug this into \eqn{eom12}, and expand in powers of $c_s^2$.  We obtain $\dot \pi_0 = \Phi$ and 
\be
\frac{1}{a^3 M_2^4} \frac{d}{dt} \left\{ a^3 M_2^4 \, \dot{ \pi}_{c_s} \right\} =   a^{-2} \partial^2 \pi_0 \sim   a^{-2} H^{-1} \partial^2 \Phi  \ ,
\ee
where we have taken time derivatives to be of order $H$.  This gives $ \pi_{c_s} \sim  \partial^2 \Phi / H$, and shows that $\dot \pi - \Phi \sim c_s^2 \, \partial^2 \Phi / H^2$.  We will often use this scaling, along with $\pi \sim \Phi / H$, to estimate the sizes of various contributions and determine the non-relativistic limit in the the rest of this paper.  For example, $\partial_i \pi \sim \partial_i \Phi / H \sim v^i$, where $v \sim 10^{-5} \, k / H$ is the characteristic velocity of the large-scale modes.  Thus, in any equation for $\partial^2 \pi \sim \theta$, the non-relativistic limit means that we ignore terms like $\partial_i \pi \, \partial_i \pi \sim v^2$ but keep terms like $\partial^2 \pi \, \partial^2 \pi \sim \theta^2$.

 In $\css \rightarrow 0$ limit, the Poisson equation \eqn{zero1} becomes (see for example \cite{Creminelli:2008wc, Creminelli:2009mu, Gubitosi:2012hu})
\begin{align} \label{poisson123}
a^{-2} \partial^2 \Psi & = \frac{3}{2 } \frac{ \Omega_{m,0} \cH_0^2 a_0}{a^3} \left(\delta_m + \frac{4 M_2^4 }{\rhom}( \dot \pi - \Phi) \right), \, 
\end{align}
where we have used that $ \rhom/(2 \mpl^2)=  3  \Omega_{m,0} \cH_0^2 a_0/ (2 a^3) $.\footnote{To make a connection to a fluid picture of quintessence, it may be useful to use the Poisson equation to define the overdensity of quintessence, $\delta_D$.  If we write $2 \mpl^2 a^{-2} \partial^2 \Psi = \sum_i \bar \rho_i \delta_i$, then this gives 
\be
\delta_D = \frac{1+w}{c_s^2} \left( \dot \pi - \Phi \right)  \label{deltadlin}\ ,
\ee to linear order and in the non-relativistic limit.}    In this paper, we are interested in computing correlation functions of the adiabatic mode, i.e. the one that sources the gravitational potential and is defined by $\delta_A = 2 M_{pl}^2 a^{-2} \partial^2 \Psi  / \rhom$.  In this case, using \eqn{poisson123}, we have 
\be 
\delta_A = \delta_m + \frac{4 a^3 M_2^4 }{a_0^3 \, \bar \rho_{m,0} }( \dot \pi - \Phi) \label{deltaalinear}\ . 
\ee

Now we are in a position to derive the relevant equations for $\delta_A$.  Looking back at the equation of motion \eqn{eom12} we see that, for $\css \rightarrow 0$, we can ignore the right hand side and obtain $\dot \pi - \Phi \propto \left( a^3 M_2^4 \right)^{-1}$, which is decaying because $M_2^4 \propto a^{-3 ( 1 + w) }$ and $w \approx -1$.  Thus, after this mode decays away, we have $\dot \pi - \Phi = 0 $, and in particular, that $\partial_i \dot \pi - \partial_i \Phi = 0 $.  Using the Euler equation \eqn{fluid2} for dark matter to linear order, this gives $\frac{d}{dt} \left[ a(  v_m^i + a^{-1} \partial_i \pi ) \right] = 0$, or
\be \label{velocities}
-a^{-1}\partial_i \pi =  v^i_m  \hspace{.3in} \text{and} \hspace{.3in} - a^{-1} \partial^2 \pi =  \theta_m,
\ee
on the growing adiabatic mode.  This means that the two species follow the same geodesics, i.e. that they are comoving.

Next, take the time derivative of the definition of $\delta_A$ to get 
\begin{align} 
\dot \delta_A & = \dot \delta_m  + \frac{4}{a_0^3 \, \bar \rho_{m,0}} \frac{d}{dt}  \left\{ a^3 M_2^4 ( \dot \pi - \Phi) \right\}  \\
& = - \frac{1}{a} \theta_m + \frac{4 \, a^3 M_2^4 }{a_0^3 \, \bar \rho_{m,0}} \css a^{-2} \partial^2 \pi = - \frac{1}{a} C(a) \theta_m \ , \label{cont1}
\end{align}
where 
\be \label{cafunction}
C(a) = 1 + \frac{4 \, a^3 M_2^4 \, \css }{a_0^3 \, \bar \rho_{m,0}} = 1 + (1+w) \frac{\Omega_{D,0}}{\Omega_{m,0}} \left( \frac{a}{a_0} \right)^{-3w} \ , 
\ee
and we have used the linear dark-matter continuity equation \eqn{fluid},  the equation of motion for $\pi$ \eqn{eom12}, and the fact that the two species are comoving.  Although the term proportional to $\css$ in \eqn{eom11} is not needed to find the relation \eqn{velocities}, it is needed to find the continuity equation \eqn{cont1} because there is a term proportional to $1/ \css$ in the definition of $\delta_A$.  Thus, writing $v^i_A \equiv v^i_m = - a^{-1} \partial_i \pi$ for the common velocity, we are led to the standard linear equations for the adiabatic mode in clustering quintessence \cite{Sefusatti:2011cm}
\begin{align}
\dot \delta_A + \frac{1}{a} C(a) \theta_A & = 0 \label{contlinear} \\
\dot \theta_A + H \theta_A + \frac{3}{2} \frac{\Omega_{m,0} \cH_0^2 a_0}{a^2}  \delta_A & = 0 \label{eullinear} \ .
\end{align}
From the combination of the above equations, we can find a single second order differential equation for $\delta_A$ which we will solve in Section \ref{Linear-sec-4}.  Note that we do not need to know the solution of $\pi$ to determine $\delta_A$.  In fact, having the solution for $\delta_A$ and therefore $\Phi$, we can solve the linear field equation for $\pi$, as shown in Appendix \ref{linearpisol}.

\subsection{Non-linear equations} \label{nonlinearequationssubsec}
There are two main non-linear effects to consider: the non-linear effects on the dynamics of $\pi$, and the effects of the non-linear definition of $\delta_A$ in terms of $\pi$.  We can discuss the former by considering the action for $\pi$ \eqn{fullaction} and using the linear solution to estimate the scale $k_{{\rm NL}, D}$ when the dark-energy sector will become non-linear.  The leading non-linear interaction term in the action in the $c_s^2 \rightarrow 0$ limit is $M_2^4 \dot \pi ( \partial \pi)^2$, and the leading quadratic term is $M_2^4 \dot \pi^2$.  Using $\dot \pi \sim \Phi \sim H v / k$ and $\partial \pi \sim v$, we have 
\be \label{knldapp}
\frac{M_2^4 \dot \pi ( \partial \pi)^2}{ M_2^4 \dot \pi^2 } \Big|_{k_{{\rm NL}, D}} \sim \frac{k v }{H}\Big|_{k_{{\rm NL}, D}} \ , 
\ee
which becomes order one at the same scale as dark matter, so we have that the scales are comparable, $k_{{ \rm NL} , D} \approx \knl$.  This was to be expected, since dark energy and dark matter have the same velocity, so that when dark matter becomes non-linear, so does dark energy.  We can also verify that $|c_s^2| \ll 1$ is not a fine tuning, i.e. that it is not significantly renormalized by higher order terms.  The coefficient of the $(\partial \pi)^2$ term is protected by the shift symmetry present for $w = -1$, so we should look for a renormalization of the $M_2^4 \dot \pi^2 $ term.  This can come from the $M_2^4 \dot \pi ( \partial \pi)^2$ term, changing the coefficient from $M_2^4 $ to something of the order $ M_2^4 \left( 1 +  \langle \delta_A(x) \delta_A(x) \rangle_{\knl } \right) $, which is at most an order one change and so cannot significantly change the speed of sound away from the $| c_s^2 | \ll 1$ value.

Next, we move on to consider the non-linear corrections to \eqn{contlinear} and \eqn{eullinear}, which depend on the non-linear definition of $\delta_A$, and for this we will have to look at the equations for motion.  As we saw in the last section, the equation of motion forces $\dot \pi - \Phi \propto c_s^2$ so that the dark-energy contribution to $\delta_A$, given by $c_s^{-2} ( \dot \pi - \Phi )$ in \eqn{deltaalinear}, scales like $c_s^0$, which is good because we do not expect any terms to blow up in the $c_s^2 \rightarrow 0 $ limit that we are considering.  To see that this is true at all orders in perturbations, start with the general form of the equation of motion for $\pi$
\be \label{eulerlag}
\nabla_\mu \frac{ \delta \cL}{ \delta \,  \partial_\mu \pi} = \frac{\delta \cL }{\delta \pi},
\ee
where $\cL$ is the Lagrange density for the action \eqn{fullaction}, after introducing $\pi$ with the St\"{u}ckelberg trick.  The coefficients $\Lambda(t + \pi )$, $c(t + \pi )$, and $M_2^4(t + \pi )$ depend only on powers of $\pi$, while $\delta g^{00}_{\rm u}$ contains derivatives of $\pi$.  Thus, we have 
\begin{align}
 \frac{ \delta \cL}{ \delta \,  \partial_\mu \pi} & = \frac{\delta \, \delta g^{00}_{\rm u}}{\delta \, \partial_\mu \pi} \frac{ \delta \cL}{\delta \, \delta g^{00}_{\rm u}}  = \frac{\delta \, \delta g^{00}_{\rm u}}{\delta \, \partial_\mu \pi} \left( - c  + M_2^4 \, \delta g^{00}_{\rm u} \right) \\
 \frac{ \delta \cL}{ \delta \pi} & = \frac{\delta ( c - \Lambda ) }{\delta \pi} - \frac{ \delta c }{ \delta \pi} \delta g^{00}_{\rm u} + \half \frac{ \delta M_2^4}{ \delta \pi } \left( \delta g^{00}_{\rm u} \right)^2 \ ,
 \end{align}
 which gives the full equation of motion as 
 \be \label{eulerlag2}
 \frac{1}{\sqrt{-g}} \partial_\mu \left( \sqrt{-g} \frac{\delta \, \delta g^{00}_{\rm u}}{\delta \, \partial_\mu \pi} \left( - c  + M_2^4 \, \delta g^{00}_{\rm u} \right)  \right) =  \frac{\delta ( c - \Lambda ) }{\delta \pi} - \frac{ \delta c }{ \delta \pi} \delta g^{00}_{\rm u} + \half \frac{ \delta M_2^4}{ \delta \pi } \left( \delta g^{00}_{\rm u} \right)^2 \ .
 \ee
We will solve \eqn{eulerlag2} perturbatively (for example writing $\delta g^{00}_{\rm u} = \delta g^{00\, (1)}_{\rm u} + \delta g^{00 \, (2)}_{\rm u} + \dots$) so in the $c_s^2 \rightarrow 0$ limit, the linear equation of motion is
\be
a^{-3} \partial_0 \left( a^3  M_2^4(t)  \delta g^{00 \, (1)}_{\rm u} \right) \propto c_s^0,
\ee
which means that $\delta g^{00 \, (1)}_{\rm u} \propto c_s^2$ when evaluated on the linear solution, as we found earlier.  Now, if we continue to expand the equation of motion to higher orders, we will always get
\be
a^{-3} \partial_0 \left( a^3  M_2^4(t)  \delta g^{00 \, (n)}_{\rm u} \right) \propto c_s^0,
\ee
because any time that a factor of $M_2^4$ shows up on the right hand side, it will be multiplied by a lower order $\delta g^{00}_{\rm u}$, which, because we are solving iteratively, is proportional to $c_s^2$.  Thus, we find that the equations of motion force $\delta g^{00}_{\rm u} \propto c_s^2$ at all orders.  Indeed, we expected this result, since $M_2^4 ( \deltagu )^2$ provides the kinetic term for the action in the limit $c_s^2 \rightarrow 0$, and so was not expected to blow up.

The reason that this is important is because this is the combination, $M_2^4 \deltagu$, that shows up in $\delta_A$, which is given by the $(00)$ component of the stress tensor, up to relativistic corrections, as
\be
\delta_A = \delta_m - \frac{ 2 a^{-3}}{\bar \rho_m} \frac{\delta \sqrt{-g} \,  \cL }{\delta g^{00}},
\ee
which in the $c_s^2 \rightarrow 0$ limit becomes
\be \label{deltaanew}
\delta_A \rightarrow \delta_m       - \frac{2}{\bar \rho_m} \frac{\sqrt{-g}}{a^3} \left(  \frac{ \delta \, \delta g^{00}_{{\rm u}}}{  \delta g^{00}}    \left( - c + M_2^4 \delta g^{00}_{\rm u} \right) - \half g_{00} \left( c - \Lambda \right) \right)   - \frac{ \bar \rho_D}{\bar \rho_m}   \ ,
\ee
and is now guaranteed to have a good limit for $c_s^2 \rightarrow 0$, as expected (note that terms like $M_2^4 ( \deltagu)^2$ are negligible because they are proportional to $c_s^2$).

From here, it is easy to see that the two species remain comoving (apart from possible counterterms, which we will discuss later) in the $c_s^2 \rightarrow 0$ limit even at higher orders.  We know that $\delta g^{00}_{\rm u} \propto c_s^2$, so in particular, we have $\partial_i \delta g^{00}_{\rm u} = 0$ for $c_s^2 \rightarrow 0$, which means that, up to relativistic corrections, 
\begin{align}
0 & = \partial_i \left(  \dot \pi - \Phi - \half a^{-2} (\partial \pi)^2 \right) \\
& = \frac{d}{dt} \left( a v_m^i + \partial_i \pi \right) + v_m^j \partial_j v_m^i - a^{-2} \partial_j  \pi \partial_j \partial_i \pi  \ . \label{velocityeq}
\end{align}
Any higher order corrections to \eqn{velocityeq} are relativistic, so in fact this equation is solved by setting $\partial_i \pi = - a v^i_m$ at all orders.  Since the velocity is the same, this means that the velocity of the adiabatic mode follows the same Euler equation as the dark-matter field, up to counterterm contributions (which we discuss later).

 Now we move on to compute the non-linear corrections to the continuity equation \eqn{contlinear}.  For that, we take the time derivative of the definition of $\delta_A$ in \eqn{deltaanew}.  In general, there are many terms contributing to $\delta_A$, even at linear level, but we are only interested in the non-relativistic limit.  In that limit, we use the linear equations to see that the only non-relativistic term is $M_2^4 \deltagu \propto H^{-2} \partial^2 \Phi$ (the rest are proportional to $\dot \pi$, $\pi$, $\dot \Phi$, $\Phi$, etc.), so \eqn{deltaanew} simplifies to\footnote{In the fluid picture, following \eqn{deltadlin}, the non-linear equations lead to a definition of the quintessence overdensity \be \delta_D = \frac{1+w}{c_s^2} \left( \dot \pi - \Phi - \half a^{-2} ( \partial \pi )^2 \right) \ . \ee} 
\be  \label{deltaaa}
\delta_A = \delta_m - \frac{2}{\bar \rho_m} M_2^4 \deltagu \ . 
\ee
The equation of motion \eqn{eulerlag2} also simplifies greatly in the non-relativistic limit
\be
 - \frac{2}{a^3} \partial_t \left( a^3   M_2^4 \,\deltagu \right) = - 2 a^{-2} \partial_i \left( \partial_i \pi \left( - c + M_2^4 \deltagu \right) \right)  \ , 
\ee
and so we see that the non-linear corrections in the non-relativistic limit enter through $\deltagu$.  Now, taking the time derivative of $\delta_A$ in \eqn{deltaaa}
\begin{align}
\dot \delta_A   = & \dot \delta_m - \frac{2}{\bar \rho_{m,0}} \partial_t \left( a^3 M_2^4 \, \deltagu \right) \\
 = &  \dot \delta_m  - \frac{2}{\bar \rho_m} a^{-2} \partial_i \left( \partial_i \pi \left( - c + M_2^4 \, \deltagu \right) \right) \\
 = & - \frac{1}{a} \theta_m - \frac{1}{a} \partial_i \left( \delta_m v^i_m \right) \\
& + \frac{2 c \, a^{-2} }{\bar \rho_m} \partial^2 \pi - \frac{2}{\bar \rho_m} a^{-2} \partial_i \left( M_2^4 \, \deltagu \partial_i \pi \right)  \\
& = - \frac{1}{a} C(a) \theta_A - \frac{1}{a} \partial_i \left( \delta_A v_A^i \right) \ , 
\end{align}
where we used the non-linear Euler equation for $\delta_m$ \eqn{fluid12} and the fact that the two species are comoving $\theta_m = - a^{-1} \partial^2 \pi \equiv \theta_A$.  This, combined with the non-linear Euler equation for the velocity, gives the system at quadratic order for clustering quintessence \cite{Sefusatti:2011cm} (apart from counterterms, which we consider in Section \ref{countertermsection})
\begin{align}
\dot \delta_A + \frac{1}{a} C(a) \theta_A & = - \frac{1}{a} \partial_i \left( \delta_A v_A^i \right)  \label{nonlincont}\\
\dot \theta_A + H \theta_A + \frac{3}{2} \frac{\Omega_{m,0} \cH_0^2 a_0}{a^2}  \delta_A & = - \frac{1}{a} \partial_i \left( v_A^j \partial_j v_A^i \right)  \label{nonlineuler} \ .
\end{align}

%%%%%%%%%%%%%%%%%%%%%%%%
%
%          Section 4
%
%
%%%%%%%%%%%%%%%%%%%%%%%%%%

\section{Solution for $c_s^2 \rightarrow 0$: clustering quintessence} \label{sec-4}

Up to now, we found the explicit form of the non-linear equations for the adiabatic mode in the presence of clustering quintessence. In this section, we solve these equations up to third order and obtain the density power spectrum up to one-loop order, and we include the one-loop counterterm to correctly describe the dark matter contribution.  In this section, we extend (and make a slight correction to) the computation done in \cite{Sefusatti:2011cm}, by including $\delta_A^{(3)}$, and most importantly, the effects of UV physics through $\delta_A^{(ct)}$.  For the one-loop computation that we present in this paper, we find it easier to use the exact perturbative time dependence (i.e. Green's functions), rather than the approximate $\delta^{(n)} \sim D^n$ which is sometimes employed.

From now on, for simplicity, we use $\delta$ instead of $\delta_A$ and it is more convenient to write the continuity and Euler equations in terms of the rescaled $\theta$ which is defined as
\be
\Theta\equiv -\frac{C}{\mathcal{H}f_{+}}\theta,
\ee
where $f_{\pm}$ are the linear growth rate, $f_{\pm}=\frac{d\ln D_{\pm}}{d\ln a}$.  Our perturbative expansion for $\delta_{\vk}$ and $\Theta_{\vk}$ (switching notation to $\delta_{\vk}$ instead of $\delta( \kvec)$ for the Fourier transform) can be written
\be \label{deltaexpansion}
\delta_{\vk}(a)=\sum^{\infty}_{n=1}\delta_{\vk}^{(n)}(a)  + \delta_{\vk}^{(ct)} ( a )  \quad \textmd{and} \quad \Theta_{\vk}(a) =\sum^{\infty}_{n=1} \Theta_{\vk}^{(n)}(a) + \Theta_{\vk}^{(ct)} ( a ) ,
\ee
where $\delta^{(n)}$ are the $n$-th order solutions in the absence of counterterms, and $\delta^{(ct)}$ is the field sourced by the effective stress tensor and effective force.  First, we will ignore the stress tensor, then in Section \ref{countertermsection} we compute the counterterm contribution.

In terms of the above and in Fourier space, equations \eqn{nonlincont} and \eqn{nonlineuler} read as
\begin{align}
&a\delta'_{\vk}-f_{+}\Theta_{\vk}=\frac{(2\pi)^{3}f_{+}}{C}\iint \frac{d^3q_1}{(2\pi)^{3}}\frac{d^3q_2}{(2\pi)^{3}} \delta_{D}(\vk-\vq_1-\vq_2)\alpha(\vq_1,\vq_2)\Theta_{\vq_1}\delta_{\vq_2},\\
&a\Theta'_{\vk}-f_{+}\Theta_{\vk}-\frac{f_{-}}{f_{+}}(\Theta_{\vk}-\delta_{\vk})=\frac{(2\pi)^{3}f_{+}}{C}\iint \frac{d^3q_1}{(2\pi)^{3}}\frac{d^3q_2}{(2\pi)^{3}}\delta_{D}(\vk-\vq_1-\vq_2)\beta(\vq_1,\vq_2)\Theta_{\vq_1}\Theta_{\vq_2} \nonumber \ , 
\end{align}
such that the continuity and Euler equations at the $n$-th order respectively are
%\label{recurvise}
\begin{align}
\label{conti-eq}
&  a \td^{(n)' }_{\vk}-f_{+}\tT^{(n)}_{\vk}=\frac{f_{+}}{C}\sum\limits_{m=1}^{n-1}    \int \frac{d^3q}{(2\pi)^3}\alpha(\vq,\vk-\vq)  \tT^{(m)}_{\vq}\td^{(n-m)}_{\vk-\vq},\\\label{Euler-eq}
& a \tT^{(n) '}_{\vk}-f_{+}\tT^{(n)}_{\vk}-\frac{f_{-}}{f_{+}}(\tT^{(n)}_{\vk}-\td^{(n)}_{\vk})=\frac{f_{+}}{C}\sum\limits_{m=1}^{n-1}    \int \frac{d^3q}{(2\pi)^3}\beta(\vq,\vk-\vq) \tT^{(m)}_{\vq}\tT^{(n-m)}_{\vk-\vq},
\end{align}
and $\alpha$ and $\beta$ are given in \eqn{alphadef} and \eqn{betadef}.  In order to determine the density correlation up to one-loop order (without counterterms), we need to determine $\td^{(1)}_{\vk}$, $\td^{(2)}_{\vk}$ and $\td^{(3)}_{\vk}$.

\subsection{Linear perturbations}\label{Linear-sec-4}

At first order in perturbations, \eqn{conti-eq} and \eqn{Euler-eq} imply that 
\be\label{linear-sol}
\delta^{\rm (1)}_{\vk}(a)=  \frac{D(a)}{D(a_i)} \delta^{\rm in}_{\vk} \quad \textmd{and} \quad \Theta^{\rm (1)}_{\vk}(a)=\delta^{\rm (1)}_{\vk}(a).
\ee
where $\delta^{\rm in}_{\vk} \equiv \delta_{\vk}( a_i ) $ is the initial value of $\delta_{\vk}$ on which we will comment later. 
The growth function $D(a)$ is given by the following equation
\be\label{delta-m34}
\frac{d^2}{d\ln a^2}\bigg(\frac{D}{H}\bigg)+\bigg(2+3\frac{d\ln H}{d\ln a}-\frac{d\ln C}{d\ln a}\bigg)\frac{d}{d\ln a}\bigg(\frac{D}{H}\bigg)=0 \ ,
\ee
where $C(a)$ is
\be\label{C-def}
C(a)\equiv\bigg(1+(1+w)\frac{\Omega_{D,0}}{\Omega_{m,0}}a^{-3w}\bigg).
\ee
The equation \eqn{delta-m34} has two solutions, one growing mode \cite{Sefusatti:2011cm} 
\be\label{D+-clus}
D_{+}(a)=\frac52 \int^a_0C(\ta)\Omega_{m}(\ta)\frac{H(a)}{H(\ta)}d\ta,
\ee
and a decaying mode which is
\be
D_{-}(a)=\frac{H(a)}{H_0\Omega_{m,0}^{1/2}} \ ,
\ee
where $H_0$ is the current value of the Hubble parameter.  We find it convenient to define the time dependent energy density ratios
\be
\Omega_{m}(a)\equiv \Omega_{m,0} \frac{H_0^2}{H(a)^2} \left( \frac{a }{a_0} \right)^{-3}  \ ,  \hspace{.3in} \Omega_D(a) \equiv \Omega_{D,0} \frac{H_0^2}{H(a)^2} \left( \frac{a}{a_0} \right)^{-3 ( 1 + w) } \ . 
\ee
The linear growth indices $f_{\pm}\equiv \frac{d\ln D_{\pm}}{d\ln a}$ are
given as
\be
f_{+} ( a ) =\bigg(\frac52\frac{a}{D_{+} ( a ) }-\frac32\Omega_{m}(a)\bigg)C(a) \ ,
\ee
and
\be
f_{-}(a)=-\frac32\Omega_{m}(a)C(a) \ .
\ee
During the matter era where $\frac{\Omega_D ( a ) }{\Omega_m ( a ) }$ is negligible and $C(a)\simeq1$, the growth functions are approximately equal to their corresponding value in the exact $\Lambda$CDM model
\be
D_{+}(a)\simeq a \quad \textmd{and} \quad D_{-}(a)\simeq a^{-\frac32} \ ,
\ee
and in the same limit, the growth indices are reduced to $f_+\simeq1$ and $f_-\simeq-\frac32$.
However, as we approach the dark energy era, the linear solutions deviate from $\Lambda$CDM and depending on the sign of $(1+w)$, they have different behaviors.  In Figure \ref{figLinear}, we show the linear behavior of the power spectra of $\delta_{A , c_s^2 \rightarrow 0}$ (the adiabatic mode in clustering quintessence), $\delta_{m , c_s^2 \rightarrow 0}$ (the matter field in clustering quintessence), and $\delta_{m , c_s^2 = 1}$ (the matter field in in the presence of smooth dark energy, described in Appendix~\ref{smoothdesection})  compared to $\Lambda$CDM.  We see that $\delta_{m , c_s^2 = 1}$ is very close to $\delta_{m , c_s^2 \rightarrow 0}$, which means that clustering quintessence fluctuations have only a small effect on the matter power spectrum (the overall deviation from $\Lambda$CDM is due to the different background expansion, which is the dominant effect of non-clustering quintessence).  The dominant effect of clustering quintessence fluctuations is the way that they change the adiabatic power spectrum, which is an effect of order $1 + w$ compared to the smooth case at $a=1$ (note that $\delta_{A , c_s^2 = 1} \approx \delta_{m , c_s^2 = 1}$).

\begin{figure}[t!]
\begin{center}
\includegraphics[width=0.5\textwidth]{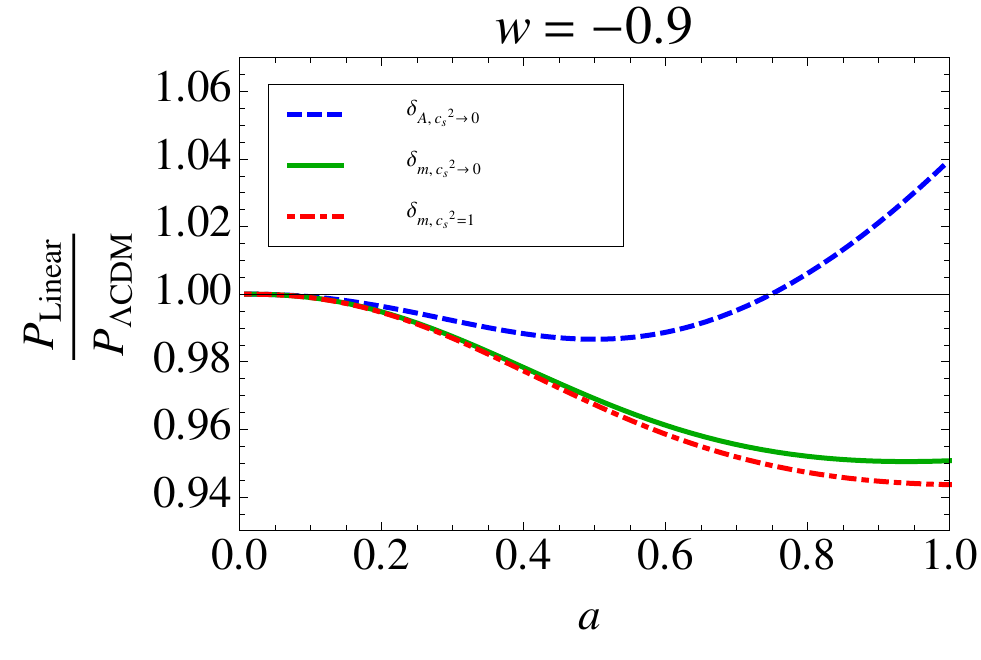}\includegraphics[width=0.5\textwidth]{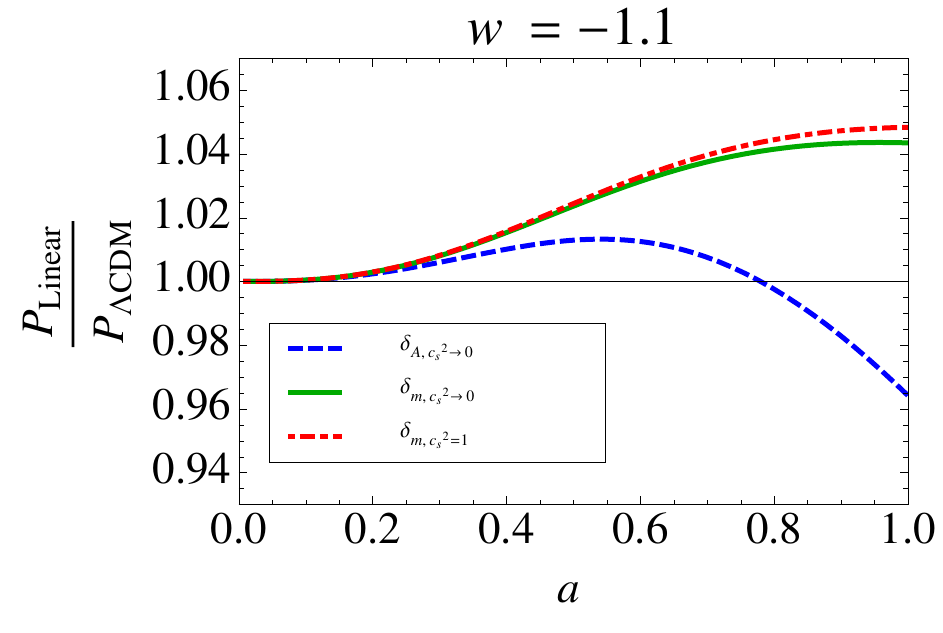}
\caption{In these plots, we compare various linear power spectra to the linear power spectrum in $\Lambda$CDM. The dashed (blue) and solid (green) lines correspond to the power spectrum of $\delta_{A}$ and $\delta_m$ in the clustering quintessence model respectively. The dot-dashed (red) line represents the power spectrum of $\delta_m$ in $w$CDM (smooth dark energy, described in Appendix~\ref{smoothdesection}). In the left panel, $w=-0.9$ and in the right panel $w=-1.1$. As we see, the power spectrum in the clustering and $w$CDM models are almost the same for redshifts $a < 0.3$, while they start to diverge as we approach the present time. Interestingly, $\delta_{m,c_s^2  \rightarrow0}$ in the clustering case remains close to its corresponding quantity in the smooth case, $\delta_{m,c_s^2 =1}$. This means that the effect of the clustering quintessence \emph{on matter} is small.  However, the total density contrast, $\delta_{A,c_s^2 \rightarrow0}$ in the clustering case is noticeably different from $\delta_m$ in the clustering and smooth dark energy.}\label{figLinear}
\end{center}
\end{figure}

Before going any further, let us take a brief moment to comment on the initial conditions. During matter domination at sufficiently early times, the linear equations are valid, and we have $\delta_{m} (a , \vk) = \frac{a}{a_i} \delta_{m} ( a_i , \vk)$, where $a_i$ is the time at which we set the initial conditions. Using the linear continuity equations for $\delta_m$ and the adiabatic mode $\delta$, we have that $\delta_m ' = \delta' / C(a)$ which gives\footnote{As an alternative approach, we solve the linear $\pi$ equations during matter era directly in Appendix \ref{lineareqs} and read $\delta$ in terms of $\delta_{m}$ in \eqn{delta-a-MD}.} \cite{Creminelli:2008wc}
\be
\delta_{\vk}^{\rm in}= \left( 1 + \frac{1+w}{1-3w} \frac{\Omega_{D,0}}{\Omega_{m,0}} \left( \frac{ a_i }{a_0} \right)^{-3w} \right) \delta_{m}^{(1)} ( a_i , \vk) \ . 
\ee

The initial power spectrum for $\delta_m^{(1)}$ can be gotten from CAMB, but we must keep in mind two subtleties.  The first is that the above expression is only valid in the matter era, so we must set the initial conditions at a time early enough so that we are in the matter era.  For this, we want $   \Omega_{D,0} a^{-3 ( 1 +w)} / ( \Omega_{m,0} a^{-3}) \ll 1$, or more specifically, $D_+(a) / a \approx 1$. For example, with $w = -0.9$, at $a = 0.167$, $\Omega_{D,0} a^{-3 ( 1 +w)} / ( \Omega_{m,0} a^{-3}) = 0.021$, and $ D_+(a) / a = 0.996$, which is well within the matter era.  On the other hand, we need to be at a late enough time such that the linear growth rate $D_+$ accurately describes the growth of structure.  At early times, when the effects of radiation are still present, our equations are incomplete (see \cite{Fitzpatrick:2009ci} for a discussion of radiation effects in the bispectrum).  Because of this, we choose to implement the initial conditions at $a = 0.167$.  Here, the difference between the linear evolution from $a = 0.167$ to $a = a_0$ with CAMB and the linear evolution with $D_+^2$ of the power spectrum is $0.2 \%$, well within our computational precision.  Alternatively, and specifically for more general dark-matter actions, one could use one of the recently developed linear codes for the EFT of dark energy \cite{Hu:2013twa, Frusciante:2016xoj, Zumalacarregui:2016pph, coopref}.

\subsection{Non-linear perturbations}

In this subsection, we present the solutions for the higher order fields $\delta^{(2)}$ and $\delta^{(3)}$, and for the counterterm contribution $\delta^{(ct)}$.  The non-linear continuity and Euler equations can be solved perturbatively in terms of four Green's functions. In particular, one can write $\delta^{(n)}_{\vk}(a)$ and $\Theta^{(n)}_{\vk}(a)$ at any perturbative order as
 \begin{align}  \label{dtGreen}
 &\delta^{(n)}_{\vk}=\int^a_0 d\ta \bigg(G^{\delta}_{1}(a,\ta)S^{(n)}_1(\ta,\vk)+G^{\delta}_{2}(a,\ta)S^{(n)}_2(\ta,\vk)\bigg) \ ,\\
  &\Theta^{(n)}_{\vk}=\int^a_0 d\ta \bigg(G^{\Theta}_{1}(a,\ta)S^{(n)}_1(\ta,\vk)+G^{\Theta}_{2}(a,\ta)S^{(n)}_2(\ta,\vk)\bigg) \ , \label{dtGreen2}
  \end{align}
where $G^{\delta}_{1}$, $G^{\delta}_{2}$ are the density Green's functions, $G^{\Theta}_{1}$, $G^{\Theta}_{2}$ are velocity Green's functions and $S^{(n)}_1(\ta,\vk)$ and $S^{(n)}_2(\ta,\vk)$ are the source terms of the continuity and Euler equations at the $n$-th order respectively.
Here we only report the final solutions and we present the details of the calculations and the explicit form of the source terms in Appendix \ref{Green-app}.

Having the formal solutions \eqn{dtGreen} and \eqn{dtGreen2}, we can find the solutions of $\delta$ and $\Theta$ at any perturbative order. Since we are interested in the one-loop power spectrum, next, we calculate the second and third order perturbations.

\subsubsection{Second-order perturbations}

At second order in the perturbations, using the linear solution in \eqn{linear-sol} along with the expression for the second-order source term in \eqn{source}, we find the source terms
 \begin{align}   \label{source-2nd}
 &S_1^{(2)}(a,\vk)=\frac{f_{+}(a)D_{+}^2(a)}{C(a)  D_+^2 ( a_i) }\int \frac{d^3q}{(2\pi)^3}\alpha_s(\vq,\vk-\vq)\delta^{\rm in}_{\vq}\delta^{\rm in}_{\vk-\vq} \ ,\\
 &S_2^{(2)}(a,\vk)=\frac{f_{+}(a)D_{+}^2(a)}{C(a) D_+ ( a_i) }\int \frac{d^3q}{(2\pi)^3}\beta(\vq,\vk-\vq)\delta^{\rm in}_{\vq}\delta^{\rm in}_{\vk-\vq} \ ,    \label{source-2nd2}
\end{align}
where $\alpha_s(\vq_{1},\vq_{2})=\frac12(\alpha(\vq_{1},\vq_{2})+\alpha(\vq_{2},\vq_{1}))$, and $\alpha$ and $\beta$ are given in \eqn{alphadef} and \eqn{betadef}.  After plugging the above in the solutions \eqn{dtGreen} and \eqn{dtGreen2}, we have\footnote{Our expressions for the second order fields differ from the analogous expressions in \cite{Sefusatti:2011cm}.  In fact, one can quickly check that the solutions Eq. (73) and Eq. (74) of \cite{Sefusatti:2011cm} are not related by the equation of motion Eq. (71) in that paper.  Inside of the parentheses of Eq. (74) of \cite{Sefusatti:2011cm}, the coefficient of $( 2 \alpha_s - 2 \beta) / 5$ should be $e^{ \tilde \eta - \eta}\partial_\eta D_- ( \eta ) / D_- ( \tilde \eta )$.  Then, the boundary conditions of the Green's functions should be set without changing the relative coefficients of the terms inside of the parentheses.}
\begin{align}\label{delta-2}
\td_{\vk}^{(2)}(a)&=\int \frac{d^3q}{(2\pi)^3}\bigg(\alpha_s(\vq,\vk-\vq)\mG^{\delta}_{1}(a)+\beta(\vq,\vk-\vq)\mG^{\delta}_{2}(a)\bigg)\delta^{\rm in}_{\vq}\delta^{\rm in}_{\vk-\vq} \ ,\\
\tT_{\vk}^{(2)}(a)&=\int \frac{d^3q}{(2\pi)^3}\bigg(\alpha_s(\vq,\vk-\vq)\mG^{\Theta}_{1}(a)+\beta(\vq,\vk-\vq)\mG^{\Theta}_{2}(a)\bigg)\delta^{\rm in}_{\vq}\delta^{\rm in}_{\vk-\vq} \ . \label{theta-222}
\end{align}
where $\mG^{\delta}_{1}$, $\mG^{\delta}_{2}$, $\mG^{\Theta}_{1}$ and $\mG^{\Theta}_{2}$ are four functions of time given as
\begin{align}    \label{2nd-solution} 
\mG^{\delta}_{\sigma}(a)&=\int^{1}_0 \frac{f_{+}(\ta)   D_{+}^2(\ta)  }{C(\ta) D_+^2 ( a_i) }G^{\delta}_{\sigma}(a,\ta)d\ta \ , \\
\mG^{\Theta}_{\sigma}(a)&=\int^{ 1}_0 \frac{f_{+}(\ta)  D_{+}^2(\ta)   }{C(\ta)  D_+^2 ( a_i ) }G^{\Theta}_{\sigma}(a,\ta)d\ta,      \label{2nd-solution2} 
\end{align}
for $\sigma = 1,2$.  Notice that, because the Green's functions depend only on time, the momentum and time integrals separate.  This means that at all orders in perturbation theory, each loop can be written as a sum over terms which are a product of a function of momentum times a function of time, which greatly reduces computational time.  Deep inside the matter era, where we can neglect the effect of dark energy, the above time functions are simply  $\mG^{\delta}_{1}(a)\simeq\frac57\big(\frac{D_+(a)}{D_+(a_i)}\big)^2$, $\mG^{\delta}_{2}(a)\simeq\frac27\big(\frac{D_+(a)}{D_+(a_i)}\big)^2$, $\mG^{\theta}_{1}(a)\simeq\frac37\big(\frac{D_+(a)}{D_+(a_i)}\big)^2$ and $\mG^{\theta}_{2}(a)\simeq\frac47\big(\frac{D_+(a)}{D_+(a_i)}\big)^2$.

\subsubsection{Third-order perturbations}

The third-order source terms (derived in \eqn{source-3rd-app}) are
 \begin{align} 
 &S_1^{(3)}(a,\vk)=\frac{f_{+}(a)D_{+}(a)}{C(a) D_+ ( a_i ) }\iint \frac{d^3p}{(2\pi)^3}\frac{d^3q}{(2\pi)^3}\bigg(\alpha^{\sigma}(\vk,\vp,\vq)\mG^{\delta}_{\sigma}(a)+\gamma^{\sigma}(\vk,\vp,\vq)\mG^{\Theta}_{\sigma}(a)\bigg)\delta^{\rm in}_{\vk-\vp}\delta^{\rm in}_{\vp-\vq}\delta^{\rm in}_{\vq}, \nonumber \\
 &S_2^{(3)}(a,\vk)=\frac{f_{+}(a)D_{+}(a)}{C(a) D_+ ( a_i ) }\iint \frac{d^3p}{(2\pi)^3}\frac{d^3q}{(2\pi)^3}\beta^{\sigma}(\vk,\vp,\vq)\mG^{\Theta}_\sigma(a)\delta^{\rm in}_{\vk-\vp}\delta^{\rm in}_{\vp-\vq}\delta^{\rm in}_{\vq},   \label{source-3rd}
\end{align}
where again $\sigma=1,2$ and summation over upper and lower indices is assumed. Note that here, $\{\alpha^{\sigma},\beta^{\sigma}, \gamma^{\sigma}\}$ are six functions of momenta made of the standard functions from dark-matter perturbation theory $\alpha(\vk_1,\vk_2)$ and $\beta(\vk_1,\vk_2)$ which we present in \eqn{momentum-fun-3} - \eqn{momentum-fun-4}.

From the combination of \eqn{dtGreen} and \eqn{source-3rd}, we find $\td^{(3)}_{\vk}$ and $\tT^{(3)}_{\vk}$ as 
 \begin{align}\label{delta-3}
 &\td^{(3)}_{\vk}(a)=  \iint \frac{d^3p}{(2\pi)^3}\frac{d^3q}{(2\pi)^3}\bigg(\alpha^{\sigma}(\vk,\vp,\vq)\mU^{\delta}_{\sigma}(a)+\beta^{\sigma}(\vk,\vp,\vq)\mV^{\delta}_{\sigma2}(a)        \nonumber \\
 & \hspace{3.2in} +   \gamma^{\sigma}(\vk,\vp,\vq)\mV^{\delta}_{\sigma1}(a)\bigg)\delta^{\rm in}_{\vk-\vp}\delta^{\rm in}_{\vp-\vq}\delta^{\rm in}_{\vq},\\
 &\tT^{(3)}_{\vk}(a)=  \iint \frac{d^3p}{(2\pi)^3}\frac{d^3q}{(2\pi)^3}\bigg(\alpha^{\sigma}(\vk,\vp,\vq)\mU^{\Theta}_{\sigma}(a)+\beta^{\sigma}(\vk,\vp,\vq)\mV^{\Theta}_{\sigma2}(a)   \nonumber \\
 & \hspace{3.2in} +\gamma^{\sigma}(\vk,\vp,\vq)\mV^{\Theta}_{\sigma1}(a)\bigg)\delta^{\rm in}_{\vk-\vp}\delta^{\rm in}_{\vp-\vq}\delta^{\rm in}_{\vq},
\end{align}
where $\mU^{\delta}_{\sigma}(a)$, $\mV^{\delta}_{\sigma\tilde{\sigma}}(a)$, $\mU^{\Theta}_{\sigma}(a)$ and $\mV^{\Theta}_{\sigma\tilde{\sigma}}(a)$ are functions of time given as 
\begin{align}
\mU^{\delta}_{\sigma}(a)=\int^{{1}}_0 \frac{f_{+}(\ta)D_{+}(\ta)}{C(\ta) D_+ ( a_i) }\mG^{\delta}_{\sigma}(\ta)G^{\delta}_{1}(a,\ta)d\ta,\\
\mU^{\Theta}_{\sigma}(a)=\int^{{1}}_0 \frac{f_{+}(\ta)D_{+}(\ta)}{C(\ta) D_+(a_i) }\mG^{\delta}_{\sigma}(\ta)G^{\Theta}_{1}(a,\ta)d\ta,\\
\mV^{\delta}_{\sigma\tilde\sigma}(a)=\int^{{ 1}}_0 \frac{f_{+}(\ta)D_{+}(\ta)}{C(\ta)  D_+(a_i)}\mG^{\Theta}_{\sigma}(\ta)G^{\delta}_{\tilde\sigma}(a,\ta)d\ta,\\
\mV^{\Theta}_{\sigma\tilde\sigma}(a)=\int^{{1}}_0 \frac{f_{+}(\ta)D_{+}(\ta)}{C(\ta)    D_+(a_i) }\mG^{\Theta}_{\sigma}(\ta)G^{\Theta}_{\tilde\sigma}(a,\ta)d\ta.
\end{align}
During the matter era when dark energy is negligible, the above time functions are proportional to $D_+^3(a)$.  Some examples are $\mU^{\delta}_1(a)\simeq\frac{5}{18}\big(\frac{D_+(a)}{D_+(a_i)}\big)^3$, $\mU^{\delta}_2(a)\simeq\frac{1}{9}\big(\frac{D_+(a)}{D_+(a_i)}\big)^3$, $\mU^{\theta}_1(a)\simeq\frac{5}{42}\big(\frac{D_+(a)}{D_+(a_i)}\big)^3$, $\mU^{\theta}_2(a)\simeq\frac{1}{21}\big(\frac{D_+(a)}{D_+(a_i)}\big)^3$, $\mV^{\delta}_{11}(a)\simeq\frac{1}{6}\big(\frac{D_+(a)}{D_+(a_i)}\big)^3$, and $\mV^{\delta}_{12}(a)\simeq\frac{1}{21}\big(\frac{D_+(a)}{D_+(a_i)}\big)^3$ .

\subsubsection{Counterterms} \label{countertermsection}

Our approach to dealing with the counterterms will be to first work in the basis of $\delta_m$ and $\pi$, and then find the contribution to the $\delta_A$ equations.  First, let us consider the response of dark matter to gravitational non-linearities; we will have to include an explicit counterterm to describe how the UV physics of dark-energy affects the large scale dark-matter field.  In general, the expansion of the dark-matter stress tensor and force term will take a form analogous to the two fluid case in \cite{Lewandowski:2014rca}
\bea \label{stresstensor10}
&&-  \left( \frac{1}{\rho_m} \partial_j \tau^{ij} \right)_s ( a , \xvec) + \gamma^i_s ( a , \xvec )  =  \\ \nonumber
&&\qquad\qquad\int d a' \Big[\kappa^{(1)}( a , a' )\,    \partial^i \partial^2 \Phi ( a' , \xvec_{\rm fl}( \xvec ; a,a') ) +\kappa^{(2)}( a , a' )\, \fr{1}{H}\partial^i \partial_j v_m^j ( a' , \xvec_{\rm fl}( \xvec ; a,a') )  \\ \nonumber
&& \qquad \qquad \qquad \qquad + \kappa^{(\rm stoch.)}_1 ( a , a' ) \bar \Delta^i_{\rm stoch.} ( a ' , \xvec_{\rm fl}( \xvec ; a,a') )     \ldots \Big ] \ .
\label{str1}
\eea
Notice that we do not include a direct coupling like $ \partial^i \partial_j \partial^j \pi \sim \partial^i \partial_j v^j_{D}$ because the two species interact only through gravity, which means that when gravity is turned off, dark matter should not feel a response from dark energy.  As discussed in \cite{Lewandowski:2014rca}, the relative velocity, $v_{\rm rel}^i  \equiv v^i_m - \partial^i \pi$, can appear with no derivatives.  For the case that we study in this paper, the species are comoving, so $v_{\rm rel}^i = 0$.  The stochastic term $\bar \Delta^i_{\rm stoch.}$ is now different from the pure dark matter case considered in Section \ref{eftreviewsection} because of the effective force $\gamma^i_s ( a , \xvec )$.  In the dark-matter only case, the stochastic contribution to the power spectrum goes like $k^4$ because of momentum conservation.  However, in the case of two species which can exchange momentum, momentum is not conserved separately in each species, so the contribution to the power spectrum can go like $k^2$.  In particular, we expect the stochastic part of $\gamma^i_s$ to be Poisson-like so that in momentum space we can have 
\be
\langle \Delta^i_{\rm stoch.} ( \kvec) \Delta^j_{\rm stoch.} ( \kvec ') \rangle = \frac{(2 \pi)^3}{\knl^3} \delta( \kvec + \kvec') C^{(1)} \delta^{ij} + \dots,
\ee
where $C^{(1)}$ is expected to be an order one number, and $\dots$ stands for terms higher order in $k/\knl$.  To get the contribution to the power spectrum, we contract the above with $k_i k_j$, and so we have a $k^2$ contribution.  However, as discussed in \cite{Lewandowski:2014rca}, this is expected to be subleading with respect to the counterterm contribution $k^2 P(k)$, so we will not study these operators in this paper.\footnote{ Stochastic terms can also be included in the Lagrangian for the dark-energy degree of freedom by coupling $\pi$ to a dissipative sector through terms like $\mathcal{O} \delta g^{00}_u$, where $\mathcal{O}$ is some composite operator of the dissipative sector \cite{LopezNacir:2011kk}.  However, as in the dark-matter sector, we expect these effects to be small for the one-loop computation that we perform, so we ignore them in this work.  }

Thus, evaluating the counterterms \eqn{stresstensor10} on the linear solutions and performing the $a'$ integral as usual, we are led to the following counterterm on the right-hand side of the Euler equation for dark matter \eqn{nleuler}:
\be
9 \, ( 2 \pi) \,H(a)^2 \frac{k^2}{\knl^2}  \left(  c_{s,\delta}^2 (a)  \delta_{\kvec} ( a )  + c_{s,v_m}^2 ( a ) \frac{1}{H} \theta_{m, \, \kvec} ( a ) \right) \ ,
\ee
which after using the linear equations of motion, and the fact that the species are comoving at linear order, becomes 
\begin{align} \label{counterform}
& 9 \, ( 2 \pi) \,H(a)^2 \frac{k^2}{\knl^2}  \left(  c_{s,\delta}^2 (a)  \delta^{(1)}_{\kvec} ( a  )  - c_{s,v_m}^2 ( a ) \frac{a^2}{C(a)} \delta^{(1)}_{\kvec} ( a )' \right) \\
& \hspace{.5in} = 9 \, ( 2 \pi) \,H(a)^2 \frac{k^2}{\knl^2} \, c_{A,m}^2 ( a )  \delta^{(1)}_{\kvec} ( a  )  \ .
\end{align}
where $ c_{A,m}^2 ( a ) =  \left(  c_{s,\delta}^2 (a)    - c_{s,v_m}^2 ( a ) \frac{a^2 D_+'(a)}{D_+(a) C(a)}  \right) $.  As the name suggests, $c_{A,m}^2$ is the contribution to the speed of sound of the adiabatic mode from the matter sector.

Next, we move on to the dark-energy sector.\footnote{In the dark-energy sector, it is worth checking that the quantum unitarity cutoff for the dark-energy action can be near or above the non-linear scale for dark mater, $\knl$.  In the small $c_s^2$ limit, the EFT for dark-energy will eventually become strongly coupled, bringing the cutoff down to smaller momentum.  Thus, we need to make sure that a cutoff near $\knl$ and a small $c_s^2$ are not contradictory assumptions.  From \cite{Cheung:2007st, Senatore:2010jy}, we know that the cutoff for the dark-energy sector is $\Lambda_U^4 \simeq 16 \pi^2 M_{2}^4 c_s^7$.  In the small $c_s^2$ limit, we have that $M_2^4 \approx \bar \rho_D ( 1 + w) / ( 4 c_s^2) $, and at the current time we have $\bar \rho_D \approx 3 H^2 \mpl^2$, so we can write the cutoff as $\Lambda_U^4 = 12 \pi^2 H^2 \mpl^2 ( 1 + w) c_s^5$.  Now, imposing that $\Lambda_U \gtrsim \alpha \, c_s \, \knl$, we find that 
\be
 |c_s| \gtrsim \frac{\alpha^4}{12 \pi^2 | 1 + w|} \left( \frac{ \knl}{H} \right)^4 \left( \frac{H}{ \mpl } \right)^2 \ , 
\ee 
or in terms of the dark matter speed of sound $ \bar c^2_{m} \sim H^2 / \knl^2$ we have 
\be
 | c_s | \gtrsim \frac{\alpha^4}{12 \pi^2 | 1 + w| } \frac{1}{\bar c_{m}^4} \left( \frac{H }{\mpl} \right)^2 \ .  
\ee
Using $\bar c_m \sim 10^{-3}$ and $H / \mpl \sim 10^{-60}$ today, the above constraint becomes $| c_s | \gtrsim  \alpha^{4} \, 10^{-110} /|1 + w|$.  If $c_s^2$ does not satisfy this, then the effective field theory would not be valid for computing at the non-linear scale.  Taking a hypothetical value of $c_s^2 \sim \bar c_m^2$, we see that the unitarity cutoff $\Lambda_U$ will be much higher than $\knl$.  This does not mean, however, that the non-linear scale determined by gravitational non-linearities in the dark-energy sector, $k_{{\rm NL}, D}$, will be so high: $\Lambda_U$ is the scale at which quantum fluctuations make the system strongly coupled, and $k_{ \rm NL,D}$ is the scale at which the non-linear couplings in the classical equations of motion become important.  }  Although it is perfectly consistent to have $|c_s^2| \ll \bar c_m^2$ (which must be true if $w < -1$), we can also consider the case that $c_s^2 \approx \bar c_m^2$ and perturbatively find the effects of a small but non-zero $c_s^2$.  When $c_s^2 \neq 0$, the two species are not comoving, and one would expect to have to solve the full system for both degrees of freedom $\delta_m$ and $\pi$, even at linear level.  However, we can get away with solving for just the adiabatic mode $\delta$ by considering this new feature perturbatively.  We start with the following equations for $\delta^{(ct)}$ sourced by $\delta^{(1)}$ in Fourier space (we suppress the $\kvec$ argument because it is the same on all fields) 
\begin{align} \label{firstone}
& \cH \delta^{(ct) '} + \frac{1}{a} \theta_{m}^{(ct)} + \frac{1}{a} ( 1 + w) \frac{\Omega_{D,0}}{\Omega_{m,0}} \left( \frac{a}{a_0} \right)^{-3w}  a^{-1} k^2 \pi^{(ct)} = 0  \\ \label{secondone}
& a\cH\theta_m^{(ct)'}  + \cH\theta_m^{(ct)}   - k^2 \Phi^{(ct)}  =  9 \, ( 2 \pi) \,H(a)^2 \frac{k^2}{\knl^2} \, c_{A,m}^2 ( a )  \delta^{(1)}  \\ \label{thirdone}
& \cH \frac{d }{d a } \left( a^3 M_2^4 \left( \cH \pi^{(ct)'}  - \Phi^{(ct)} \right) \right) = - c_s^2 a^3 M_2^4 a^{-2} k^2 \pi^{(1)} \ ,
\end{align}
where we have left off some relativistic terms proportional to $c_s^2$ in \eqn{firstone}.  Since we will need an explicit form for $- k^2 \pi^{(ct)}$ to plug into \eqn{firstone}, we integrate \eqn{thirdone} with respect to $a$ one time and take $\partial^2$ to obtain\footnote{Here, we have used the linear equations $a^{-1} k^2 \pi^{(1)} = \theta_m^{(1)} = - a \cH \delta^{(1)'} / C(a)$, and defined
\be 
f_1(a) = \frac{- k_{{\rm NL}, D}^2}{9 ( 2 \pi ) H^2 ( a )} \frac{1}{a^3 M^4_2 ( a ) } \int^a d \tilde a \, \frac{ \tilde a^3 M^4_2 ( \tilde a)}{C ( \tilde a) } \frac{D_+' ( \tilde a)}{D_+ ( a ) } \ .
\ee}
\begin{align}
- \cH k^2  \pi^{(ct)'} + k^2 \Phi^{(ct)} & = \frac{c_s^2}{a^3 M_2^4} \int^a d \tilde a \, \frac{ \tilde a^3 M_2^4 ( \tilde a ) }{ \cH ( \tilde a)} \tilde a^{-2}  k^4 \pi^{(1)} ( \tilde a )  \\
& = 9 ( 2 \pi ) H^2 ( a )\, c_s^2 \,  f_1 (a ) \, \frac{k^2 }{k_{{\rm NL}, D}^2} \delta^{(1)} ( a ) \ ,  \label{chance}
\end{align}
where we have written the correction in terms of the non-linear scale in the dark-matter sector $k_{{\rm NL},D}$, which in general could be different from the non-linear scale for dark matter $\knl$, but as shown in \eqn{knldapp} it is expected to be comparable.  However, for the rest of this paper, we will assume $k_{{\rm NL},D} = \knl$ for simplicity.  Next, add \eqn{chance} to \eqn{secondone} and integrate the result with respect to $a$ to obtain\footnote{We have defined
\be 
c_1^2 ( a ) = \frac{-1}{H(a)} \int^a d \tilde a \, \frac{H ( \tilde a)^2 }{\cH ( \tilde a) } \left( c_{A,m}^2 ( \tilde a ) + c_s^2 f_1 ( \tilde a ) \right) \frac{D_+(\tilde a )}{D_+( a )} \ . 
\ee}
\be \label{mansion}
a \theta_m^{(ct)} - k^2 \pi^{(ct)} =  -  9 ( 2 \pi ) H(a ) \, c_1^2 ( a ) \frac{k^2}{\knl^2} \delta^{(1)} ( a )  \ , 
\ee
which gives the deviation of the dark energy from the dark-matter trajectories.  In a sense, this generates an isocurvature mode, as the two species no longer move together.  Next, use \eqn{mansion} to replace $-k^2 \pi^{(ct)}$ in \eqn{firstone}, then take the time derivative of that and use \eqn{secondone} to replace $\cH \partial_a \left( a \theta_m^{(ct)} \right)$ to finally obtain\footnote{Again, we make some definitions 
\begin{align}
c_A^2(a) & = c_{A,m}^2 ( a ) + ( 1 + w) \frac{\Omega_{D,0}}{\Omega_{m,0}} \left( \frac{a}{a_0} \right)^{-3w} c_{A,2}^2 ( a ) \\
c_{A,2}^2 ( a ) & = \cH (a) \left( H(a)^2 D_+(a) a^{-3w} \right)^{-1} \frac{d}{da} \left( \frac{H(a) a^{-3w} c_1^2(a) D_+(a)}{C(a)} \right) \ .
\end{align}}
\be \label{finallygf}
- \cH(a) \frac{ d}{d a} \left( \frac{a^2 \cH (a) }{C(a)}  \frac{d \delta^{(ct)}}{ d a } \right) + \frac{3}{ 2} \frac{\Omega_{m,0} \cH_0^2 a_0 }{a} \delta^{(ct)} = 9 ( 2 \pi ) H(a)^2 c_A^2(a) \frac{k^2}{\knl^2} \delta^{(1)} \ .
\ee
Notice that the linear differential operator on the left-hand side is indeed the same as the one for the adiabatic mode in \eqn{delta-m34}.  Analogous to \eqn{csai}, we can use the Green's function of \eqn{finallygf} to define the speed of sound $\bar c_A^2 ( a )$ that enters the power spectrum\footnote{Analogous to \eqn{csai}, we define
\be
\bar c_A^2 (a)  = \int^a d \tilde a \, G_+( a ,\tilde a )  \frac{D_+(\tilde a)}{D_+(a)} 9 H( \tilde a)^2 c_A^2 (\tilde a  ) \ , 
\ee
where $ G_+ $ is the retarded Green's function for the linear equation 
\bea
&& - \cH(a) \frac{ d}{d a} \left( \frac{a^2 \cH (a) }{C(a)}  \frac{d G_+ ( a , \tilde a) }{ d a } \right) + \frac{3}{ 2} \frac{\Omega_{m,0} \cH_0^2 a_0 }{a} G_+( a , \tilde a)  = \delta^{(1)}_D(a-\tilde a) \ ,\n
&& G_+(a,a)=0\ , \qquad \left.\pd_aG_+(a,\tilde a)\right|_{ a=\tilde a}=\frac{C(\tilde a)}{\tilde a^2\cH(\tilde a)^2}\ .
\eea
 }
\be \label{countertermref}
\delta^{(ct)}_{\kvec} ( a  ) =  - ( 2 \pi ) \,  \bar c_A^2 ( a )   \frac{k^2}{\knl^2}  \frac{D_+( a ) }{D_+ ( a_i)}  \delta^{\rm in}_{\kvec}  \ . 
\ee
Thus, we see that at one-loop, both $c_s^2$ and $c_{A,m}^2$ contribute to the power spectrum with the same functional dependence.  Thus, even if $c_s^2$ is comparable to $c_{A,m}^2$, the effect is automatically included in $\bar c_A^2$.  

Of course, in this discussion we have neglected the initial isocurvature mode, which however is expected to be extremely small. In any case, its inclusion in the formalism and the calculations is straightforward, as it is identical to including the initial isocurvature mode for baryons (which is larger), as done in~\cite{Lewandowski:2014rca}.

\subsection{Power spectrum}
Up to now, we worked out the total density up to third order, including the counterterm contribution.  Now, we use this result to determine the adiabatic power spectrum in the presence of the dark energy up to one-loop order.
The equal-time power spectrum is defined in terms of the density variance as
\be
\langle\delta_{\vk}( a ) \delta_{\vk'}( a ) \rangle=  (2\pi)^3\delta_D(\vk+\vk')P(a , k) \ . 
\ee
Using the perturbative expansion of the field in \eqn{deltaexpansion}, and assuming Gaussian initial conditions,\footnote{Here, we restrict to Gaussian initial conditions, although it is straightforward to extend to non-Gaussian initial conditions \cite{Angulo:2015eqa, Assassi:2015jqa, Assassi:2015fma}.} we can write 
\be
P ( a , k ) = P_{11} ( a , k ) + P_{22} ( a , k ) + P_{13} ( a , k ) + P_{13}^{ct} ( a , k )  + \cdots
\ee
where the various contributions are given by 
\begin{align}
 \langle\delta^{(1)}_{\vk}( a ) \delta^{(1)}_{\vk'}( a ) \rangle ' & = P_{11} ( a , k) \\
 \langle\delta^{(2)}_{\vk}( a ) \delta^{(2)}_{\vk'}( a ) \rangle ' & =  P_{22} ( a , k) \\
2   \langle\delta^{(1)}_{\vk}( a ) \delta^{(3)}_{\vk'}( a ) \rangle '  & =  P_{13} ( a , k) \\
2   \langle\delta^{(1)}_{\vk}( a ) \delta^{(ct)}_{\vk'}( a ) \rangle '  & =  P_{13}^{ct} ( a , k) \ , 
 \end{align}
and $\langle \cdots \rangle ' $ means that we have removed a factor of $(2\pi)^3\delta_D(\vk+\vk')$ from the expectation value.  In particular, on the initial conditions, this means that $\langle \delta^{\rm in}_{\kvec} \delta^{\rm in}_{\kvec ' } \rangle' = P^{\rm in}_{\kvec}$.  

Then, using the linear solution \eq{linear-sol}, the second-order solution \eq{delta-2}, the third-order solution \eq{delta-3}, and the counterterm solution \eq{countertermref}, we have the following expressions for the power spectrum contributions
\begin{align}
&P_{11}(a, k)=\frac{ D^2_{+}(a)  }{  D^2_+ ( a_i ) }P^{\rm in}_{\vk},\\
&P_{22}(a , k)=2\int \frac{ d^3q}{(2 \pi)^3 } \bigg(\alpha_s(\vq,\vk-\vq)\mG^{\delta}_{1}(a)+\beta(\vq,\vk-\vq)\mG^{\delta}_{2}(a)\bigg)^2P^{\rm in}_{\vk-\vq}  \, P^{\rm in}_{\vq},\\\label{P13}
&P_{13}(a , k)= 4 \frac{ D_{+}(a)}{D_+ ( a_i)} P_{\vk}^{\rm in}\int \frac{d^3q}{ (2 \pi)^3} \bigg(\alpha^{\sigma}(\vk,\vk+\vq, \vk )\mU^{\delta}_{\sigma}(a)+\beta^{\sigma}(\vk,\vk+\vq,\vk)\mV^{\delta}_{\sigma2}(a) \nonumber \\ 
& \hspace{3.5in} +\gamma^{\sigma}(\vk,\vk+\vq,\vk)\mV^{\delta}_{\sigma1}(a)\bigg)P_{\vq}^{\rm in}  \ ,  \\
& P_{13}^{ct} ( a , k ) = - 2 \, ( 2 \pi ) \, \bar c_A^2 ( a ) \frac{k^2}{\knl^2} \left( \frac{ D_+ ( a ) }{D_+ ( a_i ) } \right)^2 P^{\rm in }_{\kvec} \ . 
\end{align}
We can also write the one-loop contributions in a more compact form.  $P_{22}$ can be written as
\be\label{P22--}
P_{22}(a, k)=2\bigg(\A^{\sigma\tilde{\sigma}}(k)\mG^{\delta}_{\sigma}(a)\mG^{\delta}_{\tilde\sigma}(a)\bigg),
\ee
where the symmetric momentum matrix $\A^{\sigma\tilde{\sigma}}(k)$ is given as
\begin{align}
&\A^{11}(k)=\int   \frac{d^3q}{(2 \pi)^3}   \bigg(\alpha_s(\vq, \vk-\vq)\bigg)^2P^{\rm in}_{\vk-\vq}P^{\rm in}_{\vq},\\
&\A^{22}(k)=\int    \frac{d^3q}{(2 \pi)^3}   \bigg(\beta(\vk-\vq,\vq)\bigg)^2P^{\rm in}_{\vk-\vq}P^{\rm in}_{\vq},\\
&\A^{12}(k) =  \A^{21} ( k )  =\int    \frac{d^3q}{(2 \pi)^3}   \bigg(\alpha_s(\vk-\vq,\vq)\beta(\vk-\vq,\vq)\bigg)P^{\rm in}_{\vk-\vq}P^{\rm in}_{\vq}.
\end{align}
Moreover, from \eqn{P13}, we can write $P_{13}$ as
\be\label{P13--}
P_{13}(a, k )= 4 \frac{ D_{+}(a) }{ D_+ ( a_i ) } \bigg(\B^{\sigma\tilde{\sigma}}(k)\mV^{\delta}_{\sigma\tilde{\sigma}}(a)+\C^ \sigma(k)\mU_{\sigma}^{\delta}(a)\bigg),
\ee
where the momentum matrix $\B^{\sigma\tilde{\sigma}}(k)$ is
\begin{align}
&\B^{11}(k)=    P^{\rm in}_{\vk}\int  \frac{d^3q}{(2 \pi)^3}   \alpha(\vk+\vq,-\vq)\alpha_s(\vk,\vq)P^{\rm in}_{\vq} ,\\
&\B^{12}(k)=   P^{\rm in}_{\vk}\int   \frac{d^3q}{(2 \pi)^3}  2  \beta(  -\vq   , \vk+\vq)\alpha_s(\vk,\vq)P^{\rm in}_{\vq} ,\\
&\B^{21}(k)=   P^{\rm in}_{\vk}\int   \frac{d^3q}{(2 \pi)^3}   \alpha ( \kvec + \qvec , - \qvec ) \beta ( \kvec , \qvec)  P^{\rm in}_{\vq},\\
&\B^{22}(k)=   P^{\rm in}_{\vk}\int   \frac{d^3q}{(2 \pi)^3}     2  \beta(-\vq   , \vk+\vq) \beta  (\vk,\vq)      P^{\rm in}_{\vq} ,
\end{align}
and $\C^{\sigma}(k)$ is given as 
\begin{align}
&\C^1(k)= P^{\rm in}_{\vk}\int      \frac{d^3q}{(2 \pi)^3}      \alpha(-\vq,\vk+\vq)\alpha_s(\vk,\vq)P^{\rm in}_{\vq}  ,\\
&\C^2(k)= P^{\rm in}_{\vk}\int   \frac{d^3q}{(2 \pi)^3}    \alpha ( - \qvec , \kvec + \qvec ) \beta ( \kvec , \qvec)    P^{\rm in}_{\vq}  .
\end{align}
Thus, the final one-loop computation requires us to compute nine integrals over momentum and ten integrals over time.

%%%%%%%%%%%%%%%%%%%%%%%
%
%
%       Biased Tracers
%
%
%%%%%%%%%%%%%%%%%%%%%%%

\section{Biased tracers}
Up to now, we studied the total underlying density contrast $\delta_A$, corresponding to the Newtonian potential $\Phi$, and computed its power spectrum to one-loop level. Those are the quantities which are measured by weak lensing (WL) surveys \cite{Amendola:2016saw}.\footnote{In fact, the weak lensing and the integrated Sachs–Wolfe (ISW) effect, are measuring the lensing potential, $(\Phi+\Psi)/2$. However since in our model we have $\Phi=\Psi$ (no anisotropic stress), the lensing and the Newtonian potentials are equal.}  The other promising cosmological observation, complementary to WL, are large-scale structure surveys which observe the overdensity of collapsed objects, rather than the underlying density contrast. Therefore, for precision cosmology, it is important to relate observable properties of tracers (e.g. density of galaxies) to the initial conditions and underlying total matter distribution. In its most general form, biased tracers can be non-local and stochastic functions of the underlying dark matter density and velocity. Over the past few years, considerable progress has been made in this direction. In particular, the formulation of biased tracers in the EFTofLSS has been worked out in \cite{Senatore:2014eva} (or equivalently, \cite{Mirbabayi:2014zca}) which generalized and completed the analysis  of \cite{McDonald:2009dh}. Recently, in \cite{Angulo:2015eqa}, the predicted biased tracers in the EFTofLSS with two species, dark matter, and baryons, has been worked out, and, when restricted to dark matter only, compared to numerical simulation. 
Here, following \cite{Senatore:2014eva, Angulo:2015eqa}, we describe the bias expansion for the density of collapsed objects in the presence of clustering quintessence.  In the case of the smooth dark energy with $c_s=1$ (described in Appendix~\ref{smoothdesection}), we expect the effect of dark energy perturbations is negligible and therefore its contribution to the bias is only through the expansion rate.\footnote{Deep inside the horizon, the divergence of the velocity grows with respect to $\delta_A$. However, deep inside the horizon, it also oscillates rapidly compared to the time scale of formation of halos, which is Hubble. Therefore, we expect that its contribution will be highly suppressed as well.}

For the purpose of this work, we consider Gaussian and adiabatic initial conditions for the total density and neglect the perturbatively suppressed effect of baryons,\footnote{It has been explicitly shown in \cite{Lewandowski:2014rca} and \cite{Angulo:2015eqa} that the effect of baryons is perturbatively suppressed.} which can be straightforwardly included.  Thus, in our setup, the final halo overdensity, $\delta_{ h}$, can only be a function of the tracers' trajectory as well as the local observables of the dark matter and dark energy. We expect that the contribution of the long wavelength perturbations of dark matter and dark energy is weighted by their density parameters which are  
\be
\Omega_{m}(t)=\Omega_{m,0}a(t)^{-3} \quad \textmd{and} \quad \Omega_{D}(t)=\Omega_{D,0}a(t)^{-3(1+w)} .
\ee
As a result, we expect that the effect of dark energy perturbations is negligible during the matter era while it can be important at low redshifts as the dark energy becomes dominant. At this point, for simplicity, we define $\phi$ which is the rescaled version of the Newtonian potential, $\Phi$, so that
\be
\partial^2\phi\equiv\delta_A.
\ee 
In the presence of clustering dark energy, we have the following generalization to the overdensity of halos in Eulerian space
\bea\label{eq-euler_bias}
&&\delta_h(\vec x,t)\simeq \int^t dt' H(t')\; \bigg(   \bar c_{\partial^2\phi}(t,t')\; \frac{\partial^2\phi(\vec{x}_{\rm fl},t')}{H(t')^2} +\bar{c}_{\delta_D}(t,t')\Omega_D(t')\delta_D(\vec{x}_{\rm fl},t')    \\    
&&\quad + \bar c_{\partial_i v^i}(t,t') \frac{\partial_i v_A^i(\vec{x}_{\rm fl},t')}{H(t')}     +\bar c_{\partial_i v^i_{\rm rel.}}(t,t') \frac{\partial_i v_{\rm rel.}^i(\vec{x}_{\rm fl},t')}{H(t')}      \nonumber \\
&& \quad  + \bar c_{\partial_i \partial_j \phi \partial^i \partial^j \phi}(t,t') \;\frac{\partial_i\partial_j \phi(\vec{x}_{\rm fl}, t')}{H(t')^2}\frac{ \partial^i \partial^j \phi(\vec{x}_{\rm fl},t')}{H(t')^2} + \ldots  \nonumber   \\  
&&\quad+ \bar c_{\epsilon_m}(t,t')\;\Omega_{m}(t')\,\epsilon_m(\vec{x},t')+ \bar c_{\epsilon_D}(t,t')\;\Omega_{D}(t')\,\epsilon_D(\vec{x},t') \nonumber \\
&&\quad+\big[\bar c_{\epsilon_m\partial^2\phi}(t,t') \;\Omega_{m}(t')\,\epsilon_m(\vec{x}_{\rm fl},t')+\bar c_{\epsilon_D\partial^2\phi}(t,t') \;\Omega_{D}(t')\,\epsilon_D(\vec{x}_{\rm fl},t')\big]\frac{\partial^2\phi(\vec{x},t')}{H(t')^2}+  \ldots \nonumber \\ \nonumber
&&\quad+  \bar c_{\partial^4\phi}(t,t')   \;\frac{\partial^2_{x_{\rm fl}}}{k_{M}^2}\frac{\partial^2\phi(\vec{x},t')}{H(t')^2}+
\ldots \bigg) \ ,
\eea
where $k_{\rm M}$ is the comoving wavenumber enclosing the mass of the halo, $\vec{x}_{\rm fl}$ is defined as
\be
\vec{x}_{\rm fl}(\vec{x},\tau,\tau')=\vec{x}-\int^{\tau}_{\tau'}d\tau'' \vec{v}_A(\tau'',\vec{x}_{\rm fl}(\vec{x},\tau,\tau')),
\ee
and the operators which are labelled by $\epsilon$ are describing the stochastic effects.
Note that in the above, we have included the possibility that the clustering depends independently on the dark energy (similar to what happens for baryons~\cite{Angulo:2015eqa}). One therefore needs to work out what are the diffeomorphism invariant combinations that one can write out of $\pi$. At linear level, there are just two such combinations: $\dot\pi-\Phi$ and $\partial_i\pi$, which, unsurprisingly, are respectively equal to the (rest frame) overdensity $\delta_D$ and the velocity $v_D^i$ (a similar construction can be carried on at higher order). However, as we discussed in Section \ref{countertermsection}, the isocurvature mode is very small, and is generated by the counterterms. In the absence of an initial isocurvature mode, the generated relative velocity is proportional to the gradient of $\delta_A$, and so it is degenerate with terms we could have written just in terms of it. If fact, without the presence of initial isocurvature modes, it is unlikely that a non-degenerate term is ever generated.

The expression \eqn{eq-euler_bias} allows us in principle to compute the correlation functions of tracers in the presence of dark energy. We leave the computation, and the inclusion of baryons and primordial non-Gaussianities, to future work.

%%%%%%%%%%%%%%%%%%%%%
%
%   Results
%
%%%%%%%%%%%%%%%%%%%%%%%

\section{Results}  \label{resultssection}

In this section, we present the results of our numerical computations.\footnote{The Mathematica notebook used for the computations in this section is available at \website}  All of our plots are done at $a=1$ and we use the cosmological parameters $\Omega_{m,0} = 0.27$, $\Omega_{D,0} = 0.73$ , $H_0 = 71 \, \text{km/s/Mpc}$, $\Delta_\zeta^2 = 2.42 \times 10^{-9}$ and $n_s = 0.963$.  For comparison, we also include dark matter evolution in the presence of a smooth dark energy with $c_s^2 = 1$, typically called $w$CDM, which is a theory that provides a familiar and simple example against which to compare our results (see Appendix~\ref{smoothdesection}).  For notational convenience, we can write the various power spectra at $a=1$ as 
\be
P^i [ w , \bar c^2_i ] = P^i_{11}[w] + P^i_{1-{\rm loop}}[w] - 2 ( 2 \pi )  \, \bar c^2_i \, \left( \frac{k}{\knl} \right)^2 P^i_{11}[w] \ , 
\ee
where $i$ stands for $\Lambda$CDM, $w$CDM, or clustering quintessence (CQ).  Furthermore, it is useful to parameterize the speed of sound of the adiabatic mode as 
\be \label{xidef}
\bar c_A^2 = \bar c_m^2 \left( 1 + \xi \, ( 1 + w ) \frac{\Omega_{D , 0}}{ \Omega_{m,0}} \left( \frac{a}{a_0} \right)^{-3w}  \right) ,
\ee
where $\bar c_m^2$ is the value of the speed of sound in $\Lambda$CDM ($w=-1$), and $\xi$ encodes the deviation due to the extra species when $w \neq -1$.  In general, we expect $\xi \sim \mathcal{O}(1)$ since we expect the effect to be of similar order as the ratio of the energy densities and proportional to $1+w$.  Although we do not discuss the time dependence in this paper, we include the time dependent factor in \eqn{xidef} so that the effect goes to zero at early times, as expected.  Because we are not comparing to simulation data in this paper, and we would just like to stress the non-linear corrections, we choose a reasonable value of $\bar c_m^2 = 0.20 \, (\knl/ \unitsk)^2$ taken from a previous comparison to non-linear dark-matter data \cite{Lewandowski:2015ziq}.  We would also like to note that while the plots in this section were made from our code with the exact $1+w$ dependence in the growth factor and loops, we have also implemented in our code an approximate computation that expands to first order in $1+w$.  In the latter case, one does not have to rerun the computations of the loops and time dependent functions for each $w$.  The trade-off is that one has to make about twice as many computations to start with, but then can freely explore different values of $w$, which overall saves computational time.

\begin{figure}[htb!] 
\begin{center}
\includegraphics[width=8cm]{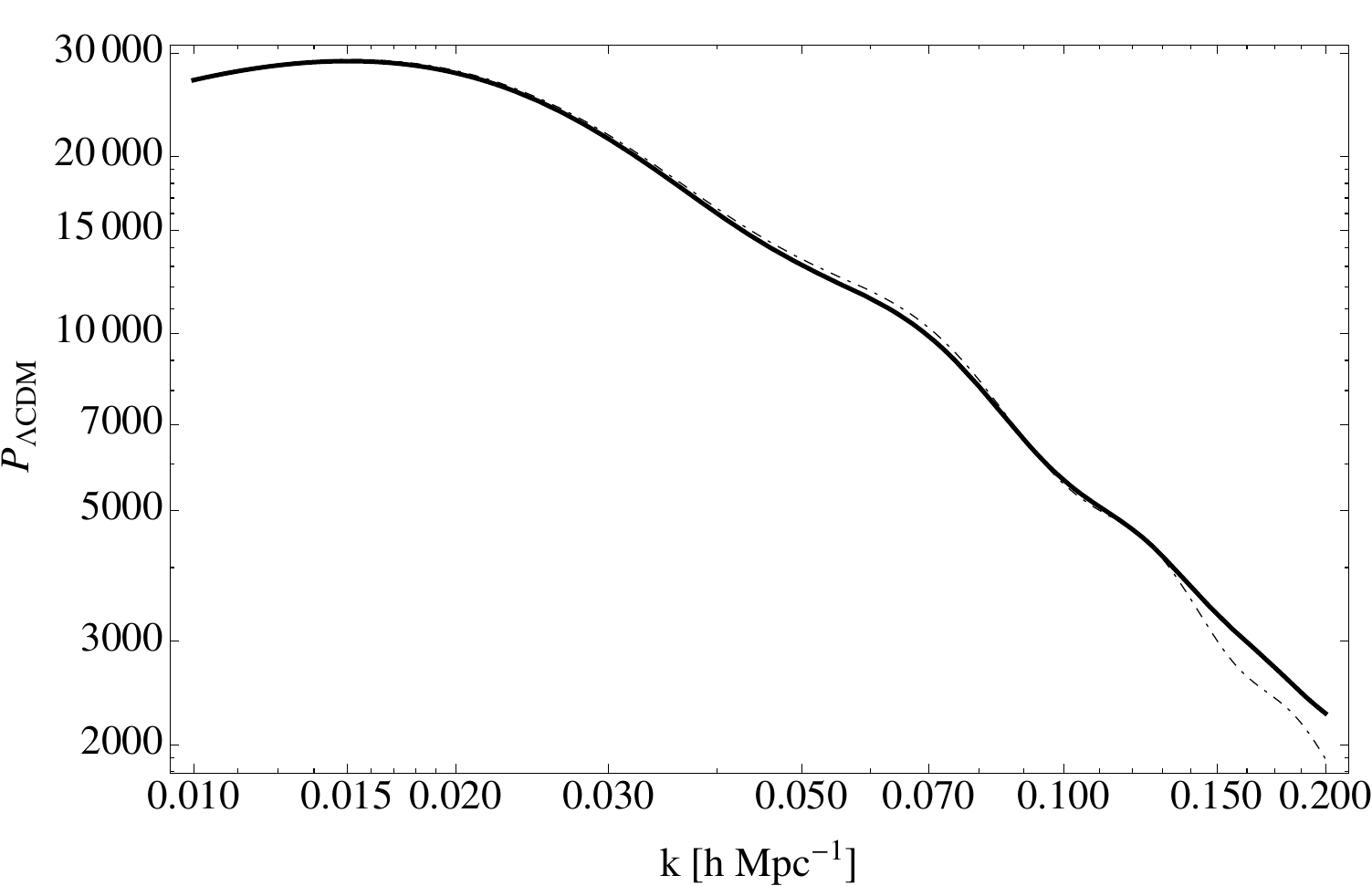} \includegraphics[width=8cm]{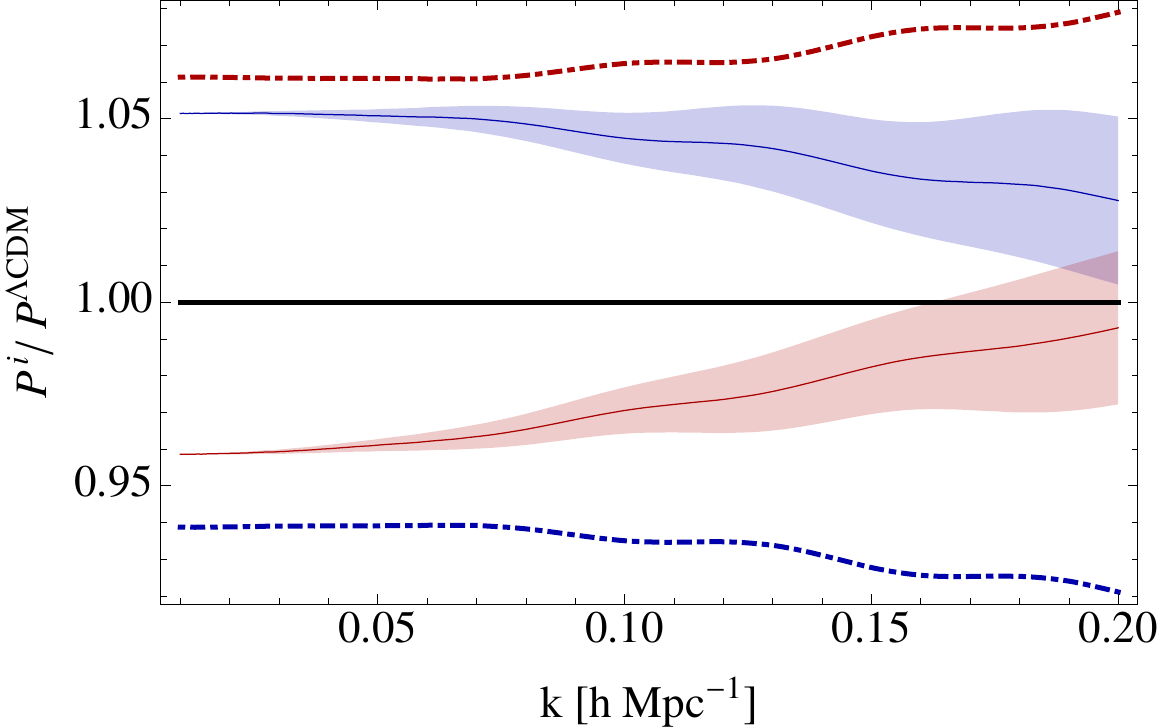} 
\caption{On the left-hand side, we show the linear (dot-dashed) and one-loop (solid) power spectra in $\Lambda$CDM (i.e. $w=-1$) for reference.  On the right-hand side, we plot the one-loop computations compared with the one-loop $\Lambda$CDM power spectrum.  The blue curves have $w = -0.9$, and the red curves have $w = -1.1$, the dot-dashed curves are the $w$CDM power spectra, the solid curves are clustering quintessence with $\xi = 0$ (defined in \eqn{xidef}), and the bands around them are $-1 \leq \xi \leq 1$.  All curves have $\bar c_m^2 = 0.20 \, (\knl/ \unitsk)^2$.      }  \label{comp1}
\end{center}
\end{figure}

In Figure \ref{comp1}, we compare clustering quintessence and $w$CDM to the $\Lambda$CDM power spectrum.  Although $w$CDM with $w < -1$ \emph{does not exist} as a consistent theory, we plot it for illustration purposes only.  As expected, both clustering quintessence and $w$CDM are different from $\Lambda$CDM even at low~$k$ because of the different background evolution for $w \neq -1$.  Also, the overall size of the corrections is of the expected order $1 + w$.  In order to isolate the non-linear corrections in each model, we find it useful in Figure \ref{comp2} to plot 
\be
R^i \equiv \frac{P^i }{P^i_{11}} \ . 
\ee
From this plot, it is clear that the size of the corrections at low $k$, when compared to the relevant linear power spectra, all go to zero as expected.  From Figure \ref{comp1}, we see that the non-linear corrections in clustering quintessence generically tend to make the power spectrum more like $\Lambda$CDM at higher values of $k$, while in $w$CDM the non-linear corrections continue to make the power spectrum different from $\Lambda$CDM.  This explains the potentially confusing fact that, for example, the blue clustering quintessence curve is above $1$ in Figure \ref{comp1} but below $1$ in Figure \ref{comp2}: in Figure~\ref{comp2}, there is no information about the relative size of the $\Lambda$CDM and clustering quintessence power spectra, and the reason that the blue curve is below $1$ in Figure~\ref{comp2} is the same as why it is decreasing in Figure \ref{comp1}.  Then, in Figure \ref{comp3}, we examine the individual corrections, for which it is useful to write 
\begin{align}
P^i_{11} & = P^{\Lambda CDM}_{11} + \Delta P^i_{11} \\
P^i_{1-{\rm loop}} & = P^{\Lambda CDM}_{1-{\rm loop}} + \Delta P^i_{1-{\rm loop}}  \ .
\end{align}

\begin{figure}[htb!] 
\begin{center}
\includegraphics[width=8cm]{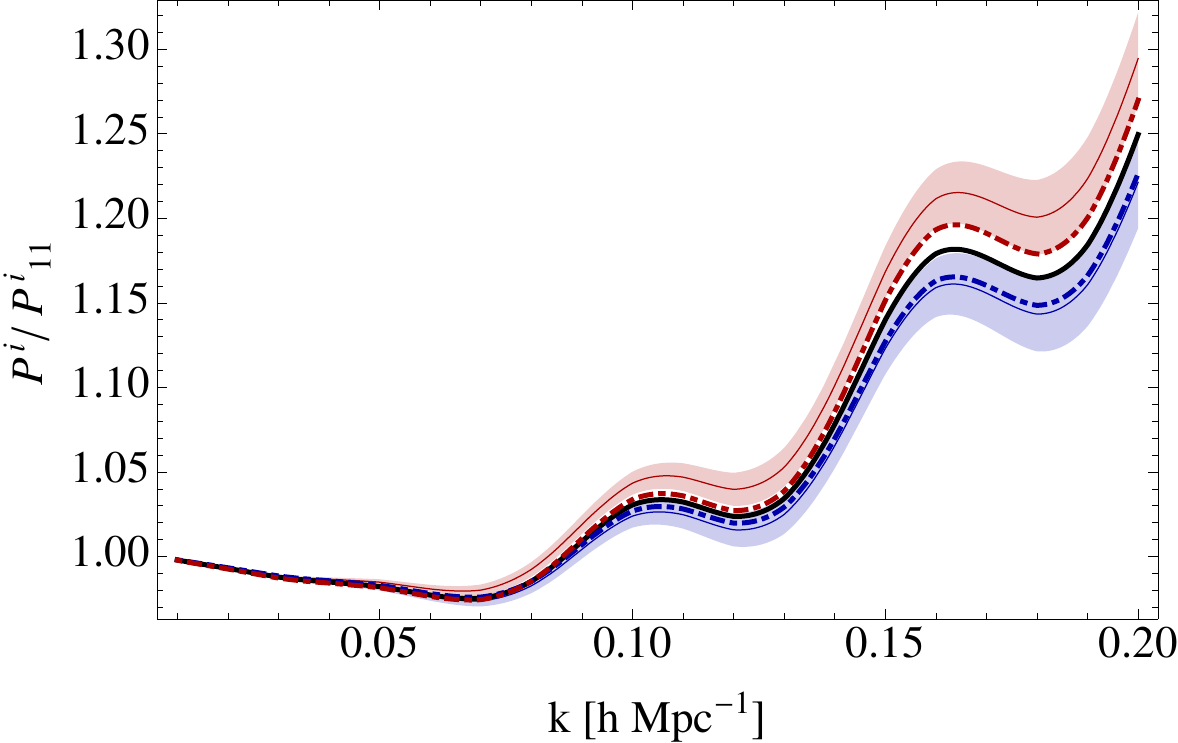} \includegraphics[width=8cm]{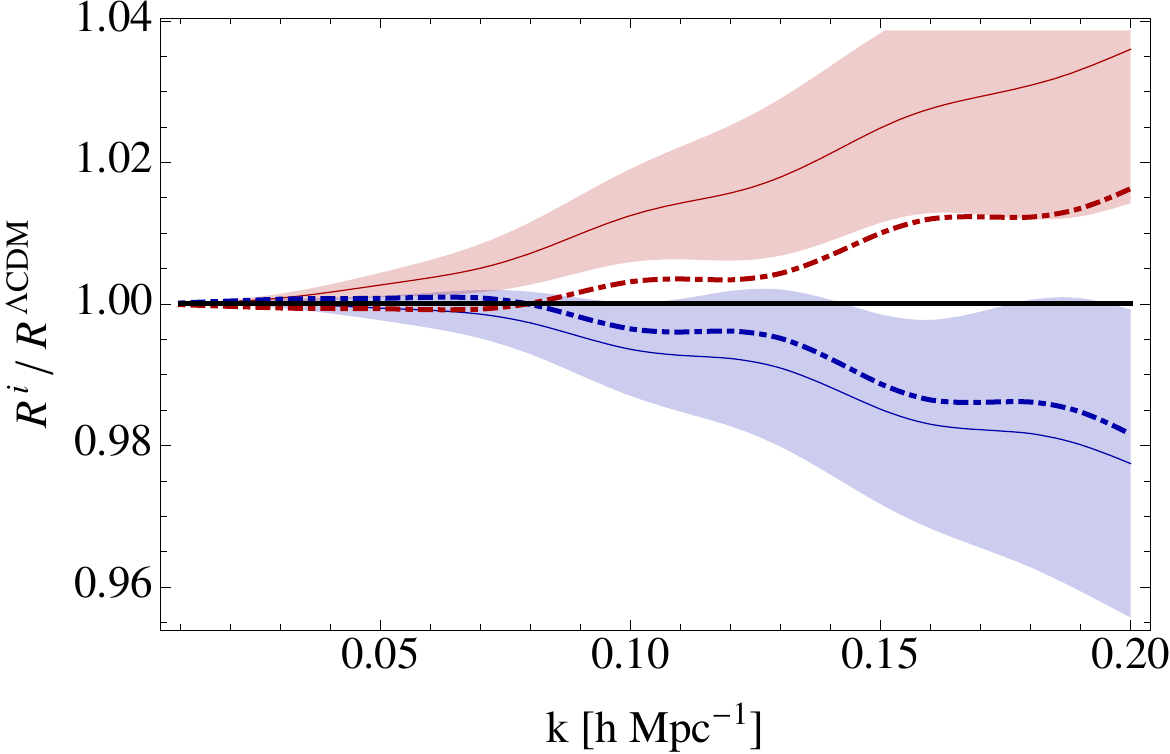}
\caption{On the left-hand side, we show the size of the non-linear corrections for the various power spectra by plotting $P^i / P^i_{11}$, where $i$ stands for $\Lambda$CDM, $w$CDM, or clustering quintessence.  On the right-hand side, we compare the size of the non-linear corrections for the various power spectra to the size of the corrections in $\Lambda$CDM by plotting $\frac{P^i / P^i_{11}}{P^{\Lambda CDM} / P^{\Lambda CDM}_{11} }$.  In both plots, the black curve is $\Lambda$CDM, blue curves have $w = -0.9$, the red curves have $w = -1.1$, the dot-dashed curves are the $w$CDM power spectra, the solid curves are clustering quintessence with $\xi = 0$, and the bands around them are $-1 \leq \xi \leq 1$.  All curves have $\bar c_m^2 = 0.20 \, (\knl/ \unitsk)^2$.      }  \label{comp2}
\end{center}
\end{figure}

\begin{figure}[htb!] 
\begin{center}
\includegraphics[width=8cm]{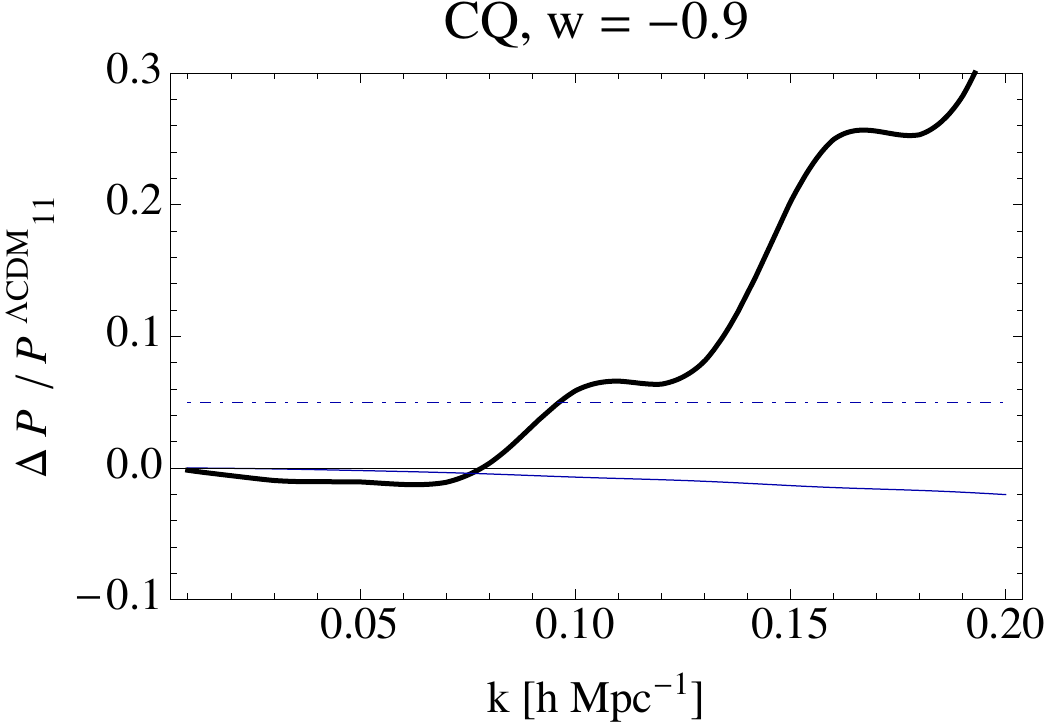} \includegraphics[width=8cm]{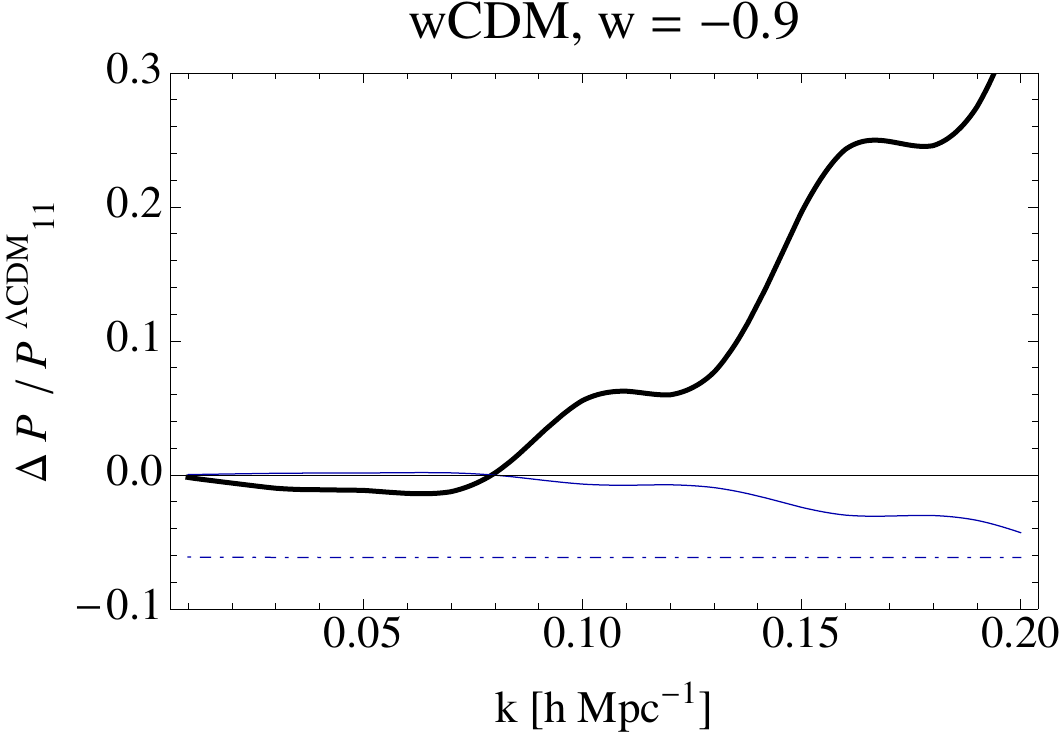}  \\ \vspace{.1in}
\includegraphics[width=8cm]{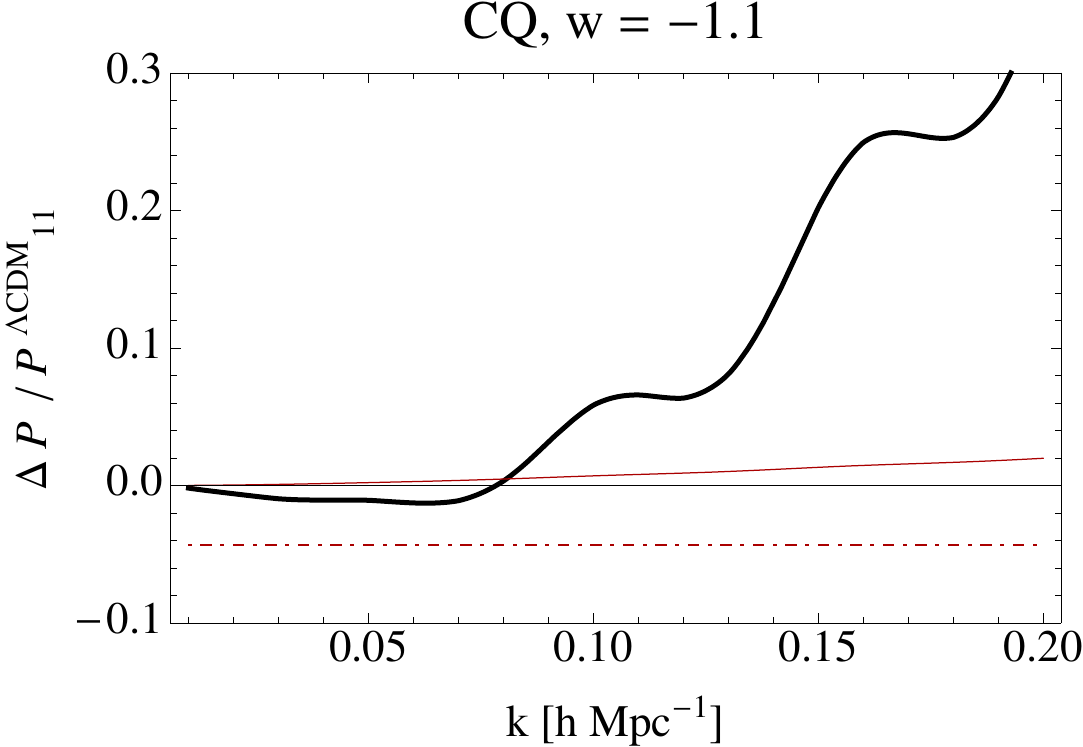} \includegraphics[width=8cm]{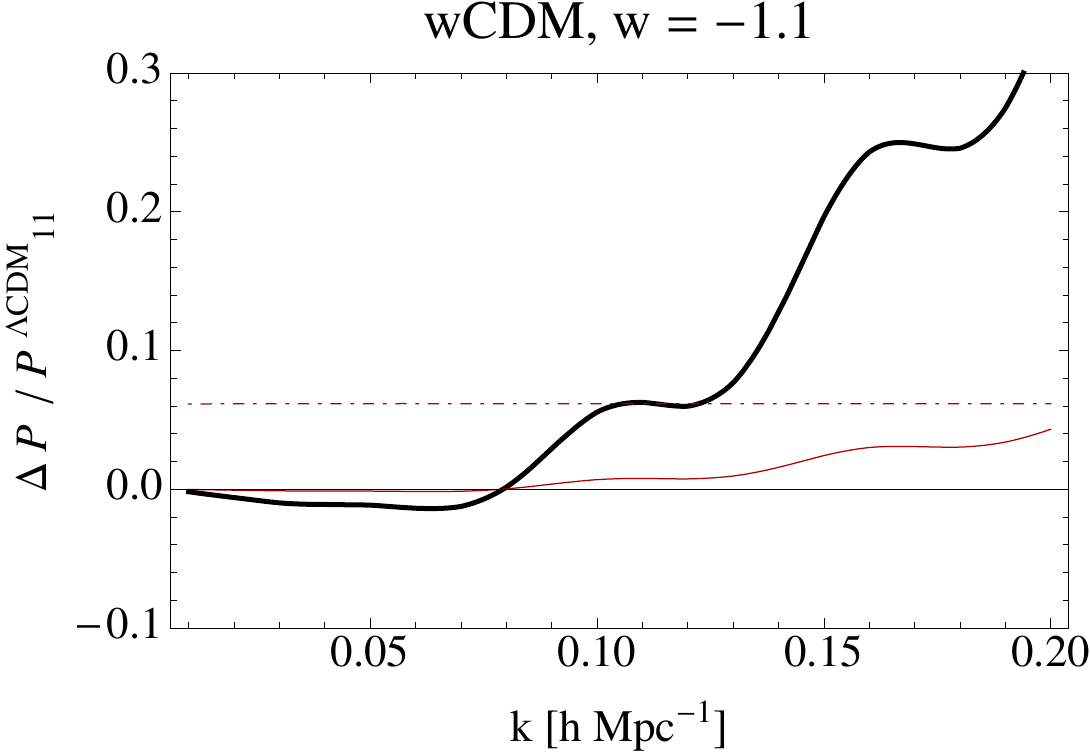}  
\caption{ In these plots, we examine the various $1+w$ corrections to the power spectrum, which we decompose as $P^i_{11}  = P^{\Lambda CDM}_{11} + \Delta P^i_{11}$ and $P^i_{1-{\rm loop}}  = P^{\Lambda CDM}_{1-{\rm loop}} + \Delta P^i_{1-{\rm loop}} $.  All of the thick black curves are $P^{\Lambda CDM}_{1-{\rm loop}}$, the dot-dashed curve is $\Delta P^i_{11}$, the thin solid curve is $\Delta P^i_{1-{\rm loop}}$, the blue curves have $w=-0.9$, and the red curves have $w = - 1.1$.      }  \label{comp3}
\end{center}
\end{figure}

Finally, as an indication of the degeneracy between $w$CDM and clustering quintessence in the adiabatic power spectrum, we present Figure \ref{degen}, which shows that the two can be brought within 1\% of each other by altering the values of $w$ and $\bar c_A^2$.  Although we recognize this possibility, we do not investigate this degeneracy further, since it is out of the scope of this paper.  We do note, however, that while there is a degeneracy in the adiabatic power spectrum, this will not be the case in other observables.  For example, non-clustering $w$CDM has $\delta_A \approx \delta_m$, but in clustering quintessence $\delta_A$ and $\delta_m$ are order $(1+w)$ different.  Also, the physics is much different in the early universe because $w$CDM is an extra relativistic species, while clustering quintessence is always non-relativistic.

\begin{figure}[htb!] 
\begin{center}
\includegraphics[width=10cm]{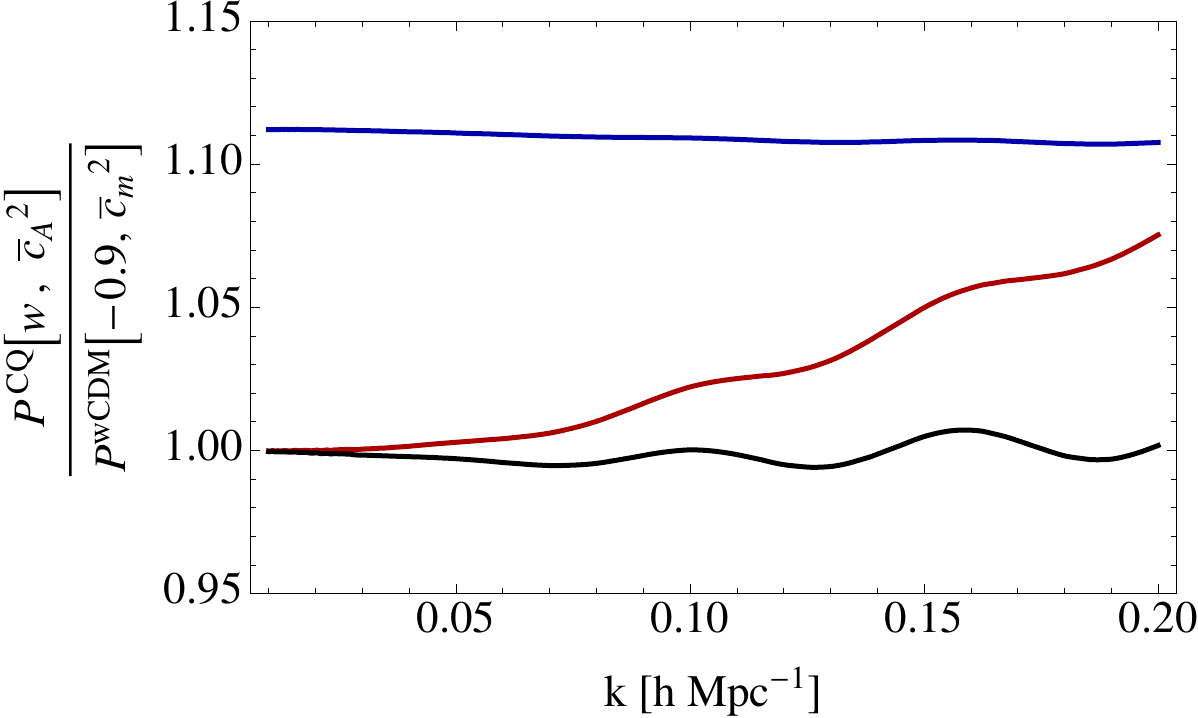} 
\caption{In this plot, we show how the adiabatic power spectrum in clustering quintessence can mimic the one from $w$CDM with different values of $w$ and $\bar c_A^2$.  In this plot, for the power spectrum in clustering quintessence, we take the indicative values of $w =-0.9$ and $\bar c_m^2 = 0.20 \, (\knl/ \unitsk)^2$ and we vary the values of $w$ and $\bar c_A^2$ in the clustering quintessence power spectrum.  The blue curve has $w=-0.9$ and $ \bar c_A^2 = \bar c_m^2$, the red curve has $w = -1.15$ and $ \bar c_A^2 = \bar c_m^2$, and the black curve has $w = -1.15$ and $\bar c_A^2 = 1.9  \,  \bar c_m^2$.   }  \label{degen}
\end{center}
\end{figure}

\section{Conclusions}\label{discussion-section}

The observational study of dark energy will make tremendous progress in the next few years thanks to the remarkable way we will be able to probe the Large-Scale structure of the universe either through galaxy surveys or through the CMB.  Since most of the information is stored at high wavenumbers, it is important to have an accurate description of the mildly non-linear regime, which is amenable to a perturbative analysis. The Effective Field Theory of Large-Scale Structure provides the formalism to perform such analytic predictions in an accurate way. In the presence of dark energy, the clustering of LSS is affected both by the background cosmology and by the perturbations of dark energy. A very useful formalism to study the phenomenology of dark energy is the so-called Effective Field Theory of Dark Energy~\cite{Creminelli:2006xe,Cheung:2007st,Creminelli:2008wc,Gubitosi:2012hu, Gleyzes:2013ooa,Gleyzes:2014rba}, which assumes that dark energy is a system that spontaneously breaks time diffeomorphisms, and the fluctuating degree of freedom is the associated Goldstone boson. The advantage of using such a Lagrangian formalism to describe dark energy instead of a more general setup where one generically parametrizes some observational signatures is that the Lagrangian formulation makes it easy to ensure that our signatures originate from a system compatible with our well-established principles of physics: locality, causality, unitarity, etc.

In this paper, we have studied the dynamics of LSS in the presence of dark energy, with particular focus on the mildly non-linear regime. We formulated the set of non-linear equations for the system, including the relevant counterterms that account for the effect of short-distance physics at long distances, and that are modified in the presence of dark energy. We have also derived the equations that describe the clustering of biased tracers. Specializing for definiteness to the case of clustering quintessence, we have then perturbatively solved the equations of motion for dark matter and dark energy. This has allowed us to produce the one-loop power spectrum of the total density contrast. We have then discussed the different behaviors in the presence of smooth and clustering quintessence, for $w\gtrless -1$, and for the linear and the non-linear solution. Finally, we have begun to discuss the effect on the predicted clustering of the modification of the numerical value of the counterterms that is expected in the presence of dark energy, and how these several effects can mimic each other.

There are several directions in which our work can be continued. One could perform calculations to higher order, to explore the $k$-reach of the theory. One can choose different parameters for the EFTofDE, effectively studying novel dark energy models: such an endeavor is already well developed at linear level~\cite{Hu:2013twa, Zumalacarregui:2016pph, coopref}, and it would be nice to construct it at the mildly-non-linear one. It would be interesting to study how much the unknown counterterms reduce the available information, and, in general, to perform numerical simulations of clustering quintessence to map out these counterterms in the case of dark energy. The EFTofLSS is in this case particularly useful because it should allow a more efficient implementation of the numerical simulations (see for example~\cite{Baldauf:2011bh,Baldauf:2015vio,Lazeyras:2015lgp}), and also allows us to map out the very large parameter space using the Taylor expansion techniques of~\cite{Cataneo:2016suz}.

\section*{Acknowledgments}

A.M. is grateful to the hospitality of Stanford University where this work was initiated and acknowledges partial support from Allameh Tabatabaii grant of Boniad Melli Nokhbegan Iran. L.S. is partially supported by DOE Early Career Award DE-FG02-12ER41854.  M.L. is partially supported by a CEA Enhanced Eurotalents fellowship.

\section*{Appendices}

\appendix

\section{Details regarding $\delta K$}\label{Gold-appendix}

%\input{A-appendix_v21} 
%\label{deltakappendix}

In this appendix, we review the details of the computation of $\delta K_{\rm u}$ and consider higher order terms, thus extending the computations done in \cite{Cheung:2007sv} and \cite{Gubitosi:2012hu}.  

\subsection{Review of the ADM formalism}
In this section, we mostly follow the notation and expressions in \cite{poisson} and \cite{wald}.  As a review of the formalism and notation, consider a space with coordinates $x^\mu$ and metric $\gmn$.  Then consider a foliation of the space by space-like hypersurfaces $\Sigma_t$, where $t$ labels the hypersurface.  This foliation can be generated by a function $t(x^\mu)$ such that $t ( x^\mu ) = t_1$ defines the three-dimensional hypersurface $\Sigma_{t_1}$.  The one-form $d t$ is normal to the surface $\Sigma_t$, so we can define a unit normal whose components in the basis $\{ d x^\mu \}$ are given by 
\be
n_\mu = -N \frac{\partial t}{\partial x^\mu},
\ee
where $N$ is the lapse function and is given by 
\be
N = \frac{1}{\sqrt{- g^{\mu \nu} \partial_\mu t \,  \partial_\nu t }} \ . 
\ee
Now, on each of the hypersurfaces, let $y^i$, for $i=1,2,3$ be coordinates.  Then, in order to use $(t , y^i)$ as coordinates on the whole space, we first introduce a time-flow vector field defined by 
\be
t^\mu = \frac{\partial x^\mu}{\partial t},
\ee
which satisfies $t^\mu \partial_\mu t = 1$.  This ensures that as we move in the direction of $t^\mu$ by an amount $\Delta t$ away from surface $\Sigma_{t_1}$, we end up on the surface $\Sigma_{t_1 + \Delta t}$.  We also have a set of three $4$-vectors which are tangent to the hypersurfaces
\be
e^\mu{}_i = \frac{\partial x^\mu}{\partial y^i },
\ee
and satisfy $e^\mu{}_i n_\mu = 0$ (this is also the transformation matrix used in pulling back the metric onto $\Sigma_t$, as we will see).  In general, $t^\mu$ will not be parallel to $n^\mu$, so we can define their relation as 
\be
t^\mu = N n^\mu + N^i e^\mu{}_i  \ , 
\ee
and $N^i$ is called the shift vector.  We can imagine the original coordinates as a function of the new, adapted coordinates: $x^\mu ( t , y^i)$, which means that the coordinate one forms are related by $d x^\mu = t^\mu dt + e^\mu{}_i dy^i$.  The metric is then given by 
\begin{align}
ds^2 = \gmn d x^\mu \, d x^\nu & = \gmn \left( t^\mu dt + e^\mu{}_i dy^i \right) \left( t^\nu dt + e^\nu{}_i dy^i \right) \\
& = - N^2 dt^2 + \hat g_{ij} \left( dy^i + N^i dt \right) \left( dy^j + N^j dt \right)  \label{admmetric1},
\end{align}
where $\hat g_{ij} \equiv \gmn e^\mu{}_i e^\nu{}_j$ is the induced spatial metric (i.e. the metric $\gmn$ pulled back to the hypersurface).  This is the standard ADM parametrization.

From \eqn{admmetric1}, we can read off that $g_{00} = -N^2 + \hat g_{ij} N^i N^j$, $g_{0i } = \hat g_{ij} N^j$, and $g_{ij} = \hat g_{ij}$.  Other important relationships between the ADM variables and the full inverse metric components $g^{\mu \nu}$ in the adapted coordinates are 
\bea
 \hat g_{ij}  = g_{ij}  && \hat \nabla_k \, \hat g_{ij}   = 0  \\
 \sqrt{ -  g }  = N \sqrt{  \hat g} && N^2  = - \frac{1}{g^{00}}  \\
 \hat g^{ij}  \equiv ( \hat g_{ij} )^{-1} = g^{ij} - \frac{g^{0i} g^{0j}}{g^{00}} \label{inver} && N^i  = - \frac{g^{0i}}{g^{00}}  \ . 
\eea
To get the expression for $\hat g^{ij}$, we use the fact that $\hat g^{ij} \equiv ( \hat g_{ij} )^{-1}$, $g_{i\mu}g^{\mu\nu}=\delta_i^{\nu}$ (which implies $g_{0i}=-\frac{g_{ij}g^{0j}}{g^{00}}$) and $\hat g_{ij}=g_{ij}$, which gives
\be
\hat g_{ik}\hat g^{ij}=g_{ik}g^{ij}+g_{0k}g^{0j}=\hat g_{ik}\bigg(g^{ij}-\frac{g^{0i}g^{0j}}{g^{00}}\bigg).
\ee

  Above, we have also introduced $\hat \nabla$, which is the spatial covariant derivative compatible with $\hat g_{ij}$, and is defined by $\hat \nabla_i \hat T_j = e^\mu{}_i \, e^\nu{}_j \nabla_\mu T_\nu$ where $\hat T_j = e^\nu{}_j T_\nu$.  

\subsection{Computation of $\delta K_{\rm u}$}
We start with the following expression for the extrinsic curvature $K_{ij}$ (see \cite{poisson} page 76)
\be
K_{ij} = \half e^\mu{}_i \, e^\nu{}_j \mathcal{L}_n \gmn,
\ee
where $\mathcal{L}_n$ is the Lie derivative in the direction of the unit normal $n^\mu$.  Then we make the following manipulations \cite{wald}
\begin{align}
K_{ij} & = \frac{1}{2N} e^\mu{}_i \, e^\nu{}_j  \left( N n^\sigma \nabla_\sigma \gmn + g_{\sigma \nu} N \nabla_\mu n^\sigma + g_{\mu \sigma} N \nabla_\nu n^\sigma \right) \\
& = \frac{1}{2N} e^\mu{}_i \, e^\nu{}_j  \left( N n^\sigma \nabla_\sigma \gmn + g_{\sigma \nu}  \nabla_\mu (N n^\sigma )+ g_{\mu \sigma} \nabla_\nu ( N n^\sigma ) \right) \\
& = \frac{1}{2N} e^\mu{}_i \, e^\nu{}_j  \left( \mathcal{L}_t  \gmn - \mathcal{L}_N \gmn \right),
\end{align}
where we have used $e^\mu{}_i n_\mu = 0$, $\mathcal{L}_t$ is the Lie derivative in the direction of $t^\mu$, and $\mathcal{L}_N$ is the Lie derivative in the direction $N^i e^\mu{}_i$.  Then, we have 
\begin{align}
e^\mu{}_i \, e^\nu{}_j  \mathcal{L}_N \gmn & = e^\mu{}_i \, e^\nu{}_j  \left( \nabla_\mu ( g_{\nu \sigma} e^\sigma{}_i N^i )  +    \nabla_\nu ( g_{\mu \sigma} e^\sigma{}_i N^i )     \right) \\
& = \hat \nabla_i N_j + \hat \nabla_j N_i,
\end{align}
where we have defined $N_i = \hat g_{ij} N^j$.  Finally, in the $(t,y^i)$ coordinates
\begin{align}
e^\mu{}_i \, e^\nu{}_j  \mathcal{L}_t \gmn & = \partial_t \hat g_{ij} \ , 
\end{align}
which can be checked by expanding both sides and realizing that 
\begin{align}
e^\mu{}_i e^\nu{}_j g_{\sigma \nu } \partial_\mu t^\sigma = \frac{ \partial x^\mu}{\partial y^i} e^\nu{}_j \, g_{\sigma \nu} \frac{\partial }{\partial x^\mu} \frac{\partial x^\sigma}{\partial t} = e^\nu{}_j \, g_{\sigma \nu} \frac{\partial }{\partial y^i } \frac{\partial x^\sigma}{\partial t} = e^\nu{}_j \, g_{\sigma \nu} \, \partial_t e^\sigma{}_i \ . 
\end{align}
This leads us to write
\be
N K_{ij} = \half \left( \partial_t \hat g_{ij} - \hat \nabla_i N_j - \hat \nabla_j N_i  \right)
\ee 
in agreement with \cite{wald}, so that 
\be
N K  = N \hat g^{ij} K_{ij} = \half \hat g^{ij} \partial_t \hat g_{ij} - \hat \nabla_i N^i \ . 
\ee

The background value of $K$ is $ 3 H$, so we have the trace of the perturbed extrinsic curvature as
\be
\delta K = \frac{1}{N }\left( \half \hat g^{ij} \partial_t \hat g_{ij} - \hat \nabla_i N^i \right) - 3 H \ . 
\ee
From here, it is straightforward to introduce $\pi$.  First of all, we will have $- 3 H \rightarrow - 3 H - 3 \dot H \pi - \frac{3}{2} \ddot H \pi^2 + \dots$. Then we have 
\begin{align}
g^{00}_{\rm u} & \rightarrow   g^{00} + 2 g^{0 \mu} \partial_\mu \pi + g^{\mu \nu} \partial_\mu \pi \partial_\nu \pi  \\
\hat g^{\rm u}_{ij}  = g^{\rm u}_{ij} & \rightarrow P^{-1}{}^\mu{}_i P^{-1}{}^\nu{}_j \,  g_{\mu \nu}  = g_{ij} + g_{00}  \frac{\partial_i \pi \partial_j \pi}{1 + \dot \pi}  \\
 g^{0i}_{\rm u}  & \rightarrow P^0{}_\mu P^i{}_\nu g^{\mu \nu}  = g^{ i j } \partial_j \pi \\
 \sqrt{- g_{\rm u}} & \rightarrow  \det P^{-1} \sqrt{-g} = (1+ \dot \pi)^{-1} \sqrt{-g}   \\
 g^{ij}_{\rm u} & \rightarrow g^{ij} \\
 \partial_0 & \rightarrow \frac{1}{1 + \dot \pi}  \partial_0 \ , \hspace{.2in} \partial_i \rightarrow - \frac{\partial_i \pi}{1 + \dot \pi} \partial_0 + \partial_i ,
\end{align}
where, because we will need to derive the contribution to the Poisson equation, we have kept the full metric dependence, but have assumed that the metric is diagonal.  We reproduce the transformation matrices here for convenience
\be
P^\mu{}_\rho = \begin{pmatrix} 1 + \dot \pi & \partial_i \pi \\ 0 & 1 \end{pmatrix}_{\mu \rho} \hspace{.5in} P^{-1}{}^\rho{}_\mu = \begin{pmatrix} \frac{1}{1 + \dot \pi} & - \frac{\partial_i \pi}{1 + \dot \pi} \\ 0 & 1 \end{pmatrix}_{ \rho \mu} \ . 
\ee

We will look at the two relevant terms individually.  First consider 
\begin{align}
\half \hat g^{ij}_{\rm u} \partial_t \hat g^{\rm u}_{ij} \rightarrow \half  \left( g^{ij} - \frac{ g^{i k} g^{j l} \partial_k \pi \partial_l \pi}{ g^{00} + 2 g^{0\mu} \partial_\mu \pi + g^{\mu \nu} \partial_\mu \pi \partial_\nu \pi } \right)     \frac{1}{1 + \dot \pi}                \partial_t \left( g_{ij} + g_{00} \frac{ \partial_i \pi \partial_j \pi}{1 + \dot \pi } \right)  \ . 
\end{align}
Next, consider
\begin{align}
\hat \nabla_i N^i = \frac{1}{\sqrt{\hat g}} \partial_i \left( \sqrt{\hat g} N^i \right) = \partial_i N^i +N^i \partial_i \log \sqrt{\hat g} \ .
\end{align} 
The first term gives 
\be
\partial_i N^i_{\rm u} \rightarrow     \left( - \frac{\partial_i \pi}{1 + \dot \pi} \partial_0 + \partial_i  \right)    \left( \frac{ - g^{i k } \partial_k \pi}{g^{00} + 2 g^{0 \mu} \partial_\mu \pi + g^{\mu \nu} \partial_\mu \pi \partial_\nu \pi }  \right) \ , 
\ee
and the second one
\begin{align}
& N^i_{\rm u} \partial_i \log \sqrt{\hat g_{\rm u}} \rightarrow \left( \frac{- g^{i k } \partial_k \pi}{g^{00} + 2 g^{0 \mu} \partial_\mu \pi + g^{\mu \nu} \partial_\mu \pi \partial_\nu \pi }  \right)    \\
& \hspace{1in} \times  \left( - \frac{\partial_i \pi}{1 + \dot \pi} \partial_0 + \partial_i  \right)     \log \left(   \sqrt{ - ( g^{00} + 2 g^{0\mu} \partial_\mu \pi + g^{\mu \nu} \partial_\mu \pi \partial_\nu \pi )} \, (1 + \dot \pi)^{-1} \sqrt{-g}  \right)  \nonumber \ . 
\end{align}
The metric perturbations are given by 
\be
g_{\mu \nu}  = \begin{pmatrix} -( 1 + 2 \Phi ) & 0  \\ 0 & a^2 ( 1 - 2 \Psi ) \delta_{ij}  \end{pmatrix}  \ . 
\ee
Finally, putting this all together, we obtain, to second order in the perturbations,\footnote{This formula differs from the one provided in \cite{Cheung:2007sv} at second order. It also differs from one of the two expression given in \cite{Gubitosi:2012hu} at linear order.}
\begin{align}
\deltaku   \rightarrow &       - 3 \dot \Psi - 3 H \Phi - a^{-2} \partial^2 \pi - 3 \dot H \pi  \nonumber \\
& +  a^{-2} \left(   ( \partial^2 \pi ) \left(\dot \pi - \Phi - 2 \Psi \right) + (\partial_i \pi ) \partial_i \left( \half H  \pi +2 \dot \pi   -2 \Phi + \Psi \right) \right)  \nn \\
& + \frac{9}{2} H \Phi^2 + 3 \Phi \dot \Psi - 6 \Psi \dot \Psi - \frac{3}{2} \ddot H \pi^2 \ . 
\end{align}

%%%%%%%%%%%%%%%%%%%%%
%
%          Linear Equations
%
%%%%%%%%%%%%%%%%%%%%%

\section{Linear equations} \label{lineareqs}

In this appendix we will work out the linear equations that we will need in the rest of the paper.  We are considering the action
\be
 \int d^4 x \, \sqrt{-g}\bigg[ \frac{\mpl^2}{2}  R - \Lambda(t) - c(t) g_{\rm u}^{00}  +  \frac{M_2^4}{2} ( \delta g_{\rm u}^{00} )^2 - \frac{\bar{m}_1^3 }{2 } \delta g_{\rm u}^{00} \delta K_{\rm u}  \bigg]   \ .
\ee
For the equation of motion of $\pi$, we will include only the terms that are relevant for the linear equation of motion for $\pi$.  This means that we have dropped all terms that do not involve a factor of $\pi$, and all terms that would contribute higher than first order in the equation of motion.  First we consider
\begin{align}
 - \sqrt{-g} \left(  c(t)  \, g^{00}_{ \rm u}  + \Lambda(t)  \right) & \approx  - a^3   \left( 1 + \Phi - 3 \Psi \right)  \Bigg\{ \Lambda - c + ( \dot \Lambda - \dot c ) \pi + \half \left( \ddot \Lambda - \ddot c \right) \pi^2    \\
& \hspace{.5in}    + 2 c ( \Phi - \dot \pi )   + c \left( 4 \Phi \dot \pi - \dot \pi^2 + a^{-2} ( \partial \pi )^2 \right) + 2 \dot c \pi ( \Phi - \dot \pi ) \Bigg\}  \nonumber \ .
\end{align}
From this expression, we see that in order for the term linear in $\pi$ to vanish, we must have that $\dot \Lambda + \dot c + 6 H c = 0$, which also means that $\ddot \Lambda + \ddot c + 6 \dot H c + 6 H \dot c =0$.  Using these two relations and $\dot c = - 3 H (1+w) c$, and letting $\pi \rightarrow \pi + \delta \pi$, we get that the change in the action due to this term is
\be
\rightarrow 2 a^3 c(t) \left( - \ddot \pi  + 3 H   w \dot \pi + a^{-2} \partial^2 \pi + 3 \dot H \pi + 3H ( 1 -w) \Phi + \dot \Phi + 3 \dot \Psi \right)  \delta \pi .
\ee
Next, we have
\begin{align}
\sqrt{-g} \frac{M_2^4}{2} \left( \delta g_{\rm u}^{00} \right)^2  \rightarrow 4 M_2^4 a^3 \left\{  \frac{ \partial_t  M_2^4}{M_2^4} ( \Phi - \dot \pi) + 3 H ( \Phi - \dot \pi ) + \dot \Phi - \ddot \pi \right\} \delta \pi \ .
\end{align}
And finally,
\begin{align}
- \sqrt{-g} \frac{\bar m_1^3}{2} \delta g_{\rm u}^{00} \delta K_{\rm u} &  \approx  a^3 \mm ( \dot \pi - \Phi ) \left( - 3 \dot \Psi - 3 H \Phi - a^{-2} \partial^2 \pi - 3 \dot H \pi  \right) \nn  \\
&\rightarrow  - \delta \pi \frac{d}{dt} \left\{  a^3 \mm \left(  - 3 \dot \Psi - 3 H  \Phi  - a^{-2} \partial^2 \pi - 3 \dot H \pi \right) \right\} \nonumber \\
& \hspace{1in} - \delta \pi \left( a^3 \mm a^{-2} \partial^2 ( \dot \pi - \Phi) + 3 \dot H a^3 \mm ( \dot \pi - \Phi ) \right) \ .
\end{align}
Dividing by $a^3 \delta \pi$, and adding the above contributions together gives the linear equation of motion
\begin{align}
& 2 c \left( - \ddot \pi + 3 H w \dot \pi + a^{-2} \partial^2 \pi + 3 \dot H \pi + 3 H (1-w) \Phi + \dot \Phi + 3 \dot \Psi \right) \nonumber \\
&- 4 M_2^4 \left( \ddot \pi - \dot \Phi + \frac{\partial_t M_2^4}{M_2^4} \left( \dot \pi - \Phi \right) + 3 H ( \dot \pi - \Phi) \right) \nonumber \\
& + (H \mm + \partial_t \mm)  a^{-2} \partial^2 \pi + \mm a^{-2} \partial^2 \Phi  \nonumber \\
& + 3 \,  a^{-3}  \frac{d}{dt} \left\{  a^3 \mm \left(    \dot \Psi   + H  \Phi \right) \right\} + 3 \dot H \mm \Phi  + 3 \, a^{-3} \frac{d}{dt} \left( a^{3} \mm \dot H \right) \pi= 0  \label{linearpieom} \ .
\end{align}
In the limit $\bar{m}_1^3 \rightarrow0$, the analogue of this equation is provided in synchronous gauge in \cite{Creminelli:2008wc}, and for $\bar m_1^3 \neq 0$ it was provided in synchronous gauge in \cite{Hu:2013twa}.  Reference \cite{Gleyzes:2013ooa} provides the linear equation of motion for $\pi$ above and the Einstein equations below, in both Newtonian and synchronous gauges, for an even broader class of dark-energy Lagrangians than we consider in this work.  For the Einstein equations at linear order, we take the expressions from \cite{Gubitosi:2012hu}, but relax the assumption of constant $\bar m_1^3$ and specialize to $\dot f = 0$.  The equation for the $(00)$ component is
\begin{align}
&2 \mpl^2 ( a^{-2}\partial^{2}\Psi  -3H\dot\Psi )   -2c\dot\pi-(\dot c+\dot\Lambda)\pi-2\Lambda\Phi+4M_{2}^{4}(\Phi-\dot \pi) \nn \\
&  +\bar m_{1}^{3}\left[3(\dot\Psi+H\Phi)   +  3\dot H\pi    +a^{-2}\partial^{2}\pi+3H (\Phi-\dot \pi)\right]= \delta T_{00} \label{zero1}\, .
\end{align}
By using the background equations of motion $\dot \Lambda + \dot c + 6 H c = 0$, \eqn{cequation1}, \eqn{lambdaequation1}, \eqn{anotherfriedman}, and $\delta T_{00} = \delta \rho _m + 2 \bar \rho_m \Phi$ we get
\begin{align}\label{T00}
&2 \mpl^2  \big( a^{-2}\partial^{2}\Psi  -3H(\dot\Psi + H \Phi) \big)
+ ( \bar \rho_D +  \bar p_D) (\Phi - \dot \pi + 3 H \pi) + 4M_{2}^{4}(\Phi-\dot \pi) \nn \\
& +\bar m_{1}^{3}\left[3(\dot\Psi+H\Phi)  + 3\dot H\pi+a^{-2}\partial^{2}\pi+3H (\Phi-\dot \pi)\right]= \delta \rho_m\, .
\end{align}
The equation for the $(ij)$ trace component is 
\begin{align}\label{p1}
& 2 \mpl^2  \left( \ddot\Psi +H \dot\Phi+3H \dot\Psi+ (3H^{2}+ 2\dot H) (\Phi+\Psi) +  \partial^2(\Phi - \Psi)/(3 a^2)\right)   \\
&+2 c(\Phi-\dot\pi)-2\Psi (\Lambda-c)+ (\dot\Lambda-\dot c)\pi - \bar m_{1}^{3} [ \dot\Phi-\ddot\pi +3H(\Phi-\dot\pi)] -\partial_t(\bar m_{1}^{3}) (\Phi-\dot\pi)  =  \delta T^k_{\ k}/(3 a^2)  \nn  \;.
\end{align}
Again using the background equations of motion, we have
\begin{align}\label{p2}
&2 \mpl^2   \left(\ddot\Psi +H \dot\Phi+3H \dot\Psi+ (3H^{2}+ 2\dot H) \Phi  +  \partial^2(\Phi - \Psi)/(3 a^2)\right) \nn \\
&- \dot{ \bar p}_D \pi  + (\bar \rho_D + \bar p_D) (\Phi- \dot \pi) - \bar m_{1}^{3} [ \dot\Phi-\ddot\pi +3H(\Phi-\dot\pi)] -\partial_t(\bar m_{1}^{3}) (\Phi-\dot\pi)= \delta p_m  \;.
\end{align}
The equations for the $(ij)$ traceless components are 
\begin{equation}
\mpl^2 \big(\partial_{i}\partial_{j} - \frac13 \delta_{ij} \partial^2 \big) (\Psi-\Phi)= \delta T_{ij}  - \frac13 \delta_{ij} \delta T^k_{\ k} = 0 \;. \label{eq_traceless}
\end{equation}
The equations for the $(0i)$ components are 
\begin{equation} \label{0i}
2 \mpl^2\partial_{i} (\dot \Psi+H\Phi)- (\bar \rho_{D}+ \bar p_{D})\partial_{i}\pi - 2\bar m_{1}^{3}\partial_{i}(\Phi-\dot\pi)=\delta T_{i0} \;.
\end{equation}
Notice that compared to \cite{Gubitosi:2012hu}, we have corrected a factor of two in \eqn{0i} in front of $( \bar \rho_D + \bar p_D) \partial_i \pi$, a minus sign in front of $3 \bar m_1^3 \dot H \pi$ in \eqn{T00},  and we have extra terms in \eqref{p1} and \eqref{p2} due to the time dependence of $\bar{m}_1^3$ in our system.

From the above action, we can read off the linear energy-momentum tensor of the dark energy as
\bea
\delta\rho_D && = 2c(t)(\dot \pi - 3 H \pi -\Phi ) + 4M_{2}^{4}(\dot \pi-\Phi) \nn \\
&& \hspace{.5in} +\bar m_{1}^{3}\left[ - 3\dot H\pi-  a^{-2} \partial^{2}\pi+3H (\dot \pi-\Phi)-3(\dot\Psi+H\Phi)\right],\\
\delta p_D  && =\dot{ \bar p}_D \pi  + 2c(t)(\dot \pi-\Phi) + \bar m_{1}^{3} [ \dot\Phi-\ddot\pi +3H(\Phi-\dot\pi)] +\partial_t(\bar m_{1}^{3}) (\Phi-\dot\pi) ,\\
\partial_i\delta q_D  &&  =-(\bar{\rho}_D+\bar p_D)\partial_{i}\pi + 2\bar m_{1}^{3}\partial_{i}(\dot\pi-\Phi),\\
\pi^D_{ij}  &&   =0,
\eea
where $\delta\rho_D$ and $\delta p_D$ are the energy and pressure density of dark energy, $\partial_i\delta q_D=\delta T^{~0}_{D~i}$ is the momentum density and $\pi^D_{ij}$ is the anisotropic stress.\footnote{The anisotropic stress is the trace-less spatial part of the energy-momentum tensor, $\pi_{ij}=T_{ij}-\frac13g_{ij}T^k_k$.}  The momentum density corresponding to the matter is also given as $\partial_i\delta q_m=\delta T^{0}_{~i}$. As we see in the above, in the limits that we consider in this paper, the dark energy pressure density is very small and a relativistic correction comparing with its energy density.\\

 Inserting the dark energy $T_{\mu\nu}$ in the linear Einstein equations, we find two constraints, 
 the anisotropic stress equation
\be
\mpl^2  \big(\partial_{i}\partial_{j} - \frac13 \delta_{ij} \partial^2 \big)   (\Psi-\Phi)=\pi^D_{ij}+\pi^m_{ij},
\ee
and (from the combination of \eqn{T002} and \eqn{0i2}) the Poisson equation\footnote{For completeness and for future reference, the full linear $G_{00}$ component of the Einstein equation including the relativistic corrections is
  \begin{align}\label{T002}
 &2 \mpl^2  \big( a^{-2}\partial^{2}\Psi  -3H(\dot\Psi + H \Phi) \big)  = \delta \rho_m+\delta \rho_D\, .
 \end{align}
 while the $G_{0i}$ is given as
 \begin{equation} \label{0i2}
 2 \mpl^2\partial_{i} (\dot \Psi+H\Phi)+\partial_i\delta q_D=-\partial_i\delta q_m \;,
 \end{equation}
 and finally the trace of $G_{ij}$ gives rise to
 \begin{align}
 &2 \mpl^2   \left(\ddot\Psi +H \dot\Phi+3H \dot\Psi+ (3H^{2}+ 2\dot H) \Phi  +  \partial^2(\Phi - \Psi)/(3 a^2)\right)= \delta p_m+ \delta p_D \;.
 \end{align}}
\be
\delta\rho_m-3H\delta q_m+\delta\rho_D-3H\delta q_D=2\mpl^2a^{-2}\partial^2\Psi.
\ee
It is noteworthy to mention that in the above, the combination $\delta\rho_{x}-3H\delta q_x$ is equal to the energy density perturbation on velocity orthogonal slicing, $\delta\rho^{(v.o.)}$ (see \cite{Kodama:1985bj,Creminelli:2009mu}).
Since the linear order anisotropic stress of our dark matter and dark energy are zero, the Bardeen potentials are equal here, $\Psi=\Phi$. The combination of dark matter continuity and Euler equations, linear equation of $\pi$ and the above constraint equations fully determine our system at linear level.

\subsection{Linear solution of $\pi$ with $\bar{m}_1^3=0$}  \label{linearpisol}

We are now ready to solve the linear equation of $\pi$ in the clustering case ($c_s^2\ll1$) analytically. Recalling that $\frac{\partial_t \M}{H\M}=-3(1+w)$ and setting $\bar{m}_1^3=0$, we have the field equation of $\pi(a,\vx)$ in \eqn{eom11} as
\be\label{pi-lin-B}
\ddot{\pi}-\dot{\Phi}-3wH(\dot{\pi}-\Phi)-\frac{c_s^2}{a^2}\partial^2\pi=0.
\ee
It is noteworthy to mention that the Newtonian potential, $\Phi$, in the above equation is of the same order as the $\pi$ terms and is \textit{not} a relativistic correction in the field equation of $\pi$.  As we see in Section \ref{sec-4}, the system of equations for dark matter and dark energy reduce to one single second order differential equation for the total density contrast, $\delta_A(a,\vk)=\delta_{m}(a,\vk)+a^{-3w}\frac{\Omega_{ D,0}}{\Omega_{ m,0}}\frac{1+w}{c_s^2}(\dot\pi-\Phi)$. That equation has only one \textit{growing} solution for $\delta_A$ which by the Poisson equation is directly related to the Newtonian potential, $\Phi= -\frac{3\Omega_{ m,0}\cH_0^2a_0}{2ak^2}\delta_A$. Working out $\delta_A$ and therefore $\Phi$ in Section \ref{sec-4}, we can determine $\pi$ by solving \eqn{pi-lin-B} in which the Newtonian potential acts as the source term. In fact, the homogeneous part of the above equation has only decaying solutions and the inhomogeneous part, which is sourced by $\Phi$, provides the growing mode.

In order to see the decaying nature of the homogeneous solution let us study its evolution during matter era.  It is straightforward to see that it is damping during the dark-energy dominated era as well.
 In Fourier space,  we can read off the homogeneous part of \eqn{pi-lin-B} in terms of the conformal time $\tau$, as
\be
\pi^{\rm h}_{\tau\tau}-(1+3w)\cH\pi^{\rm h}_{\tau}+c_{s}^2k^2\pi^{\rm h}=0,
\ee
where $\pi^{\rm h}(\tau,\vk)$ is the homogeneous solution of $\pi$. The solution of $\pi^{\rm h}(\tau,\vk)$ in the matter era, when $\cH = 2 / \tau$, can be expanded in terms of the Bessel $J_{\nu}$ and $Y_{\nu}$ functions as
\be
\pi^{\rm h}(\tau ,\vk)=\tau^{\frac32(1+2w)}\bigg(c_1J_{\nu}(c_sk\tau)+c_2Y_{\nu}(c_sk\tau)\bigg) \quad \textmd{where} \quad \nu=\frac32(1+2w),
\ee
which using that $\tau\propto a^{\frac12}$ implies that $\pi^{\rm h}\propto a^{3(1+2w)}$ is always decaying and therefore negligible. As a result, only the particular solution of $\pi$, called $\pi^{\rm p}$, which is sourced by $\Phi$ (and hence under the influence of dark matter) can be growing and important in structure formation. That is in agreement with the fluid picture observation that the energy contrast of dark energy, $\delta_{ D}$ is proportional to $\delta_{m}$ during the matter era.

In order to determine the particular solution of \eqn{pi-lin-B}, $\pi^{\rm p}$, we use the following decomposition
\be\label{ansatz}
\pi^{\rm p}(a,\vx)=\int^a  d\ta \frac{\Phi(\ta,\vx)}{\cH(\ta)}+\tp(a,\vx).
\ee
where $\tp(a,\vx)$ satisfies\footnote{Using the fact that the source term is proportional to $c_s^2$ and $c_s^2k^2\ll \cH^2$, we dropped the spatial derivative of $\tp$ in \eqref{pi-tild-a}.} 
\be\label{pi-tild-a}
a^2\cH^2(a)\tp''+a^{2+3w}\cH(a)(\cH a^{-3w})'\tp'=c_{s}^2\int^a  d\ta \frac{\partial^2\Phi(\ta,\vx)}{\cH( \ta  )}.
\ee
The above equation has the following solution
\be\label{tp-sol}
\tp(a,\vec{x})=\int^a da'\int^{a'} da'' K(a',a'')S_{\Phi}(a'',\vec{x}),
\ee
where $S_{\Phi}(a,\vec{x})$ is the source term on the RHS of \eqn{pi-tild-a} and the kernel $K(a,a')$ is 
\be\label{kernel}
K(a,a')=\frac{(a/a')^{3w}}{\cH(a')\cH(a)a'^2}.
\ee
Finally, using \eqn{ansatz} and \eqn{tp-sol}, we obtain the linear order growing solution of $\pi(a,\vx)$ as 
\be\label{pi-linear-B}
\pi (a,\vx)=\int^a  da' \frac{\Phi(a',\vx)}{\cH(   a'  )}+  c_{s}^2\int^a da'\int^{a'} da'' \int^{a''}  da''' \frac{\partial^2\Phi(a''',\vx)}{\cH(a''')} K (a',a'') ,
\ee
which is a function of the background and the Newtonian potential.

\subsubsection{$\pi$ during matter domination}

Up to now, we worked out the $\pi(a,\vx)$ field in terms of the gravitational potential $\Phi(a,\vx)$. Now we turn to find the explicit form of $\pi(a,\vx)$ during matter era. In that regime, $\Phi$ is almost constant
\be
\Phi(a,\vx)\simeq\Phi(a_i,\vx),
\ee
and we have the  Hubble parameter as $\cH(a)\simeq a^{-\frac12}\cH_0$ where $\cH_0=H_0(\Omega_{m,0}a_0^3)^\frac12$.
Going to Fourier space and using the above in \eqn{pi-linear-B}, we find the explicit form of the $\pi$ field during the matter era 
\be\label{pi-solution-DM}
\pi(a,\vk)\simeq \frac{2a}{3\cH(a)}  \Phi(a_i,\vk)\bigg(1-  \frac{2}{5(1-3w)}\frac{c_s^2k^2}{\cH^2}\bigg) \ .
\ee
From this solution, we see that $\dot \pi - \Phi \sim c_s^2 H^{-2} \partial^2 \Phi$.  Although this solution is only valid during matter domination, as shown in \eqn{pi-linear-B}, the qualitative features and the scalings with $c_s^2$ are not different at other times.

Throughout this work, we choose the $\pi$ language for describing the dynamics of the dark energy sector and its gravitational interactions with the dark matter. At this point, we determine the corresponding dark energy density contrast, $\delta_{ D}$, and velocity, $\theta_{ D}$, generated by $\pi$ in the fluid picture. During the matter era, the dark energy has negligible contribution to the Possion equation and we can read $\Phi$ as
\be
-k^2\Phi(a,\vk)\simeq\frac32\cH^2\delta_m(a,\vk).
\ee
Using the above in \eqn{pi-solution-DM} and recalling that $\delta_m=-\theta_m/\cH$, we find $\delta_{ D}$ and $\Theta_{ D}$ as
\bea
&&\delta_{ D}(a,\vk)=\frac{(1+w)}{c_s^2}(\dot{\pi}-\Phi)=\left(\frac{1+w}{1-3w}\right)\delta_{m}(a,\vk),\\
&&\theta_{ D}(a,\vk)=\frac{k^2}{a}\pi=\theta_{m}(a,\vk).
\eea
Moreover, the total density contrast $\delta_A$ in \eqn{deltaalinear} is
\be\label{delta-a-MD}
\delta_A(a,\vk)=\delta_{m}(a,\vk)+a^{-3w}\frac{\Omega_{ D,0}}{\Omega_{ m,0}}\frac{1+w}{c_s^2}(\dot\pi-\Phi)=\bigg(1+a^{-3w} \frac{\Omega_{ D,0}}{\Omega_{ m,0}}\frac{1+w}{1-3w}\bigg)\delta_{m}(a,\vk).
\ee
The above are the well-known linear relations in the fluid picture.
%%%%%%%%%%%%%%%%%%%%%%%%%%%%%%%%%%%%%%%%%%%%%

\subsection{Linear solution of $\pi$, including $\motb$}
In this part, we consider $\motb\neq0$ and solve the linear equation of $\pi$ in the limit that $| c_s^2 | \ll1$. From \eqref{eom11}, we find the speed of sound in the presence of $\motb$ as
\be
c_s^2=\frac{c(t)+ \half a^{-1} \partial_t ( a \,  \bar m_1^3(t) )}{2M_2^4(t) +c(t) } \ .
\ee
Moreover, in the limit that $| c_s^2 |  \ll  1$, the field equation of $\pi$ is given as
\be\label{pi-lin-B2}
\frac{1}{a^3 M_2^4} \frac{d}{d t } \left\{ a^3 M_2^4 \left( \dot \pi - \Phi \right) \right\} -  c_s^2 a^{-2}\partial^2\pi- c_s^2 \, \alpha_{\bar{m}}  a^{-2}  \frac{\partial^2\Phi}{H}=0,
\ee
where $\alpha_{\bar{m}}$ is a dimensionless quantity of order one, given as
\be
\alpha_{\bar{m}}\equiv \frac{H\motb}{4M_2^4 c_s^2}  \simeq \frac{H\motb}{2c(t)+a^{-1}\partial_t(a\motb)}  \ , 
\ee
where in the last passage we used $| c_s^2 | \ll 1$.  From \eqn{pi-lin-B2}, we can easily see that in the $c_s^2 \rightarrow 0$ limit, we will again have the two species comoving, i.e. $\partial_i \pi = - a v^i_m$ (as in \eqn{velocities}).  

Next, we can use the Poisson equation \eqn{zero1} to get 
\be
\delta_A = \delta_m + \frac{ 4 a^3 M_2^4}{a_0^3 \bar \rho_{m,0}} \left( \dot \pi - \Phi \right) - \frac{ a^3 \bar m_1^3}{a_0^3 \bar \rho_{m,0}} a^{-2} \partial^2 \pi \ , 
\ee
in the non-relativistic limit.  Taking the time derivative of this and using the equation of motion for $\pi$ and the Euler equation for dark matter gives
\begin{align}
\dot \delta_A & = \dot \delta_m + \frac{4}{\bar \rho_{m,0}} \frac{d}{dt} \left( a^3 M_2^4 ( \dot \pi - \Phi) \right) - \frac{1}{\bar \rho_{m,0}} \frac{d}{dt} \left( a \mm \right) \partial^2 \pi - \frac{a \mm}{\bar \rho_{m,0}} \partial^2 \dot \pi \\ 
& = \dot \delta_m + \frac{4 a^3 M_2^4}{\bar \rho_{m,0}} \left( c_s^2 a^{-2} \partial^2 \pi + \frac{\mm}{4 M_2^4} a^{-2} \partial^2 \Phi \right)  - \frac{1}{\bar \rho_{m,0}} \frac{d}{dt} \left( a \mm \right) \partial^2 \pi - \frac{a \mm}{\bar \rho_{m,0}} \partial^2 \dot \pi  \\
& = \dot \delta_m + \frac{ 2 a c }{\bar \rho_{m,0}} \partial^2 \pi + \frac{a \mm}{\bar \rho_{m,0}} \partial^2 ( \Phi - \dot \pi) \\
& = - \frac{1}{a} \theta_m - \frac{1}{a} \frac{2 a^3 c}{\bar \rho_{m,0}} \theta_m = -\frac{1}{a} C(a) \theta_A , \label{anothercont}
\end{align}
where we used $\partial^2 ( \dot \pi - \Phi ) \propto c_s^2$, $\theta_A \equiv \theta_m$, and we have called $C(a) = 1 + \frac{2a^3 c}{\bar \rho_{m,0} } = 1 + (1+w) \frac{\Omega_{D,0}}{\Omega_{m,0}} a^{-3w}$.  Thus, \eqn{anothercont} along with the fact that the two species are comoving, $\partial_i \pi = - a v^i_m$, means that the linear equations for $\delta_A$ are the same as in the $\bar m_1^3 = 0$ case studied in the main text, \eqn{cafunction}
\begin{align}
\dot \delta_A + \frac{1}{a} C(a) \theta_A & = 0   \label{dumbledor}\\
\dot \theta_A + H \theta_A + \frac{3}{2} \frac{\Omega_{m,0} \cH_0^2 a_0}{a^2}  \delta_A & = 0 \label{snape}  \ .
\end{align}

As before, because the species are comoving, there is only one growing-mode degree of freedom.  Thus, we can solve for $\delta_A$ in \eqn{dumbledor} and \eqn{snape}, and express $\Phi$ in terms of $\delta_A$ in the equation of motion for $\pi$ \eqn{pi-lin-B2}.  Thus, we are able to treat $\Phi$ as a source in the equation of motion for $\pi$.  Again we use the following decomposition
\be\label{ansatz-22}
\pi^{\rm p}(a,\vx)=\int^a  d\ln\ta \frac{\Phi(\ta,\vx)}{H(\ta)}+\tp_{\bar{m}}(a,\vx).
\ee
where $\tp_{\bar{m}}(a,\vx)$ satisfies 
\be\label{pi-tild-a-2}
a^2\cH^2(a)\tp_{\bar{m}}''+a^{2+3w}\cH(a)(\cH a^{-3w})'\tp_{\bar{m}}'=c_{s}^2\int^a  d\ln\ta \frac{\partial^2\Phi(\ta,\vx)}{H( \ta  )}+ c_s^2 \, \alpha_{\bar{m}}\frac{\partial^2\Phi}{H}.
\ee
The above equation has the following solution
\be\label{tp-sol-2}
\tp_{\bar{m}}(a,\vec{x})=\int^a da'\int^{a'} da'' K(a',a'')S^{\bar{m}}_{\Phi}(a'',\vec{x}),
\ee
where $S^{\bar{m}}_{\Phi}(a,\vec{x})$ is the source term on the RHS of \eqref{pi-tild-a-2} and the kernel $K(a,a')$ is given in \eqref{kernel}. 
Finally, using \eqref{ansatz-22} and \eqref{tp-sol-2}, we obtain the linear order growing solution of $\pi(a,\vx)$ as 
\be\label{pi-linear-B-2}
\pi (a,\vx)=\int ^a  da' \frac{\Phi(a',\vx)}{\cH(   a'  )}+  c_{s}^2\int^a da'\int^{a'} da'' \bigg(       \alpha_{\bar{m}} \frac{\partial^2\Phi(a'',\vx)}{H(a'')} + \int^{a''}  da''' \frac{\partial^2\Phi(a''',\vx)}{\cH(a''')}   \bigg)K(a',a''),
\ee
which is a function of the background expansion rate and the Newtonian potential.

%%%%%%%%%%%%%%%%%%%%%
%
%         Greens Function
%
%%%%%%%%%%%%%%%%%%%%%%%

\section{The density and velocity Green's functions}\label{Green-app}

The non-Linear continuity and Euler equations are two coupled inhomogeneous differential equations which can be solved analytically in terms of four Green's functions, two density Green's functions, $G^{\delta}_{1}$, $G^{\delta}_{2}$, and two velocity Green's functions, $G^{\Theta}_{1}$ and $G^{\Theta}_{2}$. Therefore, at each perturbative order, $\delta^{(n)}_{\vk}(a)$ and $\Theta^{(n)}_{\vk}(a)$ are
 \begin{align}
 &\delta^{(n)}_{\vk}=\int^1_0 d\ta \bigg(G^{\delta}_{1}(a,\ta)S^{(n)}_1(\ta,\vk)+G^{\delta}_{2}(a,\ta)S^{(n)}_2(\ta,\vk)\bigg),    \label{dtGreen-app}  \\
  &\Theta^{(n)}_{\vk}=\int^1_0 d\ta \bigg(G^{\Theta}_{1}(a,\ta)S^{(n)}_1(\ta,\vk)+G^{\Theta}_{2}(a,\ta)S^{(n)}_2(\ta,\vk)\bigg),
  \end{align}
where $S^{(n)}_1(\ta,\vk)$ and $S^{(n)}_2(\ta,\vk)$ are the source terms of the continuity and Euler equations at the $n$-th order respectively 
\begin{align}
&S_1^{(n)}(a,\vk)=\frac{f_{+}(a)}{C(a)}\sum\limits_{m=1}^{n-1}\int \frac{d^3q}{(2\pi)^3}  \alpha(\vq,\vk-\vq)  \tT^{(m)}_{\vq}\td^{(n-m)}_{\vk-\vq},   \label{source}   \\
&S_2^{(n)}(a,\vk)=\frac{f_{+}(a)}{C(a)}\sum\limits_{m=1}^{n-1}\int \frac{d^3q}{(2\pi)^3} \beta(\vq,\vk-\vq) \tT^{(m)}_{\vq}\tT^{(n-m)}_{\vk-\vq}.
\end{align}

Using \eqn{dtGreen} in \eqn{conti-eq} and \eqn{Euler-eq}, we find that the four Green's functions are specified by the following equations
 \begin{align}
 &a \frac{d G^{\delta}_{\sigma}(a,\ta)}{da}-f_{+}(a)G^{\Theta}_{\sigma}(a,\ta)=\lambda_{\sigma}\delta(a-\ta),  \label{Green} \\
 &a \frac{d G^{\Theta}_{\sigma}(a,\ta)}{da}-f_{+}(a)G^{\Theta}_{\sigma}(a,\ta)-\frac{f_{-}(a)}{f_{+}(a)}\bigg(G^{\Theta}_{\sigma}(a,\ta)-G^{\delta}_{\sigma}(a,\ta)\bigg)=(1-\lambda_{\sigma})\delta(a-\ta),
 \end{align}
where $\lambda_\sigma$ is given as $$\lambda_1=1 \quad \textmd{and} \quad \lambda_2=0,$$ $\sigma=1,2$, and $\delta(a-\ta)$ is the Dirac delta function. The retarded Green's functions satisfy the boundary conditions 
\begin{align}  \label{bound}
& G^{\delta}_\sigma(a,\tilde a)  =  0 \quad \quad  \text{and}   \quad\quad G^{\Theta}_\sigma(a, \tilde a)=0  \quad \quad \text{for} \quad \quad \tilde a > a  \ , \\
 &G^\delta_\sigma ( \tilde a , \tilde a ) = \frac{\lambda_\sigma}{\tilde a}  \quad \hspace{.06in} \text{and} \hspace{.2in}  \quad G^{\Theta}_{\sigma} ( \tilde a  , \tilde a ) = \frac{(1 - \lambda_\sigma)}{\tilde a}  . \label{bound2}
\end{align}
We can then construct the Green's functions in the usual way using the linear solutions and the Heaviside step function, $\rm{H} (a-\tilde a)$, and imposing the boundary conditions \eqn{bound} and \eqn{bound2}.  This gives 

\begin{align}
&G^{\delta}_1(a,\ta)=\frac{1}{\ta W(\ta)}\bigg(\frac{d D_{-}(\ta)}{d\ta}D_{+}(a)-\frac{d D_{+}(\ta)}{d\ta}D_{-}(a)\bigg) {\rm H}(a-\ta)  \label{gdelta} \ ,\\
&G^{\delta}_2(a,\ta)=\frac{f_{+}(\ta)/\ta^2}{W(\ta)}\bigg(D_{+}(\ta)D_{-}(a)-D_{-}(\ta)D_{+}(a)\bigg){\rm H}(a-\ta) \ , \\
&G^{\Theta}_1(a,\ta)=\frac{a/\ta}{f_{+}(a)W(\ta)}\bigg(\frac{d D_{-}(\ta)}{d\ta}\frac{d D_{+}(a)}{d a}-\frac{d D_{+}(\ta)}{d\ta}\frac{d D_{-}(a)}{d a}\bigg) { \rm H}(a-\ta) \ ,\\
&G^{\Theta}_2(a,\ta)=\frac{f_{+}(\ta)a/\ta^2}{f_{+}(a)W(\ta)}\bigg(D_{+}(\ta)\frac{d D_{-}(a)}{d a}-D_{-}(\ta)\frac{d D_{+}(a)}{d a}\bigg) {\rm H}(a-\ta) \ ,   \label{gtheta}
\end{align}
where $W(\ta)$ is the Wronskian of $D_+$ and $D_-$ 
\be
W(\ta)=\frac{dD_{-}(\ta)}{d\ta}D_{+}(\ta)-\frac{d D_{+}(\ta)}{d\ta}D_{-}(\ta) \ .
\ee

Having the formal solutions \eqn{dtGreen} and the Green's functions in \eqn{gdelta} - \eqn{gtheta}, we are ready to find the solutions of total $\delta$ and $\Theta$ at any perturbative order.  For later convenience, it is useful to define the following independent combinations of the functions of momenta \eqn{alphadef} and \eqn{betadef} as
\begin{align}  \label{momentum-fun-3} 
\mathcal{\alpha}^1(\vk_{1},\vk_2,\vk_3)&\equiv\alpha(\vk_{1}-\vk_{2},\vk_{2})\alpha_s(\vk_{3},\vk_{2}-\vk_3),  \\ \mathcal{\alpha}^2(\vk_{1},\vk_2,\vk_3)&\equiv\alpha(\vk_{1}-\vk_{2},\vk_{2})\beta(\vk_{3},\vk_{2}-\vk_3), \\
\mathcal{\beta}^1(\vk_{1},\vk_2,\vk_3)&\equiv2\beta(\vk_{1}-\vk_{2},\vk_{2})\alpha_s(\vk_{3},\vk_{2}-\vk_3), \\ \mathcal{\beta}^2(\vk_{1},\vk_2,\vk_3)&\equiv2\beta(\vk_{1}-\vk_{2},\vk_{2})\beta(\vk_{3},\vk_{2}-\vk_3), \\
\mathcal{\gamma}^1(\vk_{1},\vk_2,\vk_3)&\equiv\alpha(\vk_{2},\vk_{1}-\vk_{2})\alpha_s(\vk_{3},\vk_{2}-\vk_{3}),  \\ \mathcal{\gamma}^2(\vk_{1},\vk_2,\vk_3)&\equiv\alpha(\vk_{2},\vk_{1}-\vk_{2})\beta(\vk_{3},\vk_{2}-\vk_{3}). \label{momentum-fun-4}     
\end{align}
The form of the source terms at third order can be most simplified in terms of the above functions of momenta. In particular, the source terms at third order are given as
 \begin{align}
 &S_1^{(3)}(a,\vk)=\frac{f_{+}(a)D_{+}(a)}{ C(a) D_+(a_i) }\int  \frac{d^3 p }{(2 \pi)^3}  \bigg(\alpha(\vk-\vp,\vp)\delta^{(2)}_{\vp}(a)+\alpha(\vp,\vk-\vp)\Theta^{(2)}_{\vp}(a)\bigg)\delta^{\rm in}_{\vk-\vp},   \label{source-3rd-1}  \\
 &S_2^{(3)}(a,\vk)=\frac{f_{+}(a)D_{+}(a)}{  C(a)  D_+(a_i)   }\int  \frac{d^3 p }{(2 \pi)^3}    2   \beta(\vk-\vp,\vp)\Theta^{(2)}_{\vp}\delta^{\rm in}_{\vk-\vp},
\end{align}
which after using \eqn{delta-2} and \eqn{theta-222} and in terms of \eqn{momentum-fun-3} - \eqn{momentum-fun-4}
 \begin{align} 
 &S_1^{(3)}(a,\vk)=\frac{  f_{+}(a)D_{+}(a)}{  C(a)   D_+(a_i)    }\iint \frac{d^3 p }{(2 \pi)^3   }  \frac{ d^3 q }{(2 \pi)^3} \bigg(\alpha^{\sigma}(\vk,\vp,\vq)\mG^{\delta}_{\sigma}(a)+\gamma^{\sigma}(\vk,\vp,\vq)\mG^{\Theta}_{\sigma}(a)\bigg)\delta^{\rm in}_{\vk-\vp}\delta^{\rm in}_{\vp-\vq}\delta^{\rm in}_{\vq}, \nonumber \\
 &S_2^{(3)}(a,\vk)=\frac{ f_{+}(a)D_{+}(a)}{   C(a)   D_+(a_i) }\iint \frac{d^3 p }{(2 \pi)^3 }  \frac{ d^3 q }{(2 \pi)^3}  \beta^{\sigma}(\vk,\vp,\vq)\mG^{\Theta}_\sigma(a)\delta^{\rm in}_{\vk-\vp}\delta^{\rm in}_{\vp-\vq}\delta^{\rm in}_{\vq},  \label{source-3rd-app}
\end{align}
where summation over upper and lower indices is assumed.

%%%%%%%%%%%%%%%%%%%%%%%%
%
%        smooth de
%
%
%%%%%%%%%%%%%%%%%%%%%%%%%

\section{Non-linear evolution with smooth dark energy} \label{smoothdesection}

In this section, we briefly discuss dark matter evolution in the presence of a smooth dark energy with $c_s^2 = 1$, typically called $w$CDM, which is a model that provides a familiar and simple example against which to compare our results.  Because it has been thoroughly discussed in the literature (see for example \cite{Sefusatti:2011cm, Sapone:2009mb, Amendola:2016saw}), we only give a brief explanation here.  The linear equation of a dark energy density with an arbitrary speed of sound, $c_s^2$, and equation of state $w$, is given as (neglecting relativistic corrections)
\be
\partial^2_{\tau}\delta_D+(1-3w)\cH\partial_{\tau}\delta_D+c_s^2k^2\delta_D+3(c_s^2-w)\bigg(\partial_{\tau}\cH+(1-3c_s^2)\cH^2\bigg)\delta_D=-(1+w)k^2\Phi  \  , 
\ee
where the conformal time $\tau$ is given by $ d \tau = d a / ( a \cH )$.  As we see, the sound speed introduces another natural momentum scale in the theory, the sound horizon $\cH / c_s$, besides the cosmological horizon,~$\cH$:
\begin{figure}[t!]
\begin{center}
\includegraphics[width=0.5\textwidth]{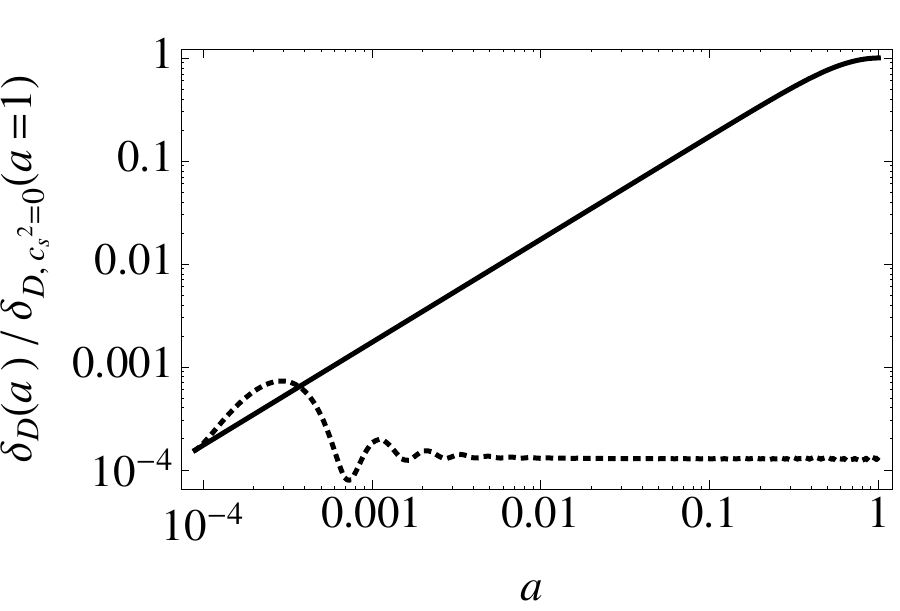}
\caption{In this plot, we show the growing modes for the dark-energy linear density contrast as a function of scale factor for two models, one with $c_s^2 = 0$ (clustering quintessence, the solid black curve), and the other with $c_s^2 = 1$ (smooth dark energy, the dotted black line), normalized to the value of the clustering density at the current time $a =1$.  The density contrast in the clustering case grows as $a \rightarrow 1$, while in the smooth case any initial fluctuations oscillate and are quickly damped.  In order to compare the growing modes, we assume that the two modes have the same size fluctuations at $a = 0.9 \times 10^{-4}$.  This time was chosen because, for a typical mode of interest for large scale structure, $k = 0.05 \unitsk$, we have $k \, \tau \approx 5$ (where $\tau$ is conformal time), which means that we are justified in dropping relativistic effects.} \label{fig-cluster-smooth} 
\end{center}
\end{figure}
\begin{itemize}
\item{Deep inside the sound horizon where $\frac{c_sk}{\cH}\gg1$, the density contrast has a damped oscillatory behavior and there is no growing solution. In this regime, sometimes called smooth dark energy, we can neglect $\delta_D$ compared to dark matter clustering.} 
\item{At super sound horizon scales $\frac{c_sk}{\cH}  \ll1$, however, the dark energy is in the clustering regime and $\delta_D$ has growing solutions. Since we are in the Newtonian regime in which $\frac{k}{\cH}\gg1$, dark energy clustering can only happen in the interval $1\ll\frac{k}{\cH}\ll\frac{1}{c_s^2}$ which requires a very small sound speed. This is the regime that we exhaustively studied in the previous section.}
\end{itemize}
In the limit that $c_s^2\sim1$, the sound horizon is very close to the cosmological horizon and the dark energy is spatially smooth.  In this regime, dark energy only affects gravitational growth of structure through changing the expansion rate (up to relativistic corrections). In the other limit with $| c_s^2| \ll1$, dark energy perturbations can cluster and contribute to the Poisson equation  as well. In Figure \ref{fig-cluster-smooth}, we present the time evolution of dark energy, $\delta_D$, in the clustering and smooth limits at linear level. This confirms that dark-energy fluctuations can be ignored, and that dark energy only affects the dark-matter clustering through changing the expansion rate.

 In $w$CDM, we have $\delta_A \approx \delta_m$, so that the equation of motion of the adiabatic mode is simply the standard one for dark matter (writing the adiabatic mode in $w$CDM as $\delta_w$) 
\begin{align} \label{nlcontinuityw}
& a \cH  \delta_w ( a, \kvec  )'  +  \theta_w ( a ,  \kvec )  = - \int \frac{d^3 q}{(2 \pi)^3}  \, \alpha ( \vq,\vk-\vq)  \theta_w ( a ,  \vq  ) \delta_w (a , \vk-\vq)  \\
& a \cH  \theta_w ( a , \kvec  ) '  + \cH \theta_w ( a , \kvec  ) + \frac{3}{2} \frac{ \Omega_{m,0} \cH_0^2 a_0 }{a}  \delta_w ( a ,  \kvec  )  =   9 \, ( 2 \pi) \, c_{s,w}^2 (a) H(a)^2 \frac{k^2}{\knl^2} \delta_w ( a , \kvec  )   \nn \\
& \hspace{2in} -   \int \frac{d^3 q}{(2 \pi)^3} \beta(\vq ,  \vk-\vq ) \theta_w ( a , \vk-\vq  ) \theta_w ( a , \vq ) .    \label{nleulerw}
\end{align}
The most important difference at late times between this model and $\Lambda$CDM is that $H(a)$, and therefore the linear growth rate $D_w$, has $w \neq -1$.  The equation for the growth factor is 
\be
\frac{d^2}{d\ln a^2}\bigg(\frac{D_w}{H}\bigg)+\bigg(2+3\frac{d\ln H}{d\ln a}   \bigg)\frac{d}{d\ln a}\bigg(\frac{D_w}{H}\bigg) + \frac{ d \ln H }{ d \ln a} \left( \frac{d \ln C }{ d \ln a}  - 1 + \frac{1}{ C } \right) \frac{ D_w }{ H } =0 \ ,
\ee
whose numerical solution can be obtained by either using CAMB or your favorite numerical solver.  Loops can then be computed in the same way as in the dark-matter case.  For the plots in Section \ref{resultssection}, we use the EdS approximation, which is as valid as it is in the dark-matter case: for example, with $w=-0.9$, $ ( \Omega_{m,0} \cH_0^2  a_0 / ( a \cH^2) ) / (a D_w' / D_w)^2 $ is unity at early times and is around $1.17$ near $a=1$.

{\footnotesize
\bibliography{references}}

\end{document}